\documentclass[twocolumn,times,trackchanges]{aastex63}
\usepackage{amssymb,amsmath, amsthm, amsfonts, graphics, graphicx}
\usepackage{subfiles}
\usepackage{float}

\usepackage{xcolor}

\listofchanges

\begin{document}

\title{Measuring Optical Extinction Towards Young Stellar Objects Using Diffuse Interstellar Bands}  
\author{Adolfo Carvalho}
\affiliation{Department of Astronomy; California Institute of Technology; Pasadena, CA 91125, USA}

\author{Lynne Hillenbrand}
\affiliation{Department of Astronomy; California Institute of Technology; Pasadena, CA 91125, USA}

\begin{abstract}
    Line-of-sight extinction estimates to well-studied young T Tauri and Herbig Ae/Be stars are based on many different measurements and analysis methods. This has resulted in wide scatter among the published $A_V$ values for the same star. In this work, we discuss the challenges in measuring extinction to actively accreting and especially outbursting young stellar objects (YSOs). We then explore a method not previously applied to young stars utilizing diffuse interstellar bands (DIBs).  In early-type stars, narrow correlations exist between DIB equivalent widths and the column density of interstellar material, and therefore the line-of-sight extinction. Here, we measure equivalent widths of the 5780 \AA\ and 6614 \AA\ DIB features in a sample of actively accreting YSOs, and apply a DIB-reddening calibration to estimate reddening and subsequently extinction. Our calibration is newly derived from a composite of available literature data and fully accounts for the scatter in these measurements. We also compare the DIBs-inferred optical line-of-sight extinction values with previous extinction estimates for our sample stars.  
\end{abstract}

\section{Introduction} \label{sec:intro}

Young stars begin their lives as deeply embedded objects. In addition to the molecular cloud and the collapsed cloud core out of which the star formed, there is an envelope of gas and dust. Eventually the star emerges, but the sightlines to young stars are complex, consisting of the low-density foreground interstellar medium, the higher density molecular cloud, and any circumstellar material \citep[see ][for a thorough discussion of the individual components of these sightlines]{mcjunkin_direct_2014}.  As a result, measuring the total line-of-sight extinction to the partially obscured underlying stellar source is challenging.

Although they are difficult to obtain, credible line-of-sight extinction measurements are critical for interpreting observations of young stellar systems. Extinction corrections are necessary for many important analyses of YSO properties, e.g., SED modelling, bolometric luminosity calculations, emission line flux measurements  which go into accretion rate and wind mass loss rate estimates, and X-ray luminosity estimates. Without good extinction corrections, the precision with which stellar and circumstellar properties can be determined suffers. Difficulty in measuring extinction precisely for both YSOs and their older cousin T Tauri and Herbig Ae/Be stars, contributes significantly to the historic inconsistencies in the estimated parameters of pre-main-sequence populations.

In addition to being situated along complex sightlines, YSOs and young stars are highly active.  Their spectra are dominated by contributions from large spots and often bright continuum veiling produced during accretion from hot inner disk gas. YSO behavior includes frequent flares \citep[e.g.][]{carvalho_JK_H4_2021}, ongoing accretion, and potentially large-amplitude outbursts \citep{Fischer_variability_ppvii_2022}. The time-variability in the continuum and line emission from a system can make it even more difficult to characterize the underlying photosphere and hence determine the line-of-sight extinction using traditional methods. 
Even quiescent-state measurements of extinction can be challenging, and uncertainties are large. 
A study of the literature on relatively calm and inactive T Tauri stars shows that extinction values for any given T Tauri star can vary by 50$\%$ or more, depending on the measurement method \citep[e.g.][see also \S\ref{sec:lit_ext} below]{mcjunkin_direct_2014, mcjunkin_empirically_2016}. 

One extinction measurement method that has not yet been applied to young stars uses diffuse interstellar bands \cite[see][for a review]{krelowski_diffuse_2018}. Hundreds of individual DIB absorption features have been identified in the optical and near-infrared spectral range. Among them, several bands are strong and are therefore well-known and well-studied, e.g., $\lambda$5780, $\lambda$5797, $\lambda$6270, $\lambda$6614. Although the DIBs carriers remain unidentified, the line strengths have been shown to correlate well with the column density of interstellar material along a given sight line \citep{friedman_studies_2011, vos_diffuse_2011, lan_exploring_2015}.  Such correlations have been demonstrated for many of the strongest lines listed above, and may be used to estimate interstellar extinction. 

A principal benefit of DIB measurements calibrated to extinction is that the DIB strengths are independent of continuum changes in the SED. Thus the DIBs can be usefully employed as extinction probes in quiescent, albeit still highly variable, as well as in flaring or outbursting YSOs. Another main advantage of DIBs measurements is avoidance of the need to know the source of the continuum, and specifically the fraction arising in the stellar photosphere vs in circumstellar contributions.

We have been able to identify strong DIBs in the spectra of about two dozen known outbursting young stellar objects, including many FU Ori stars. 
We present below the first application to young stellar objects of the established DIBs-based methodology for estimation of line-of-sight extinction. 
While the DIB line features are themselves not fully understood, they seem a promising way to estimate the extinction to these challenging young stellar objects for which few reliable $A_V$ measurements exist. The present effort highlights the potential utility of DIBs, in light of the difficulty in estimating YSO extinction through other methods. Our success with DIBs-based extinction measurements may serve as a guide for future work.  

The paper is organized as follows: 
\S \ref{sec:DIBsLit} provides  background on DIBs and their empirical correlation with interstellar reddening and extinction.
\S \ref{sec:lit_ext} examines the variety of extinction measurement methods in use for common T Tauri targets and the wide range of extinction values found in the literature for the same star. \S \ref{sec:sample} and \S \ref{sec:data} describe our young stellar object sample and the spectral data we used, \S \ref{sec:eqWs} describes the DIBs we identified and the equivalent width measurements we performed, and \S \ref{sec:Extincs} describes the conversion steps to reddening and subsequently extinction estimates. In \S \ref{sec:litcomp} we place our derived $A_V$s in the context of existing literature extinction measurements for our targets.  Finally, in \S \ref{sec:discussion} we discuss the implications of our work, and in \S \ref{sec:conclusion} we summarize our results.

Appendix \ref{appendix:Extincts} contains a summary of some of the major T Tauri star extinction efforts in the literature, elaborating on our discussion in \S \ref{sec:lit_ext}. 
In Appendix \ref{appendix:EqW} we show the DIBs regions of all of our targets and the limits of integration for our equivalent width measurements. Appendix \ref{sec:dib5797} contains measurements of the 5797 \AA\ DIB and a discussion of the sightlines to our targets.  
In Appendix \ref{appendix:rno1} we present a new determination of the spectral type of RNO 1.

\section{Background on DIBs and Their Strength in Different Environments}\label{sec:DIBsLit}

The DIBs have been studied extensively in early spectral type main sequence stars \citep[][etc.]{herbig_diffuse_1993, munari_diffuse_2008, hobbs_catalog_2008, friedman_studies_2011, vos_diffuse_2011, kos_diffuse_2013,kos_properties_2013,lan_exploring_2015, zasowski_mapping_2015,fan_behavior_2017}. Their carriers, however, are yet to be definitively identified. DIBs have not been studied much in the spectra of stars later than spectral type A, including in any low-mass pre-main sequence stars. This may be due to the larger number of stellar features that are present in the cooler, later spectral types, whereas for the earlier types, stellar line contamination is relatively minor compared to the broad DIBs features \citep[][]{hobbs_catalog_2008,friedman_studies_2011}.

Many studies correlate DIBs strength to column densities of interstellar gases such as $N(H)$ or $N(H_2)$, or directly to the reddening measurements of the targets \citep{friedman_studies_2011, vos_diffuse_2011, kos_properties_2013, lan_exploring_2015}. Comparing the DIBs strengths with the understood properties of the ISM along well-studied sight lines can illuminate how they behave in different environments. This can be critical for interpreting DIBs strength measurements in spectra of stars that do not lie along extensively studied sight lines, or in well-understood environments. 

An example of this is \citet{fan_behavior_2017}, which studies the molecular hydrogen fraction, $f_{H_2}$, in different environments where DIBs have been measured. \citeauthor{fan_behavior_2017} find that the strengths of some of the most well-studied DIBs, including those at 5780 $\text{\AA}$, 5797 $\text{\AA}$, and 6614 $\text{\AA}$, demonstrate a peak $W_{DIB}/E(B-V)$ ratio at around $f_{H_2} = 0.2$. This indicates that there may be a favorable $f_{H_2}$ for certain DIB carrier formation. They also find that in environments with greater values of $f_{H_2}$, the $W_{DIB}/E(B-V)$ ratio can be quite weak. 

This dependence of $W_{DIB}/E(B-V)$ on the local $f_{H_2}$ may help explain some of the outliers in the aforementioned correlations. It may also help explain the behavior of the $\sigma-\zeta$ effect, which has been identified in large DIBs studies of the ISM \citep{vos_diffuse_2011, kos_properties_2013, lan_exploring_2015}. The effect was first noted when \citet{krelowski_diffuse_1989} found that the sightlines to $\sigma$ Sco and $\zeta$ Oph had drastically different $W_{DIB}(5780)/W_{DIB}(5797)$ ratios. It has since been found that sight lines with varying $W_{DIB}(5780)/W_{DIB}(5797)$ ratios have different $W_{DIB}/E(B-V)$ ratios \citep{vos_diffuse_2011, kos_properties_2013, lan_exploring_2015}. The \citeauthor{fan_behavior_2017} result implies that the strength of the $\sigma-\zeta$ effect may scale with the sensitivity of the $\lambda$5780 and $\lambda$5797 line carriers to different $f_{H_2}$ environments. 
The sensitivity of the $W_{DIB}(5780)/W_{DIB}(5797)$ ratio to the molecular hydrogen fraction provides further evidence that the carriers of the two DIBs differ in their preferred environments. This may be a marker of the sensitivity of some carriers to their radiation environments, where denser environments which are also richer in molecular hydrogen may shield the carriers better. It may also or otherwise be an indication that some of the DIBs simply require denser environments for their formation mechanisms.

In light of these complex relationships between the DIBs and their environments, determining ISM properties directly from DIBs measurements may not be straightforward.
Nevertheless, DIBs equivalent widths may be able to provide extinction estimates  when there are none otherwise available.  This is often the case specifically for YSOs, which are particularly challenging in this regard. 
Indeed, for most of the sources in our sample, the DIBS method may be the best available.

\section{Established Methods for Estimating Extinction Towards Young Stars} \label{sec:lit_ext}

It is well known that 
extinction values reported in the literature for individual young stars can vary greatly for the same object, among the different sources of information.  Some of the purported differences are caused by astrophysical variability in the observational diagnostic 
used to infer extinction, e.g. photometric color.  Others are induced by systematics in the different methods that have been adopted to derive extinction estimates. 

As an illustration of the conflicting information that is in circulation for youn stars, we present in Table \ref{tab:TTSExts} a compilation of literature extinction values for a set of T Tauri stars. 
This sample is chosen for the repeated appearance of these same objects across varied works, and includes many of the best-studied young stars in one of the closest star forming regions, Taurus. 
A thorough discussion of the methods used to measure the extinctions in each reference given in Table \ref{tab:TTSExts}, is provided in Appendix \ref{appendix:Extincts}.

\begin{figure*}
    \centering
    \hspace{-0.15\linewidth}
    \includegraphics[origin=c, width= 1.05\linewidth]{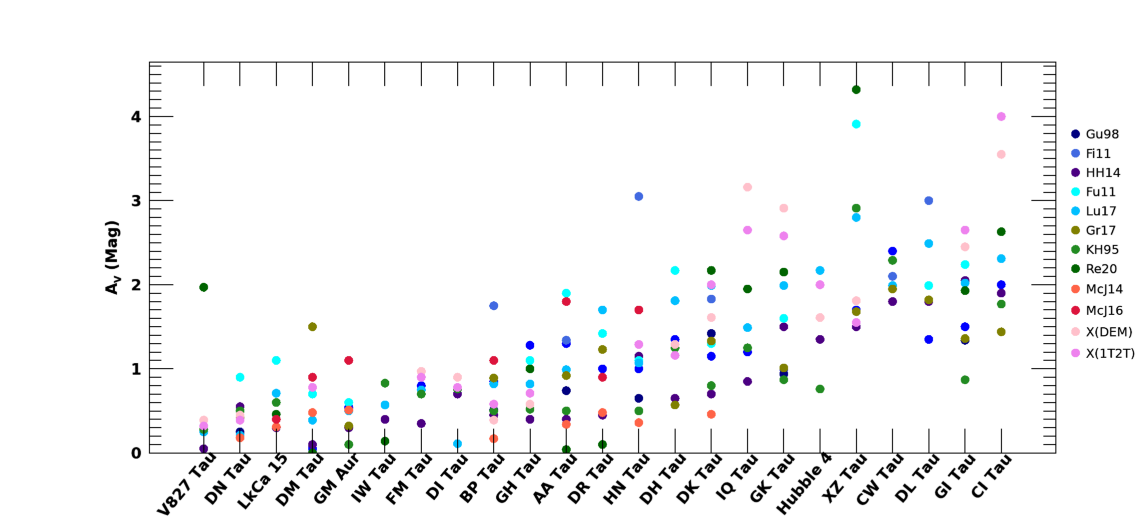}
	\caption{The data from Table \ref{tab:TTSExts}, with extinction source abbreviations explained in the table note.
	The objects are ordered left-to-right by increasing median extinction value.
	Similar hues connote similar methods used for the extinction measurement.    
	A wide range in the extinction estimates is exhibited for each of these stars, with about half the sample having a factor of two or more spread in $A_V$.
	}
	\label{fig:TTSLitExtsChart}
\end{figure*}

We present the results graphically in Figure \ref{fig:TTSLitExtsChart} to highlight the amplitude of the
discrepancies for individual sources. For the sample selected, 57 \% (13) of the targets have $\sigma(A_V) > 0.5$ mag, and 48 \% (11) of the targets have $\sigma(A_V) / \text{median}(A_V) > 0.5$.
While 30\% of the sample are known close ($<1$\arcsec) binary systems, the binaries are indicated in Table \ref{tab:TTSExts} and do not demonstrate significantly greater scatter in their literature $A_V$ values than the single component systems. 
Only 4 of the 13 targets with $\sigma(A_V) > 0.5$ mag and 3 of the 11 targets with $\sigma(A_V) / \text{median}(A_V) > 0.5$ are close binaries.

\begin{deluxetable*}{c|cccc|cc|ccc|cccc}
	\caption{Literature Measurements of Extinction to T Tauri Stars. Columns are ordered according to the measurement method.}\label{tab:TTSExts}
	\tablewidth{0pt}
	\tablehead{
	    \colhead{} & \multicolumn{4}{c}{Dust Column from Optical SED } & \multicolumn{2}{c}{Dust Column from IR SED} & \multicolumn{3}{c}{Opt/IR Color Excess} & \multicolumn{4}{c}{Gas Column Absorption}  \\ 
		\colhead{Target\tablenotemark{a}} & \colhead{V93} & \colhead{Gu98} 
		& \colhead{Fi11} & \colhead{HH14} & \colhead{Fu11} &
		\colhead{Lu17} & \colhead{Gr17} &  \colhead{KH95} & \colhead{Re20} & \colhead{McJ14} & \colhead{McJ16} & \colhead{X(DEM)} & \colhead{X(1T2T)}
	}
	\startdata
    AA Tau & 1.30 & 0.74 & 1.34 & 0.40 & 1.90 & 0.99 & 0.92 & 0.50 & 0.04 & 0.34 & 1.80 & 7.04 & 6.07 \\
    BP Tau & 0.85 & 0.51 & 1.75 & 0.45 & 1.10 & 0.82 & 0.89 & 0.50 & $\cdots$ & 0.17 & 1.10 & 0.39 & 0.58 \\
    CI Tau & 2.00 & $\cdots$ & $\cdots$ & 1.90 & 2.31 & 2.31 & 1.44 & 1.77 & 2.63 & $\cdots$ & $\cdots$ & 3.55 & 4.00 \\
    CW Tau & 2.40 & $\cdots$ & 2.10 & 1.80 & 1.99 & 1.99 & 1.95 & 2.29 & $\cdots$ & $\cdots$ & $\cdots$ & $\cdots$ & $\cdots$ \\
    DH Tau & 1.35 & $\cdots$ & $\cdots$ & 0.65 & 2.17 & 1.81 & 0.57 & 1.25 & $\cdots$ & $\cdots$ & $\cdots$ & 1.29 & 1.16 \\
    DI Tau$^*$ & $\cdots$ & $\cdots$ & $\cdots$ & 0.70 & $\cdots$ & 0.11 & 0.76 & 0.76 & $\cdots$ & $\cdots$ & $\cdots$ & 0.90 & 0.78 \\
    DK Tau$^*$ & 1.15 & 1.42 & 1.83 & 0.70 & 1.30 & 1.99 & 1.33 & 0.80 & 2.17 & 0.46 & $\cdots$ & 1.61 & 2.00 \\
    DL Tau & 1.35 & $\cdots$ & 3.00 & 1.80 & 1.99 & 2.49 & 1.82 & $\cdots$ & $\cdots$ & $\cdots$ & $\cdots$ & $\cdots$ & $\cdots$ \\
    DM Tau & 0.05 & $\cdots$ & $\cdots$ & 0.10 & 0.70 & 0.39 & 1.50 & 0.00 & 0.00 & 0.48 & 0.90 & $\cdots$ & 0.78 \\
    DN Tau & 0.50 & 0.25 & $\cdots$ & 0.55 & 0.90 & 0.21 & 0.42 & 0.50 & $\cdots$ & 0.18 & $\cdots$ & 0.45 & 0.39 \\
    DR Tau & 1.00 & $\cdots$ & $\cdots$ & 0.45 & 1.42 & 1.70 & 1.23 & $\cdots$ & 0.10 & 0.48 & 0.90 & $\cdots$ & $\cdots$ \\
    FM Tau & 0.80 & $\cdots$ & $\cdots$ & 0.35 & 0.70 & 0.75 & $\cdots$ & 0.70 & $\cdots$ & $\cdots$ & $\cdots$ & 0.97 & 0.90 \\
    GH Tau$^*$ & 1.28 & $\cdots$ & $\cdots$ & 0.40 & 1.10 & 0.82 & $\cdots$ & 0.52 & 1.00 & $\cdots$ & $\cdots$ & 0.58 & 0.71 \\
    GI Tau & 1.50 & 1.34 & $\cdots$ & 2.05 & 2.24 & 2.02 & 1.36 & 0.87 & 1.93 & $\cdots$ & $\cdots$ & 2.45 & 2.65 \\
    GK Tau & 1.00 & 0.94 & $\cdots$ & 1.50 & 1.60 & 1.99 & 1.01 & 0.87 & 2.15 & $\cdots$ & $\cdots$ & 2.91 & 2.58 \\
    GM Aur & 0.54 & 0.31 & $\cdots$ & 0.30 & 0.60 & 0.50 & 0.32 & 0.10 & $\cdots$ & 0.51 & 1.10 & $\cdots$ & $\cdots$ \\
    HN Tau$^*$ & 1.00 & 0.65 & 3.05 & 1.15 & 1.10 & 1.07 & $\cdots$ & 0.50 & $\cdots$ & 0.36 & 1.70 & $\cdots$ & 1.29 \\
    Hubble 4$^*$ & $\cdots$ & $\cdots$ & $\cdots$ & 1.35 & $\cdots$ & 2.17 & $\cdots$ & 0.76 & $\cdots$ & $\cdots$ & $\cdots$ & 1.61 & 2.00 \\
    IQ Tau & 1.20 & $\cdots$ & $\cdots$ & 0.85 & 1.49 & 1.49 & 1.25 & 1.25 & 1.95 & $\cdots$ & $\cdots$ & 3.16 & 2.65 \\
    IW Tau$^*$ & $\cdots$ & $\cdots$ & $\cdots$ & 0.40 & $\cdots$ & 0.57 & $\cdots$ & 0.83 & 0.14 & $\cdots$ & $\cdots$ & $\cdots$ & $\cdots$ \\
    LkCa 15 & $\cdots$ & $\cdots$ & $\cdots$ & 0.30 & 1.10 & 0.71 & $\cdots$ & 0.60 & 0.46 & 0.31 & 0.40 & $\cdots$ & $\cdots$ \\
    V827 Tau & $\cdots$ & $\cdots$ & $\cdots$ & 0.05 & $\cdots$ & 0.25 & $\cdots$ & 0.28 & 1.97 & $\cdots$ & $\cdots$ & 0.39 & 0.32 \\
    XZ Tau$^*$ & 1.70 & $\cdots$ & $\cdots$ & 1.50 & 3.91 & 2.80 & 1.68 & 2.91 & 4.32 & $\cdots$ & $\cdots$ & 1.81 & 1.55 \\
\enddata
    \tablecomments{Sources: V93: \citet{valenti_tts_1993}, Gu98: \citet{gullbring_disk_1998}, Fi11: \citet{fischer_characterizing_2011}, HH14: \citet{herczeg_survey_2014}, Fu11: \citet{furlan_spitzer_2011}, Lu17: \citet{luhman_survey_2017}, Gr17: \citet{Grankin_reliable_2017}, KH95: \citet{kenyon_hartmann_1995}, Re20: \citet{rebull_rotation_2020}, McJ14: \citet{mcjunkin_direct_2014}, McJ16: \citet{mcjunkin_empirically_2016}, X(DEM, 1T2T): \citet{gudel_xest_2007} using their DEM and 1 or 2 thermal source (1T2T) models.}
    \tablenotetext{a}{Targets marked with an asterisk are known close ($<1$\arcsec) binaries.}
\end{deluxetable*}

The standard extinction techniques can be divided into two general types: those based on dust continuum extinction measurements and those using gas column measurements. Dust-extinction measurements rely on assumptions about the emission source that is used to establish a truth spectrum against which to measure a color or a flux excess. Gas column measurements use gas line strengths measured against a continuum, and gas-to-dust conversion relations to compute the corresponding extincting dust column.

The dust-based extinction measurements are interpretations of continuum emission, and share the common flaw that they rely heavily on assumptions made about the underlying emitting source. In addition to the stellar photosphere, which may be young enough that the spectrum, colors, and even spectral energy distribution could exhibit gravity effects, there may be activity effects ranging from spottedness to accretion.  The former can alter spectra and colors depending on the spot temperature contrast and covering fraction.  The latter can add a hot (ultraviolet and optical wavelengths) or cool (infrared) continuum. 
Placing extincted, reddened young stars both accurately and precisely on the HR diagram has long proved difficult due to the significant consequence of extinction estimate uncertainties, mainly on the derived source luminosities. 

The gas-based measurements, by contrast, are more robust to assumptions regarding the underlying, unextincted source. They rely only on the accuracy of the gas line measurement against a continuum.
For some species such as the \ion{Na}{1} D lines discussed by \cite{Pascucci_gasline_2015},
there is a need to disentangle stellar photospheric from interstellar contributions.
There is a strong dependence for gas-based extinction estimates on assumptions about the interstellar medium along the line of sight to the target, and the relationship between the gas being measured and the well-mixed extincting dust. Along lines of sight where these relationships are less well known, the assumptions can break down and the extinction measurements become less reliable. 

While DIBs-based extinction measurements are similar to the gas line-based measurements described above, both being line-of-sight absorption techniques, there are some important differences. Most notable among these is the fact that although there can be contamination from other gas species absorbing at the same wavelengths (as discussed in detail below; see  \S\ref{sec:StelContam}), the DIBs themselves do not arise within any known stellar photosphere. Thus the column measured by the DIBs is exclusively circumstellar or interstellar. This is a significant advantage for extinction measurements to YSOs because of the previously-discussed challenges to modeling the underlying photospheric spectrum.

\section{Motivation and Definition of Sample}\label{sec:sample}

Our initial interest in the topic of this investigation was driven by the observation 
that DIBs features are often prominent in outbursting YSOs -- sources which have recently increased their brightness by factors of ten to hundreds.
This included several objects studied and published as YSO outbursts by the second author, as well as 
noting the presence of DIBs in YSO spectra published by others,  e.g. \cite{SA2020}.  
Motivated by these findings, we initiated a more systematic investigation into the presence of DIBs features in young stars.

For purposes of testing and calibrating our methodology,
we first sought to identify DIBs features in spectra of well-studied T Tauri stars. However, we were unable to distinguish and hence measure the DIBs due to the strong stellar contamination from photospheric \ion{Fe}{1} and other lines near the known DIBs wavelengths that are common in later spectral types. We found that for spectral types later than G0, the contamination is not negligible and requires careful analysis to remove (see \S \ref{sec:StelContam} below). Due to the aforementioned difficulty in establishing the true photospheric spectrum for young stars, in addition to the difficulties involved in continuum-fitting highly reddened T Tauri stars with numerous line complexes, such a procedure would introduce a large amount of uncertainty in a DIB equivalent width measurement. 

A brief search for DIBs in hotter young stars, specifically Herbig Ae/Be stars, turned up some reports of them in the hottest members of the class \citep{ellerbroek_DIBs_2013, ramirez-tannus-HAeBeDIBs_2018}, but not many others.  Furthermore, the DIBs appearing in Ae/Be spectra to which we do have access were extremely weak.

Despite the apparent rarity and/or weakness of DIBs in well-studied YSOs, 
we have been able to identify strong DIBs in the spectra of many outbursting YSOs, including FU Ori type objects.
Our assembled DIBs-bearing sample includes both ``classical" FU Ori stars
(e.g. FU Ori, V1057 Cyg, V1515 Cyg, BBW 76), and more recent 
photometric outbursts that have been identified with young stars.
Several of these are indeed bona fide FU Ori events
(e.g. V2493 Cyg / HBC 722, V899 Mon, V960 Mon, Gaia 17bpi).
Others represent different flavors of episodic accretion in young stars
(e.g. LkHa 225 S, V582 Aur, V1331 Cyg, iPTF 15afq).
Still others may not be young stars at all (e.g. ASASSN 15qi, PTF 14jg). We have included one non-outburster in the sample, RNO 1A. 

The full sample of sources for which we measured DIBs is listed in Table \ref{tab:OutburstersLocs}. We include the signal-to-noise ratios (SNRs) of our spectral data, common aliases of the targets, and their locations in J2000 coordinates. The stellar properties of targets in the sample are amassed from previous literature and listed in Table \ref{tab:OutburstersProperties} for reference. For many of the targets, there is disagreement in the literature about spectral type and measured bolometric luminosity. This is in large part due to the difficulty in identifying the extinction to the sources, which is driven largely by the challenges in disambiguating the contributions of circumstellar emission and absorption, from the underlying stellar photospheric emission and absorption.

\section{Data} \label{sec:data}
\subsection{Target Spectra} \label{sec:targ_spec}
Spectra used in this analysis were acquired at the W.M. Keck Observatory
Keck I telescope with the HIRES \citep{vogt_hires_1994} high dispersion spectrograph.
Several different spectrograph setups were used, as dictated by the main science program 
for each observing night. All setups covered both primary DIBs regions under
investigation here.  The 5780 \AA\ feature was usually in the ``blue" CCD
of HIRES, while the 6614 \AA\ feature was always in the ``green" CCD of HIRES.
The spectral resolution ranged between 25,000 and 60,000 depending
on the decker width. Data were taken over several years, between 2008 and 2020.

\begin{figure*}
    \centering
    \includegraphics[width=0.485\linewidth]{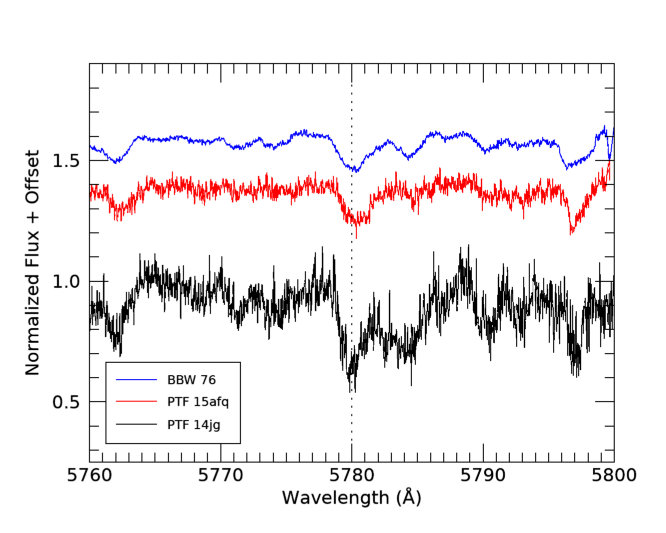}
    \includegraphics[width=0.485\linewidth]{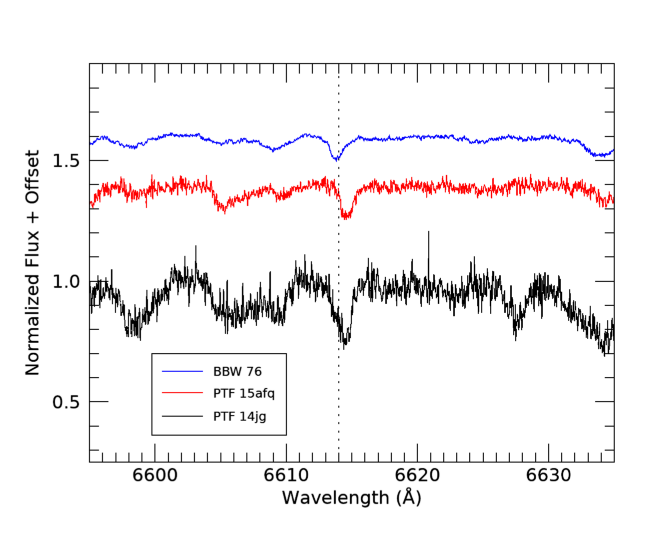}
    \caption{The spectral vicinity of the two strong DIBs features analyzed in this work, shown for BBW 76, iPTF 15afq, and PTF 14jg. These three sources represent the SNR range in our spectra, with high (180), medium (40), and low (8) SNR values respectively. The vertical dotted lines mark nominal DIB line centers at 5780 \AA\ and 6614 \AA, though DIBs profiles are known in the literature to vary slightly from source to source. The $\lambda$5797 DIB is also visible in these examples, with further discussion of this DIB provided in Appendix \ref{sec:dib5797}.}
    \label{fig:HIRESSamples}
\end{figure*}

All spectra were processed with the MAKEE pipeline reduction package (the 2008 version for older epochs and the 2019 version for newer data)\footnote{ {\url{https://sites.astro.caltech.edu/~tb/makee/}}} 
written by Tom Barlow.   The final spectra are in units of counts or ADU vs wavelength, in the heliocentric frame.
 The SNR, reported in Table \ref{tab:OutburstersLocs}, was measured by choosing a region of continuum near the DIBs of interest and dividing the median of the spectrum in that region by the standard deviation. The two values were then averaged. The regions selected were 5765-5770 \AA\ and 6620-6625 \AA\, though the exact range varied from target-to-target due to the diverse spectral-types present in the sample. The SNR varies greatly among the spectra, ranging from 8 (PTF 14jg) to 180 (BBW 76). 
Individual orders were normalized by fitting a 5th order polynomial to the regions around large absorption or emission lines. 
Examples of each of the two DIBs regions are shown in Figure \ref{fig:HIRESSamples}, showcasing high, typical, and low SNRs.

We apply our best estimate of a stellar-frame RV correction where possible. In systems with literature values, we use those. For targets without literature values, we use stellar lines near the DIBs as a reference, but do not apply a correction to the spectra. In practice, the accounting for RV affects only the figure presentation, as all DIB and stellar features were measured over their full range and not using fixed wavelength ranges.

\subsection{Standard Spectra} \label{sec:stand_spec}
The G2, K0 and K4 spectral standard stars used to assess stellar contamination to the DIBs features in Section \ref{sec:StelContam}, were acquired at the McDonald Observatory's 2.7m Harlan J. Smith Telescope using the Robert G. Tull cross-dispersed coud\'e echelle spectrograph \citep{Tull_spec_1995}. The $\sim$110 spectra of each standard were taken from 2004-2019, roughly once per night over 5-10 night observing runs per year. 

A spectral resolution of $R \equiv \lambda/\delta\lambda \sim60000$ was achieved using a 1.2\arcsec\ slit. The spectra were cross-dispersed into 54-55 orders and recorded on a Tektronix 2080x2048 CCD. Wavelength scales and instrumental radial velocity shifts determined by reference to a Thorium-Argon comparison lamp spectrum taken before and after stellar spectra. The spectra were reduced using a custom IDL routine similar to the procedures described in \citet{valenti_spec_1994} and \citet{hinkle_arcturus_2000}. 

The final continuum normalization for individual orders is accomplished by fitting a second order polynomial to the regions surrounding the stellar lines of interest. To boost the overall SNR of the standard spectra, we coadd all observations of the same source achieving final SNR values $>500$ for the reference stars.

\begin{deluxetable*}{ccccc}
	\tablecaption{Sample names, aliases, J2000 coordinates, and SNRs \label{tab:OutburstersLocs}}
	\tablewidth{0pt}
	\tablehead{
        \colhead{Target} & \colhead{Alias} & \colhead{RA} & \colhead{Dec} & \colhead{SNR} \\
	}
	\startdata
    \multicolumn{5}{c}{Young Star Outbursters } \\
    \hline
    BBW 76 &  V646 Pup & 07h50m35.5s & -33d06m24.00s & 181 \\
    FU Ori & Nova Ori 1939, HBC 186 & 05h45m22.4s & +09d04m12.30s & 144 \\
    Gaia 17bpi & $\cdots$ & 19h31m05.6s & +18d27m52.23s & 20 \\
    Gaia 19ajj & $\cdots$ & 08h08m53.0s & -35d55m33.14s & 16 \\
    IRAS 06068-0641\tablenotemark{*} & V899 Mon & 06h09m19.2s & -06d41m55.89s & 128 \\
    LkHa 225 S & V1318 Cyg S & 20h20m30.6s & +41d21m26.34s & 38 \\
    2MASS J06593158-0405277\tablenotemark{*}  & V960 Mon  &06h59m31.6s & -04d05m27.76s & 148 \\
    PTF 10qpf  & V2493 Cyg, HBC 722 & 20h58m17.0s & +43d53m43.34s & 106 \\
    iPTF 15afq & Gaia 19fct & 07h09m21.4s & -10d29m34.51s & 41 \\
    RNO 1B & V710 Cas A & 00h33m51.8s & +63d12m22.91s & 6 \\
    V582 Aur & $\cdots$ & 05h25m52.0s & +34d52m30.10s & 92 \\
    V733 Cep & Persson's star & 22h51m36.6s & +62d16m24.80s & 7 \\
    V1057 Cyg &  LkH$\alpha$ 190 & 20h58m53.7s & +44d15m28.38s & 112 \\
    V1331 Cyg & LkH$\alpha$ 120 & 21h01m09.2s & +50d21m44.80s & 55 \\
    V1515 Cyg &  HBC 692 & 20h23m48.0s & +42d12m25.78s & 85 \\
    V900 Mon & $\cdots$ & 06h57m22.2s & -08d23m17.68s & 10 \\
    Z CMa &  MWC 165, HBC 243 & 07h03m43.1s & -11d33m06.23s & 180 \\
    \hline
    \multicolumn{5}{c}{Other Photometric Variables with DIBs } \\
    \hline
    ASASSN 15qi & $\cdots$ & 22h56m08.8s & +58d31m04.13s & 28 \\
    PTF 14jg & $\cdots$ & 02h40m30.1s & +60d52m45.50s & 8 \\
    PTFS1821n & [NSW2012] 284 & 21h38m39.8s & +57d08m47.11s & 50 \\
    RNO 1A & RNO1, KW97 3-17 & 00h33m57.5s & +63d13m00.12s & 60 \\
    VES 263 & Gaia 18azl & 20h31m48.8s & +40d38m00.06s & 40 \\
\enddata
\tablenotetext{*}{We refer to these by their more common aliases hereafter.}
\end{deluxetable*}

\begin{deluxetable*}{c|cc|cc|ccc|cc|cc}
	\tablecaption{Literature properties of outbursters in our sample.\label{tab:OutburstersProperties}}
	\tablewidth{0pt}
	\tablehead{
	    \colhead{Target} & \colhead{System RV} & 
		\colhead{Ref} & \colhead{Optical SpT} & \colhead{Ref} & \colhead{$A_V$} & \colhead{Ref} & \colhead{ X-Ray $A_V$\tablenotemark{a}} & \colhead{$L_{\text{quiescent}}$} & \colhead{Ref} & \colhead{$L_{\text{burst}}$} & \colhead{Ref} 
	}
	\startdata
	\multicolumn{12}{c}{Young Star Outbursters } \\
    \hline
    BBW 76 & 30 & 5 & G0-G2I\tablenotemark{b}& 5 & 2.1 & 5 & $\cdots$ & $\cdots$ & $\cdots$ & 453\tablenotemark{f} & 5 \\
    FU Ori & 28 & 6 & G0 & 8 & 1.4 & 6 & 81 & $\cdots$ & $\cdots$ & 66 & 4 \\
    Gaia 17bpi & 4.5 & 10 & GK\tablenotemark{d} & 10 & 3 & 10 & 4.1 & 0.3 & 10 & 7.5 & 10 \\
    Gaia 19ajj & 35.8 & 26 & $\cdots$ & $\cdots$ & 6 & 26 & $\cdots$ & $\cdots$ & $\cdots$ & $\cdots$ & $\cdots$ \\
    V899 Mon & 0.92 & 3 & K/M & 21 & 2.6-4.5 & 3 & $\cdots$ & 162 & 3 & 419 & 3 \\
    LkHa 225 S & -13.5 & 32 & A2V & 9 & 7.2 & 9 & $\cdots$ & $\cdots$ & $\cdots$ & 750 & 9 \\
    V960 Mon & 38.1 & 14 & G2-K1 & 11 & 1.5 & 4 & 2.6 & 4.8 & 15 & 48 & 4 \\
    PTF 10qpf & ~1.6 & 13 & GI & 14 & 3.4 & 16 & 4.1 & 2.69 & 16 & 12 & 13 \\
    iPTF 15afq & $\cdots$ & $\cdots$ & F5-G0 & 12 & $\cdots$ & $\cdots$ & $\cdots$ & $\cdots$ & $\cdots$ & 111 & 2 \\
    RNO 1B & $\cdots$ & $\cdots$ & F8II & 27 & 14.5 & 4 & $\cdots$ & 527 & 28 & 1652 & 4 \\
    V582 Aur & $\cdots$ & $\cdots$ & G0I & 19 & 6.5,5.6 & 18,4 & $\cdots$  & $\cdots$ & $\cdots$ & 500\tablenotemark{g} & 18\\
    V733 Cep & $\cdots$ & $\cdots$ & G5-G6Ib\tablenotemark{b} & 30 & 11.5 & 4 & $\cdots$  & $\cdots$ & $\cdots$ & 43\tablenotemark{h} & 4\\
    V1057 Cyg & -16 & 6 & F7-G3 & 6 & 2.4-3.9 & 6,4 & 4.1  & $\cdots$ & $\cdots$ & 100\tablenotemark{h} & 4 \\
    V1331 Cyg & $\cdots$ & $\cdots$ & G5 & 20 & 2.4 & 25 & $\cdots$ & $\cdots$ & $\cdots$ & 53 & 22 \\
    V1515 Cyg & -12 & 23 & G2-G5 & 24 & 3.5 & 4 & 4.1 & $\cdots$ & $\cdots$ & 200 & 22 \\
    V900 Mon & $\cdots$ & $\cdots$ & $\cdots$ & $\cdots$ & 13.5 & 4 & 4.1 & $\cdots$ & $\cdots$ & 99 & 4 \\
    Z CMa & -30 & 33 & G2I & 21 & 7.1 & 4 & 4.1 & $\cdots$ & $\cdots$ & 3548\tablenotemark{h} & 4 \\
    \hline
    \multicolumn{12}{c}{Other Photometric Variables with DIBs } \\
    \hline
    ASASSN 15qi & -58.1 & 1 & F8 & 7 & 3.6 & 7 & $\cdots$ & 18 & 7 & 1000 & 7 \\
    PTF 14jg & -38.1 & 17 & A \tablenotemark{b}/late F \tablenotemark{c}/G-K \tablenotemark{d}  & 17 & 4.75 & 17 &  0.1-0.3 & $\cdots$ & $\cdots$ & 130 & 17  \\
    PTFS1821n & $\cdots$ & $\cdots$ & B1\tablenotemark{e} & $\cdots$ & $\cdots$ & $\cdots$ & $\cdots$ & $\cdots$ & $\cdots$ & $\cdots$ & $\cdots$ \\
    RNO 1A & 6.61 & 29 & F5III\tablenotemark{*} & 27 & 3.6 & 27 & $\cdots$ & $\cdots$ & $\cdots$ & $\cdots$ & $\cdots$ \\
    VES 263 & -4.1 & 31 & B1 & 31 & 5.6 & 31 & $\cdots$ & 14000 & 31 & $\cdots$ & $\cdots$ \\
\enddata
    \tablecomments{Literature Sources: (1) \citet{hillenbrand_optical_2015}, (2) \citet{sewilo_identifying_2019}, (3) \citet{ninan_v899_2015}, (4) \citet{connelley_near-infrared_2018}, (5) \citet{reipurth_evolution_2002}, (6) \citet{herbig_high-resolution_2003}, (7) \citet{herczeg_eruption_2016}, (8) \citet{kenyon_flickering_2000}, (9) \citet{magakian_new_2019}, (10) \citet{hillenbrand_gaia_2018}, (11) \citet{park_high-resolution_2020}, 
    (12) \citet{hillenbrand_afq_2019}, 
    (13) \citet{miller_evidence_2011}, (14) \citet{takagi_spectroscopic_2018}, (15) \citet{kospal_progenitor_2015}, (16) \citet{cohen_observational_1979}, (17) \citet{hillenbrand_ptf14jg_2019}, (18) \citet{zsidi_weakening_2019}, (19) \citet{semkov_photometric_2013}, (20) \citet{hamann_persson_1992III}, 
    (21) \citet{Hartmann_zcma_1989}, 
    (22) \citet{sandell_similarity_2001}, (23) \citet{herbig_eruptive_1977}, (24) \citet{kolotilov_studies_1983}, (25) \citet{chavarria_study_1981}, (26) \citet{hillenbrand_gaia_2019}, (27) \citet{staude_rno_1991}, (28) \citet{skinner_chandra_2020}, (29) \citet{gaia_collaboration_gaia_2018}, (30) \citet{reipurth_v733_2007}, (31) \citet{munari_2018_2019}, (32) \citet{Hillenbrand_2022_lkha225s}, (33) \citet{SA2020}}
    \tablenotetext{*}{We derive a spectral type closer to A0 or F8 for RNO 1A, see Appendix \ref{appendix:rno1} for more details.}
    \tablenotetext{a}{$A_V$s from \citet{kuhn_comparison_2019}, derived from XMM-Newton X-Ray spectra or adopted for their analysis.}
    \tablenotetext{b}{According to characteristics of a blue spectrum}
    \tablenotetext{c}{According to characteristics of a green spectrum}
    \tablenotetext{d}{According to characteristics of a red spectrum}
    \tablenotetext{e}{Determined by similarity to the optical spectrum VES 263 in the 5000-7500 \AA\ range.}
    \tablenotetext{f}{\citet{reipurth_evolution_2002} assume a distance of 1800 pc to the source. New Gaia parallax measurements give the distance to BBW 76 as 1092 pc, which decreases the value to 166 $L_\odot$}
    \tablenotetext{g}{May be 1050 $L_\odot$, depending on the association, whether V582 Aur belongs to Auriga OB1 or Auriga OB2 \citep{zsidi_weakening_2019}}
    \tablenotetext{h}{Work by \citet{kuhn_comparison_2019} using Gaia DR2 parallaxes places V733 Cep at 825 pc, so the updated luminosity is 40$L_\odot$. For V1057 Cyg at 795 pc, the updated luminosity is 209$L_\odot$, and for Z CMa at 1120 pc, the updated luminosity is 4540$L_\odot$.}
\end{deluxetable*}

\section{DIBs Equivalent Widths} \label{sec:eqWs}
We focus our analysis on the $\lambda$5780 and $\lambda$6614 DIBs, as they appear most strongly and consistently in our sample. The $\lambda$5797 DIB also appears in our targets but is on the edge of a spectral order and therefore measurements of its equivalent width are less reliable. We discuss our measurement of the $\lambda$5797 DIB equivalent width further in Appendix \ref{sec:dib5797}. The $\lambda$5780 and $\lambda$6614 DIBs equivalent width measurements were made in two primary steps, which are explained in detail below. First, we describe how we performed the spectral equivalent width measurements by direct integration. Second, we discuss the stellar contamination correction we derived and applied to the equivalent widths.  Corrections are significant in 25\% of the sample and, with the exception of 4 strongly contaminated targets (FU Ori, Gaia 19ajj, PTF 14jg, and V1515 Cyg), amount to a $\lesssim$ 40\% effect.

\subsection{Equivalent Width Measurements} \label{sec:eqws_meas}
DIBs lines are generally broader than stellar lines, with complex asymmetric profiles. 
As such, considerable care is needed in measuring the DIBs line strengths.  
To avoid assumptions about the profiles (such as the common gaussian approximation to an irregular shape), 
and  to prevent over- or under-estimating the line strength in a fitting process, we 
measure the line equivalent widths manually, through direct integration. 
We rely on our previous normalization to establish the continuum level, which we assume to be at 1.0.
For the noisiest spectra (SNR $\lesssim$ 20), we apply a $\sigma = 10$ pixel Gaussian smoothing kernel to identify the integration limits for our equivalent widths, but we then perform the integrations on the non-smoothed spectra.

\begin{figure*}
    \centering
    \includegraphics[width=0.485\linewidth]{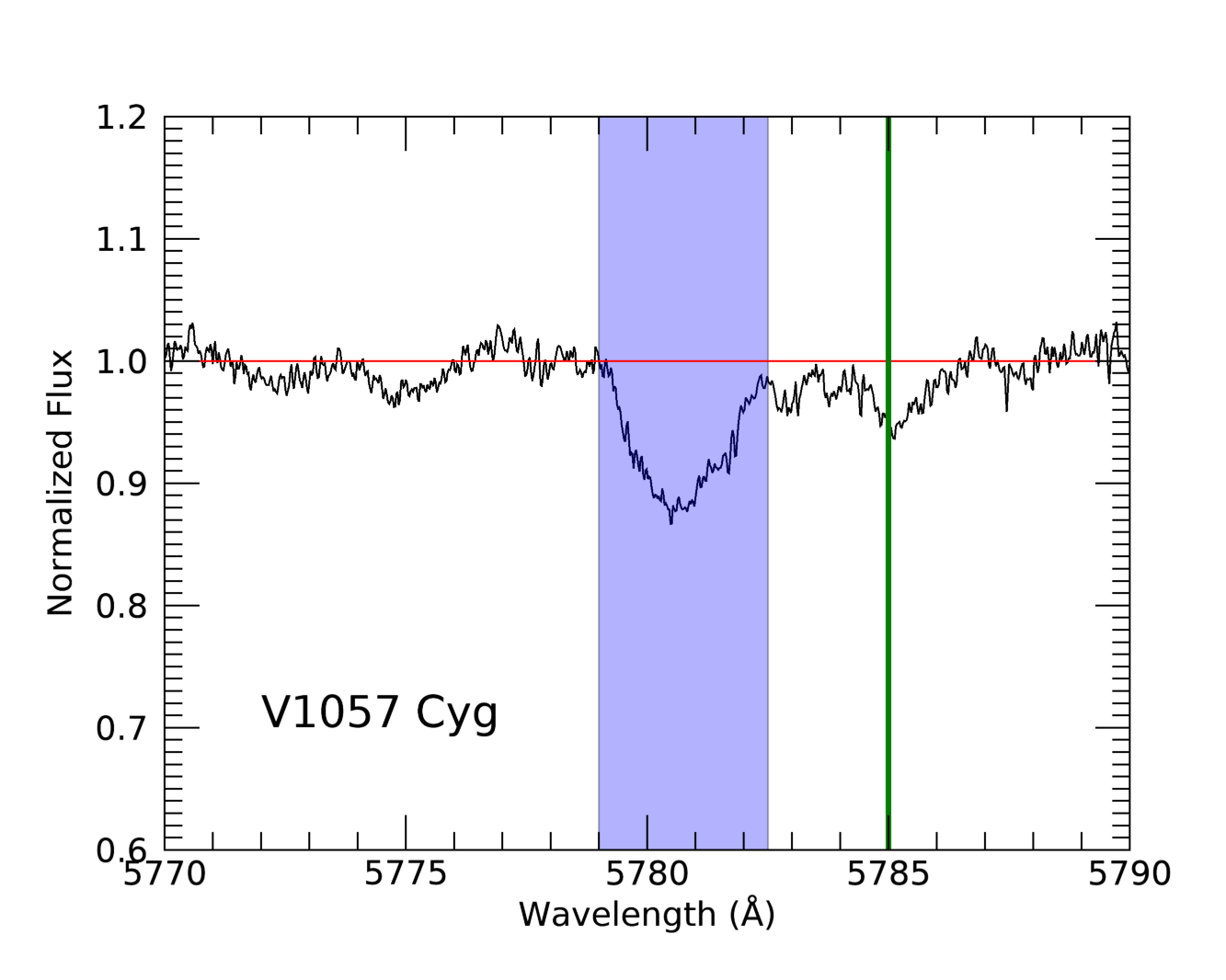}
    \includegraphics[width=0.485\linewidth]{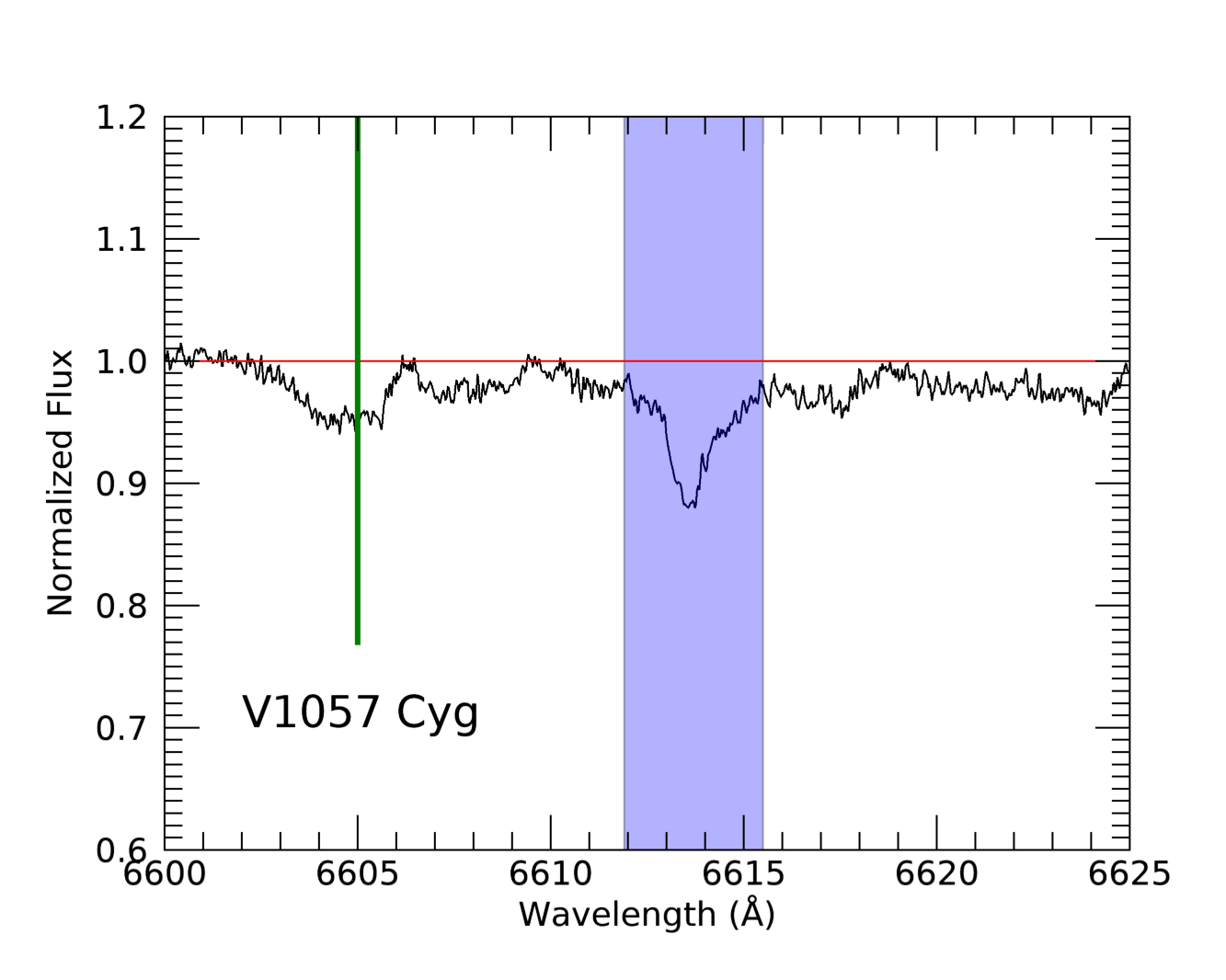}
    \caption{The integration limits for the $\lambda$5780 and $\lambda$6614 DIBs in V1057 Cyg. The blue band shows the integration limits for the equivalent width measurement. 
    The red line marks the continuum for reference. The green vertical line shows the location of the stellar line complexes used for stellar contamination corrections within the DIB region integration limits. Similar plots for the rest of the sample are shown in Appendix \ref{appendix:EqW}.}
    \label{fig:eqWLimitsV1057Cyg}
\end{figure*}

In Figure \ref{fig:eqWLimitsV1057Cyg}, we present one example of the equivalent width measurement, shown for V1057 Cyg and each of the two DIBs features studied in this work. 
We illustrate the DIBs profiles for our entire sample in corresponding plots that appear in Appendix \ref{appendix:EqW}.

We adopt an uncertainty that assumes our lines are relatively weak, so that the average flux in the integration band is comparable to the continuum flux. Although this may not be true for the strongest DIBs, the sense of the assumption is that it overestimates our uncertainties. In this weak line limit, we can then use Equation 3 from \citet{vollmann_remarks_2006},

\begin{equation}
    \sigma(W_\lambda) = \sqrt{2} \frac{(\Delta \lambda - W_\lambda)}{SNR}    
\end{equation}
where $\Delta \lambda$ is the width of the integration band and $SNR$ is the signal to noise ratio of the continuum. 

\begin{deluxetable*}{c|cccc|cccc}
	\caption{Equivalent width measurements for the $\lambda$5780 and $\lambda$6614 DIBs}\label{tab:DIBsEqWs}
	\tablewidth{0pt}
	\tablehead{
		\colhead{Target} & \colhead{$W_\lambda(5780)$} & \colhead{$\sigma_\lambda(5780)$} & \colhead{$W_*(5780)$} & \colhead{$W_{DIB}(5780)$} &  \colhead{$W_\lambda(6614)$} 	& \colhead{$\sigma_\lambda(6614)$} & \colhead{$W_*(6614)$} & \colhead{$W_{DIB}(6614)$}  \\
		\colhead{} & \colhead{\AA} & \colhead{\AA} & \colhead{\AA} & \colhead{\AA} & \colhead{\AA} &  \colhead{\AA} 	& \colhead{\AA} & \colhead{\AA} } 
	\startdata
    \multicolumn{9}{c}{Young Star Outbursters } \\
    \hline
    BBW 76 &                0.365 &                0.033 &                \textit{0.139} &                0.226 &                0.133 &                0.023 &                \textit{0.030} &                0.103 \\ 
    FU Ori &                0.272 &                0.031 &                \textit{0.228} &                0.044 &                0.059 &                0.036 &                \textit{0.057} &                0.023 \\ 
    Gaia 17bpi &                0.709 &                0.299 &                0.214 &                0.495 &                0.365 &                0.184 &                0.121 &                0.181 \\ 
    Gaia 19ajj &                0.168 &                0.228 &                0.160 &                0.008 &                0.051 &                0.115 &                0.018 &                0.033 \\ 
    V899 Mon &                0.251 &                0.046 &                \textit{0.075} &                0.176 &                0.083 &                0.023 &                0.015 &                0.068 \\ 
    LkH$\alpha$ 225 S &                0.458 &                0.137 &                0.092 &                0.366 &                0.163 &                0.083 &               (-0.003) &                0.080 \\ 
    V960 Mon &                0.600 &                0.046 &                \textit{0.137} &                0.463 &                0.231 &                0.026 &                \textit{0.077} &                0.154 \\ 
    PTF 10qpf &                0.492 &                0.050 &                \textit{0.241} &                0.238 &                0.069 &                0.035 &                0.014 &                0.055 \\ 
    iPTF 15afq &                0.381 &                0.136 &                0.071 &                0.310 &                0.168 &                0.091 &                \textit{0.097} &                0.071 \\ 
    RNO 1B &                1.099 &                0.742 &                0.433 &                0.666 &                0.354 &                0.738 &                0.054 &                0.300 \\ 
    V1057 Cyg &                0.256 &                0.041 &                \textit{0.063} &                0.193 &                0.134 &                0.041 &                0.034 &                0.100 \\ 
    V1331 Cyg &                0.074 &                0.071 &               (-0.013) &                0.087 &                0.076 &                0.059 &               (-0.015) &                0.091 \\ 
    V1515 Cyg &                0.518 &                0.047 &                \textit{0.315} &                0.203 &                0.148 &                0.041 &                \textit{0.056} &                0.092 \\ 
    V582 Aur &                0.503 &                0.057 &                \textit{0.127} &                0.376 &                0.147 &                0.043 &                0.012 &                0.135 \\ 
    V733 Cep &                0.918 &                0.746 &                0.280 &                0.638 &                0.354 &                0.585 &                0.064 &                0.290 \\ 
    V900 Mon &                0.618 &                0.607 &                0.265 &                0.353 &                0.297 &                0.411 &                0.064 &                0.233 \\ 
    Z CMa &                0.142 &                0.023 &                0.002 &                0.140 &                0.048 &                0.014 &               (-0.013) &                0.061 \\ 
    \hline
    \multicolumn{9}{c}{Other Photometric Variables with DIBs} \\
    \hline
    ASASSN 15qi &                0.642 &                0.182 &                0.052 &                0.590 &                0.311 &                0.171 &                0.049 &                0.262 \\ 
    PTF 14jg\tablenotemark{*} &                0.723 &                0.409 &                \textit{0.792} &                0.0 &                0.364 &                0.604 &                0.208 &                0.156 \\ 
    PTFS1821n &                0.713 &                0.142 &                0.052 &                0.661 &                0.273 &                0.102 &                0.021 &                0.252 \\ 
    RNO 1A &                0.444 &                0.097 &                0.027 &                0.417 &                0.215 &                0.053 &                0.014 &                0.201 \\ 
    VES 263 &                0.828 &                0.139 &                0.054 &                0.774 &                0.290 &                0.095 &               (-0.002) &                0.292 \\ 
\enddata
    \tablecomments{The values of $W_*$ marked in italics are those for which the estimated stellar contamination exceeds the 1$\sigma$ uncertainties reported in the previous column. For these sources, the correction to the measured equivalent width is significant.}
    \tablenotetext{*}{The $W_{DIB}(5780)$ value would be negative here, possibly due to overestimation of the stellar contamination to the DIB. The large positive $W_{DIB}(6614)$ value indicates this may be the case. }
\end{deluxetable*}

\subsection{Correcting for Stellar Line Contamination} \label{sec:StelContam}
\begin{figure*}
    \centering
    \includegraphics[width = 0.485\linewidth]{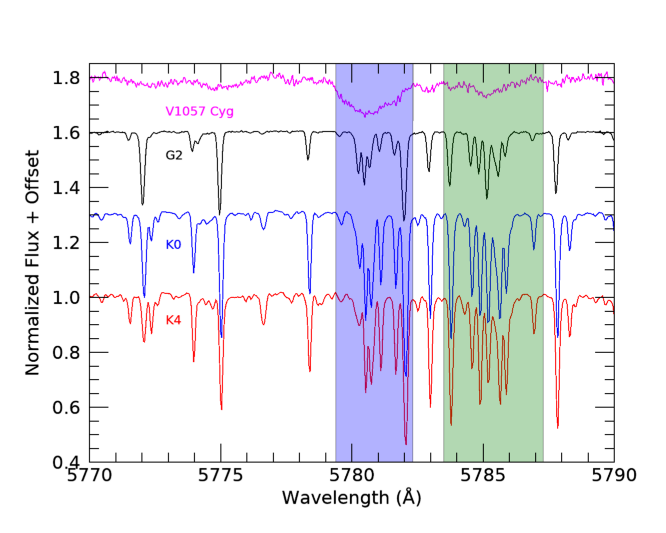}
    \includegraphics[width = 0.485\linewidth]{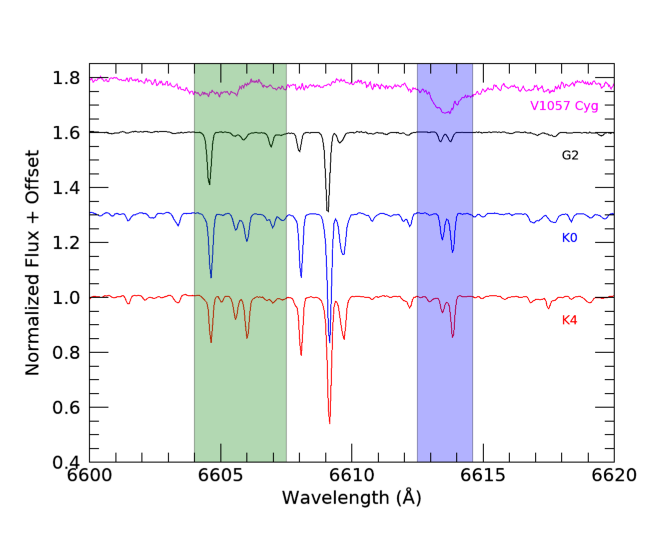}
    \caption{Demonstration of the potential for stellar contamination in the $\lambda$5780 (left) and $\lambda$6614 (right) DIBs. The blue regions show the equivalent width integration limits of the DIBs and the green regions show the reference regions containing the stellar line complexes that are used to compute the contamination ratios. The black, red, and blue spectra are those of HD 88371, HD 80367, and HD 65277 respectively, with spectral types as labelled. 
    The integration limits in blue and green are chosen to capture the stellar lines most likely to contribute to the reference line complexes and DIBs, and are those used to compute the ratios given in Table \ref{tab:ContamWidths}. The magenta spectrum at the top is V1057 Cyg, shown in more detail in Figure \ref{fig:eqWLimitsV1057Cyg}. The integration limits in the standards do not exactly match those shown in Figure \ref{fig:eqWLimitsV1057Cyg} due to the rotational broadening and blending of the stellar lines and the breadth of the DIBs features themselves.}
    \label{fig:StandardLineContSpec}
\end{figure*}

Early spectral types such as B and A stars have relatively clean spectra, and it is for this reason that they have been the traditional ``lightbulbs" in DIBs studies.
As mentioned in Section \ref{sec:DIBsLit}, for spectral types much later than G0, there is significant contamination in the DIBs regions from the stellar photosphere.  Specifically, the $\lambda$5780 DIB is contaminated by \ion{Fe}{1} (5780.5994, 5780.8036), \ion{Cr}{1} (5780.9050, 5781.1670, 5781.1790, 5781.7510), and \ion{Si}{1} (5780.3838). The $\lambda$6614 DIB is contaminated by \ion{Fe}{1} (6613.8245) and \ion{Cr}{1} (6612.1880). The stellar contamination from these lines in late spectral types can be seen in the standard spectra presented in Figure \ref{fig:StandardLineContSpec}, shown for each of the two DIBs regions.

 The YSOs in our sample generally present strong, relatively line-free continuum; their spectra are probably dominated by a hot accretion shock component. However, many outbursters do show some weak GK-type spectral signatures as well. 
  For any individual object, the optical spectral presentation depends on the accretion rate and the vertical temperature structure of the accretion disk.
  Appendix \ref{appendix:EqW} illustrates the potential for contamination in each of our sources.  To determine the extent of any photospheric line contamination in our DIBs line strength measurements, we use the 3 standard stars HD 88371 (G2), HD 80367 (K0), and HD 65277 (K4), which provide a reasonable bracketing of the most common temperature range in the mid-optical range of outbursting YSOs. None of these standards exhibit any DIBs. 

The amplitude of any possible contamination to the DIB features can be assessed by measuring the spectral line strength within the DIB region relative to a nearby spectral line not within the DIB integration band.
The comparison lines are neighboring stellar lines, where we are specifically looking at line complexes, that is, multiple closely spaced lines. These complexes are strong enough that they appear as broad features in the outburster spectra due to the rapid rotation and consequent line blending. We study the 5785 \AA\ complex, which spans roughly 2-3 \AA\, and the 6605 \AA\ complex, which has a similar 2-3 \AA\ broadness (highlighted in green in Figure \ref{fig:StandardLineContSpec}). Both line complexes contain lines from the same species as those contaminating the DIBs regions. Specifically, the 5785 \AA\ complex comprises lines from \ion{Fe}{1} (5783.8933, 5784.6576, 5785.2699) and \ion{Cr}{1} (5783.0650, 5783.85, 5784.9690, 5785.7350, 5785.9158, 5785.9449, 5787.0210). The 6605 \AA\ complex comprises lines from \ion{Fe}{1} (6604.5854), \ion{Cr}{1} (6505.5420), and \ion{Sc}{2} (6604.6010). 
As demonstrated in Figure \ref{fig:StandardLineContSpec},  in the standard star spectra, these lines are easily resolved into individual features but become blended for $v$ sin $i$ values greater than 30 km s$^{-1}$. 

In order to assess the potential for stellar line contamination in DIBs measurements,
we first compute the total equivalent widths of the 5780 \AA\ and 6614 \AA\ stellar line complexes in the DIBs-free spectral standards. We then divide those equivalent widths by the total equivalent widths of the 5785 \AA\ and 6605 \AA\ stellar line complexes in the standards. Both sets of integration ranges are shown for the standards and for V1057 Cyg in Figure \ref{fig:StandardLineContSpec}. Since the spectral types of our young star sample are not well established, and in many of the cases in fact mixed or composite disk spectra with signatures near the DIBs ranging from early G to late K, we average the line complex ratios we compute from our three spectral-type standards. 

We then measure the equivalent width of the non-DIBs 5785 \AA\ and 6605 \AA\ line complexes in our targets. 
Multiplying these equivalent widths by the ratios derived from our standards, we find the approximate stellar contamination present in the wavelength range over which we compute the DIBs equivalent widths, as detailed below. 
The stellar contamination in the DIBs bands increases for later spectral types. 


We apply corrections for stellar contamination to the directly measured DIBs equivalent widths as:

\begin{eqnarray}\label{eq:cont_wids}
    W_*(5780) = W_\lambda(5785) \times r_\text{5780~Ref}
    , \\ W_*(6614) = W_\lambda(6605) \times r_\text{6614~Ref}
\end{eqnarray}
Here,  $W_*(5780)$ is the computed stellar contamination in the 5780 \AA\ DIB band. $W_\lambda(5785)$ is the measured equivalent width of the 5785 \AA\ line complex in our target, and $r_\text{Ref} = \frac{W_\text{Ref}(5780)}{W_\text{Ref}(5785)}$ is the ratio of the equivalent widths in the reference standard spectra. An analogous correction method is applied to the 6614 \AA\ DIB band.
Hereafter we refer to the corrected measurements as $W_{DIB}(5780)$ and $W_{DIB}(6614)$, and we take these to be the equivalent width measurements of the DIBs. The observed and corrected measurements for our targets are recorded in Table \ref{tab:DIBsEqWs}. The ratios for the individual standard stars appear in Table \ref{tab:ContamWidths} 
and for the corrections we use the mean of the reported ratios. 

We note that this correction is an attempt to approximate the stellar contamination in targets for which we cannot simply subtract a stellar template. There may be some mismatch between the true contamination in the DIB and our estimate due to the unknown spectral type of the target, and the unknown amount of veiling due to accretion.   In some cases, no contamination correction may be necessary or inferred.  In others, our contamination estimates can overestimate the true contamination. This situation is seen in the case of the negative ``corrected" $W_{DIB}(5780)$ of PTF 14jg, whereas the $W_{DIB}(6614)$ we compute is positive and relatively large. The difference between the two $W_{DIB}$ values indicates that there may be some wavelength-dependent change in the stellar contamination.

Among our sample, the mean contamination from photospheric absorption within the $\lambda$5780 and $\lambda$6614 DIBs is 0.171 \AA\ and 0.045 \AA\, respectively, with medians of only 0.127 \AA\ and 0.036 \AA.   The standard deviations around the mean values are 0.179 \AA\ and 0.050 \AA\ for the two bands. 
Fractionally, the contamination ranges from 0-50\% and 0-40\% in all but 4 targets, with a median value of 28\% in the $\lambda$5780 DIB and 18\% in the $\lambda$6614 DIB. The contamination is larger than the formal measurement errors in 40\% of the $\lambda$5780 and 25\% of  $\lambda$6614 DIBs measurements, and is therefore less than or comparable to the uncertainty in the equivalent width measurement for most of our targets.
The sources for which the corrections are greater than the measurement error are marked in Table \ref{tab:DIBsEqWs}.

\begin{deluxetable*}{cccccccc}
	\tablecaption{Stellar contamination equivalent widths and ratios for different GK spectral types \label{tab:ContamWidths}}
	\tablewidth{0pt}
	\tablehead{
	    \colhead{Target} & \colhead{SpT} & \colhead{$W_\lambda (5780)$} & \colhead{$W_\lambda (5785)$} & \colhead{Ratio} & \colhead{$W_\lambda (6614)$} & \colhead{$W_\lambda (6605)$} & \colhead{Ratio} 
	}
	\startdata
    HD 88371 & G2V &                0.186 &                0.181 &                1.025 &                0.016 &                0.050 &                0.312 \\ 
    HD 80367 & K0V &                0.385 &                0.420 &                0.917 &                0.046 &                0.090 &                0.511 \\ 
    HD 65277 & K4V &                0.326 &                0.405 &                0.805 &                0.045 &                0.092 &                0.494 \\ 
\enddata
\end{deluxetable*}

\section{Converting to Extinction Estimates} \label{sec:Extincs}
To convert our equivalent width measurements of the two DIBs features into extinction measurements, we use a linear fit to the $W_{DIB}$ and $E(B-V)$ data from three literature sources \citep{friedman_studies_2011, vos_diffuse_2011, kos_properties_2013}. The data are collected from a variety of sightlines in diverse ISM environments primarily involving stars with spectral types no later than early A. For the targets that the studies have in common, there is good agreement and we take the uncertainty-weighted mean of their $W_{DIB}$ and $E(B-V)$ values.


We compute linear fits using the IDL procedure $\mathtt{robust\_linefit}$, which uses 6$\sigma$ outlier resistant bisquare weighting. We find that ordinary least-squares (OLS) linear fitting results in a slope that is much too steep, and would overestimate the extinction at higher $W_{DIB}$ values. The OLS fit residuals have significant structure in them, implying they are skewed by their sensitivity to the large density of points at smaller $W_{DIB}$ values. The residuals from fits produced by $\mathtt{robust\_linefit}$ show no structure overall and are centered within 10\% of 0.

\begin{figure}[!htb]
    \centering
    \includegraphics[width=1.05\linewidth]{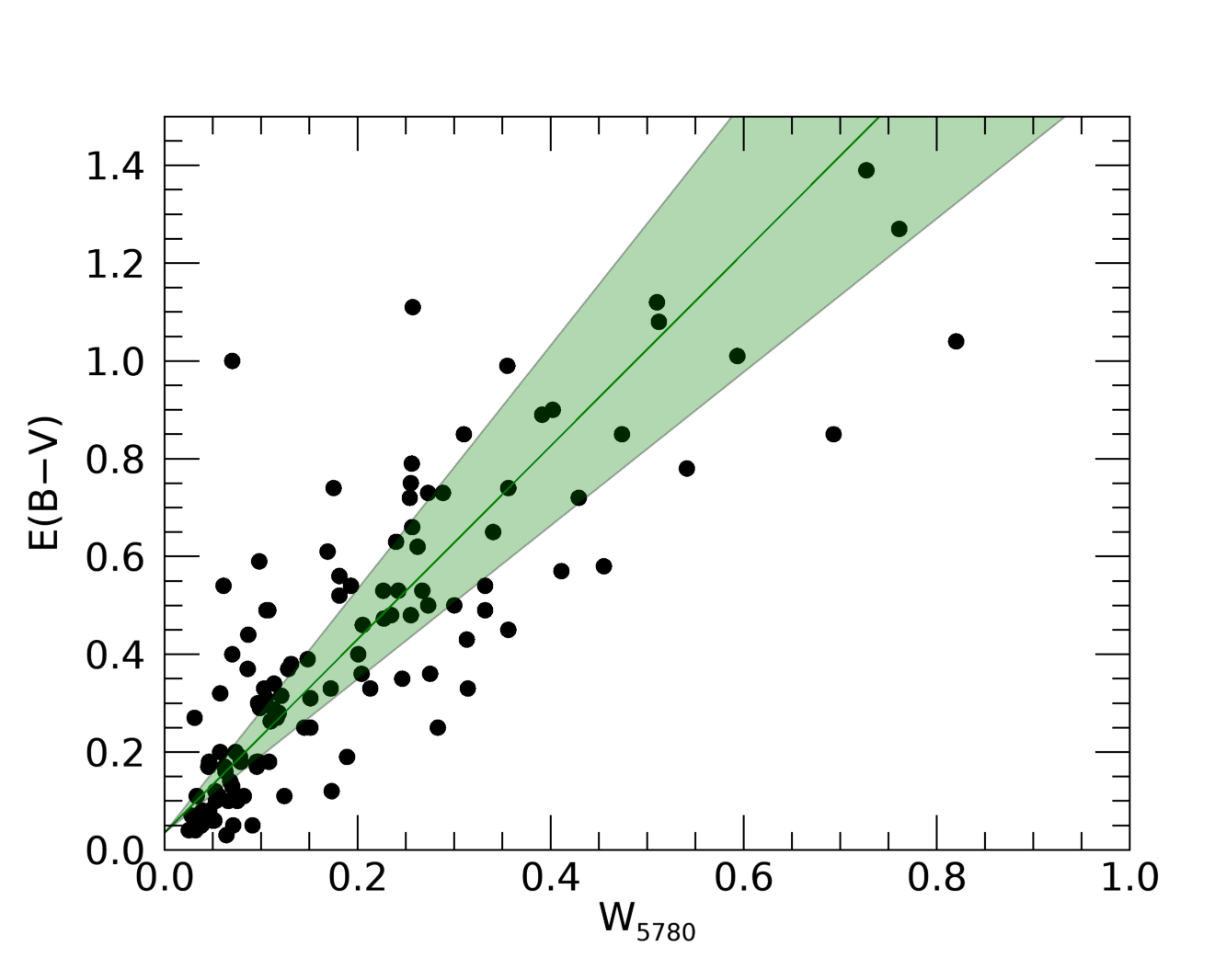}
    \includegraphics[width=1.05\linewidth]{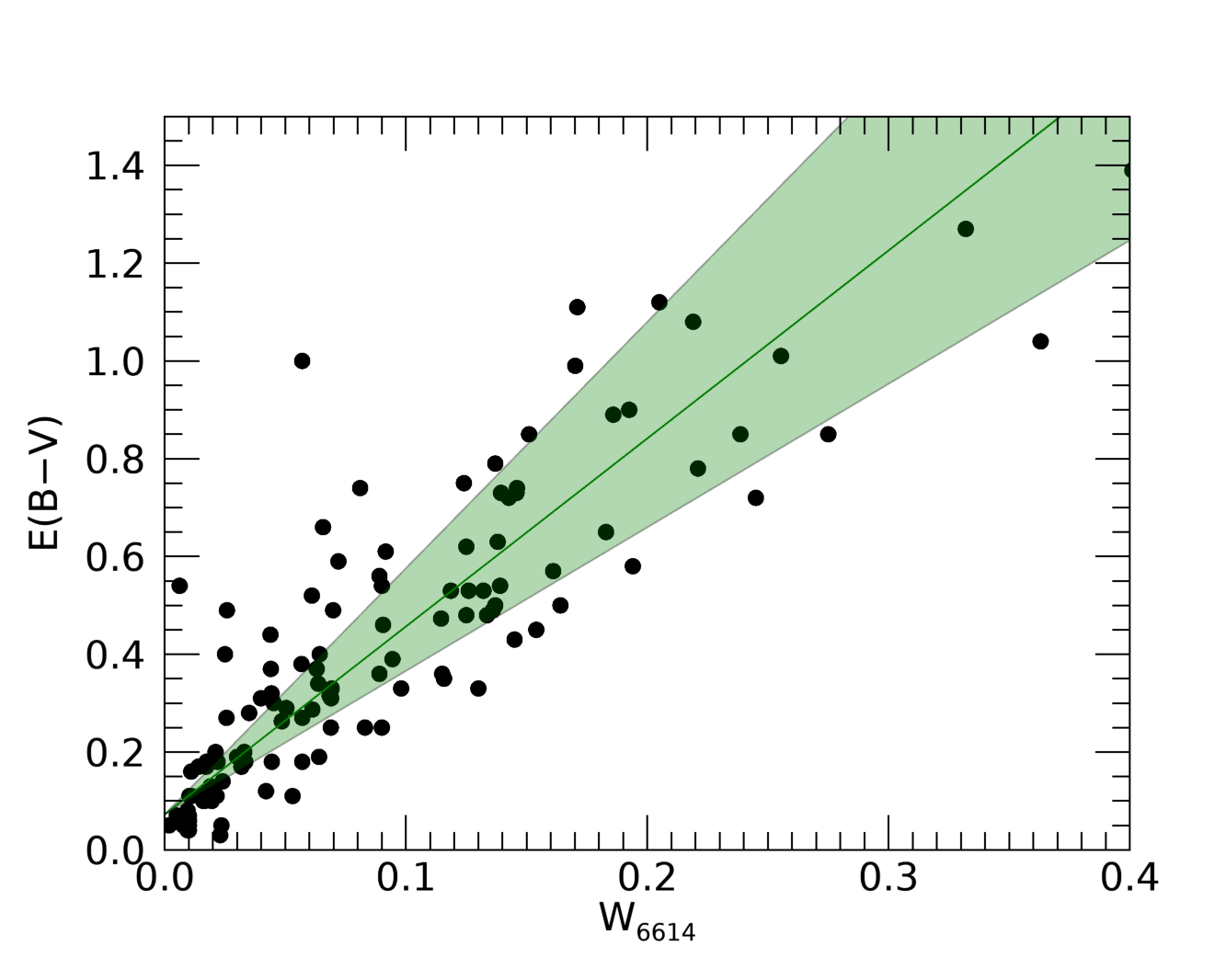}
    \caption{Linear fits to the $\zeta$ sightline $E(B-V)$ and $W_{DIB}$ measurements from \citet{friedman_studies_2011}, \citet{vos_diffuse_2011}, and \citet{kos_properties_2013} for the $\lambda$5780 and $\lambda$6614 DIBs. }
    \label{fig:LinearFits}
\end{figure}

We restrict our fits to the points that lie along $\zeta$ sightlines, as defined by having $W_{5780}/W_{5797} < 0.3$, because almost all of our targets lie along $\zeta$ sightlines; see Appendix \ref{sec:dib5797} for discussion. Although the transition between sightlines may not be a discrete point, varying the location of the boundary did not significantly affect our fits. The best-fit line is shown compared to the $\zeta$ sightline data in Figure \ref{fig:LinearFits}.

The robust fitting method does not produce errors on the best-fit parameters, so we use the IDL function $\mathtt{robust\_sigma}$ to calculate the fractional dispersion of the residuals. We then take half of the fractional dispersion and multiply by the best-fit parameters for the uncertainty in the slope and y-intercept. The resulting linear models that we adopt are: 
\begin{equation} \label{eq:5780_model}
\begin{aligned}
    E(B-V) = \left(0.035 \pm 0.009 \right) \\ 
    + \left[(1.978 \pm 0.514) \times W_{DIB}(5780) \right]
\end{aligned}
\end{equation}
and 
\begin{equation} \label{eq:6614_model}
\begin{aligned}
        E(B-V) = \left( 0.072 \pm 0.022 \right) \\
        + \left[(3.846 \pm 1.192) \times W_{DIB}(6614) \right].
\end{aligned}
\end{equation}

We also present a version of our analysis for fits to the the $\sigma$ sightline in Appendix \ref{sec:dib5797}.

An important consideration for our method is the range of $E(B-V)$ values spanned by the literature measurements. The data from \citet{vos_diffuse_2011} and \citet{kos_properties_2013} only extend to to $W_{DIB}(5780) \sim 0.32$ and $W_{DIB}(5780) \sim 0.6$ respectively. \citet{friedman_studies_2011} extend futher, out to $W_{DIB}(5780) \sim 0.8$, which provides better coverage of our own measurements. Combining the data is a means to better sample the unknown column of interstellar medium through which we observe our targets.

We summarize the previous literature fits and our newly derived combined fit of the $E(B-V)$ to DIB strengths in Table \ref{tab:DIBsFits}. 
The uncertainties we report are much larger than those in previous works, but they better account for the scatter in the data.

We are suggesting here that for continuum-dominated young stars, or those to which corrections can be applied for any stellar contribution to the DIB features (as above), the new relation could be applied in order to estimate $E(B-V)$ and extinction. In order to do so at the larger line strengths, we admittedly need to extrapolate the fit beyond existing data. 

A final step is to convert the $E(B-V)$ values to $A_V$ values.  To do so, we assume a total-to-selective reddening law $R_V$ and the equation $A_V = R_V \times E(B-V)$, which is the fiducial relation that is generally adopted in stellar astronomy.  As is well known, the inferred $A_V$ is thus sensitive to the assumed value of $R_V$, which is difficult to know without more understanding of the sightline. The \citeauthor{BessellBrettRv_1988} value $R_V=3.1$ is that most commonly used in literature. However, there is significant evidence \citep{cardelli_relationship_1989, Indebetouw_GIMPSE_Rv_2005, decleir_rvs_2022} that in denser regions of the ISM the $R_V$ value can be as high as around 5. 

\begin{figure*}
    \centering
    \includegraphics[width=0.325\linewidth]{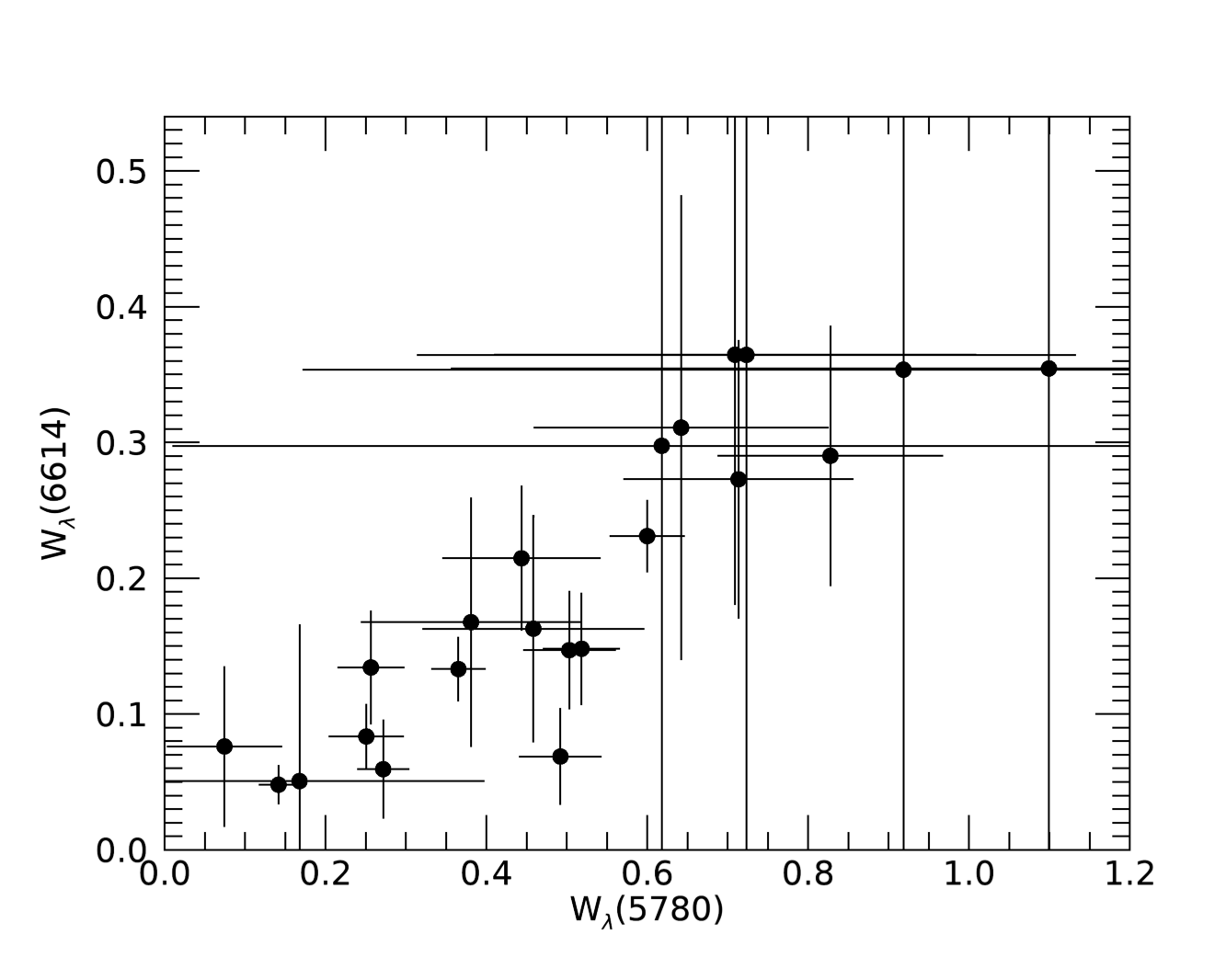}
    \includegraphics[width=0.325\linewidth]{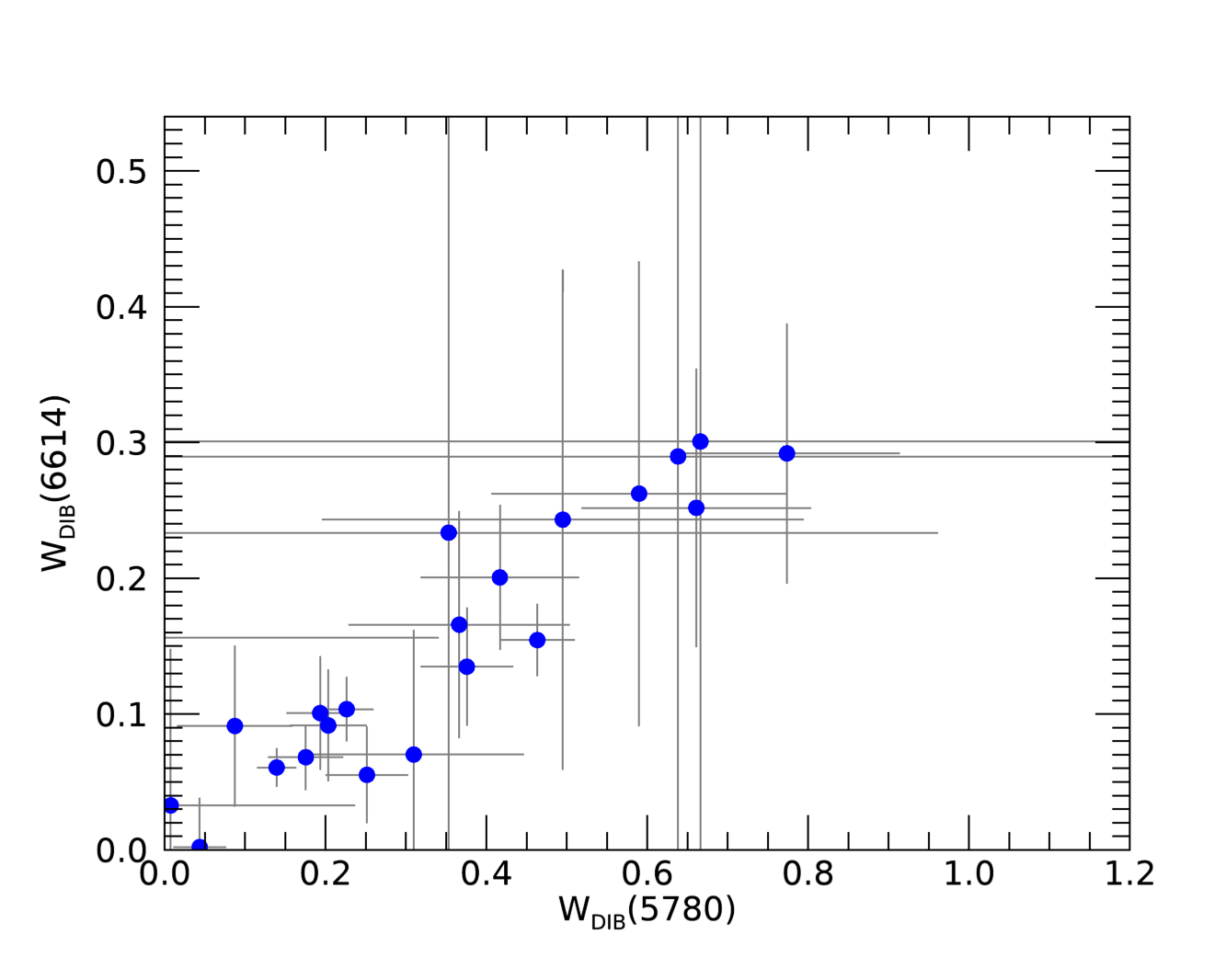}
    \includegraphics[width=0.325\linewidth]{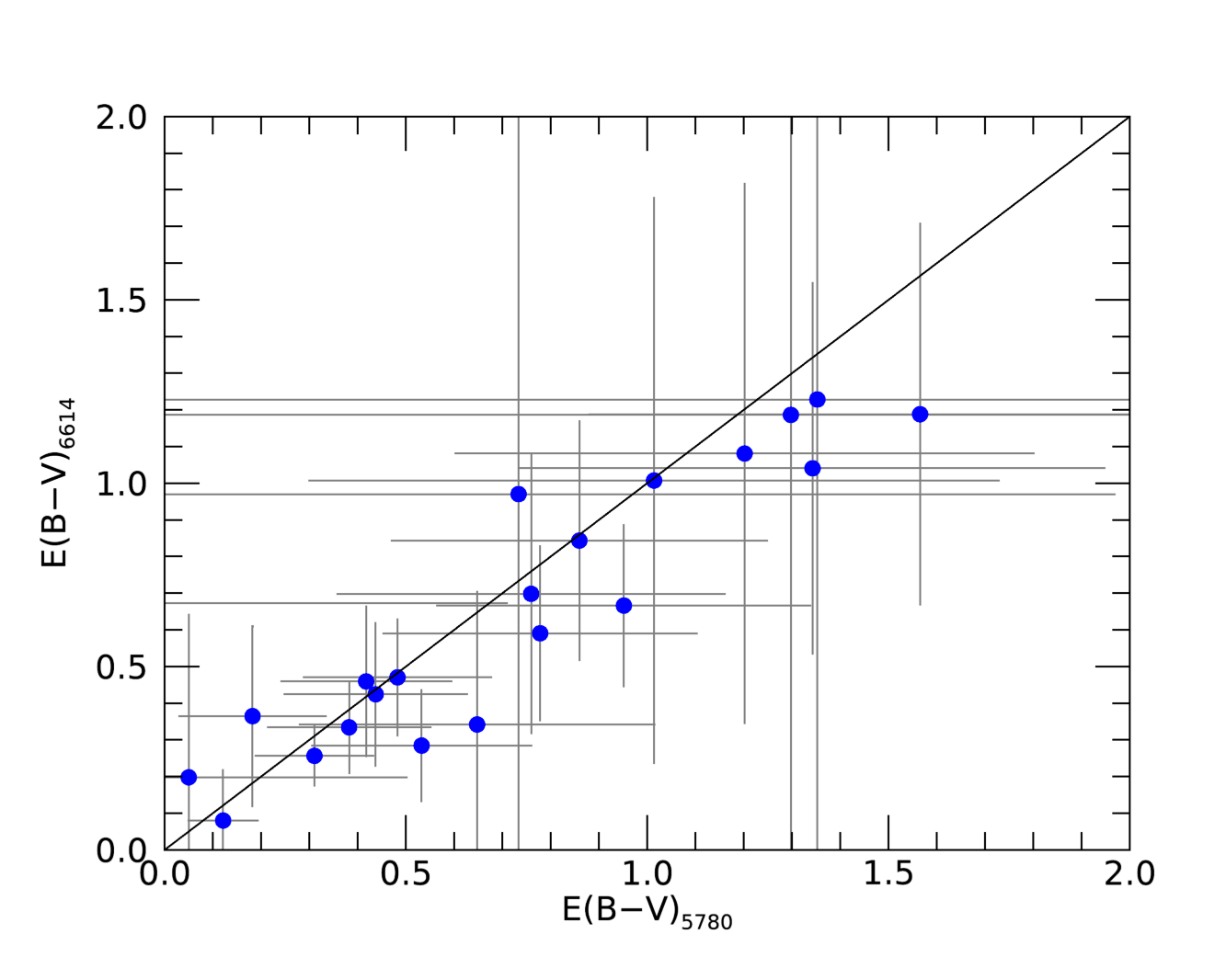}
    \caption{Correlation between our measurements in the two different DIB features studied here. \textbf{Left:} The uncorrected measured equivalent widths. \textbf{Center:} The stellar contamination corrected $W_{DIB}$ values, shown in blue. \textbf{Right:} The $E(B-V)$ values, shown in blue, computed from the stellar contamination corrected $W_{DIB}$ values and our best-fit model, with a $1:1$ correlation line shown for reference. PTF 14jg cannot be seen in the stellar-contamination-corrected plots since its estimated stellar contamination in the $\lambda$5780 DIB is greater than the measured width, placing it at a negative "corrected" $E(B-V)$ value.}
    \label{fig:5780vs6614Avs}
\end{figure*}

The final $W_{DIB}(5780)$ and $W_{DIB}(6614)$ values are plotted against one another in Figure \ref{fig:5780vs6614Avs}. The equivalent width measurements are well-correlated with one another, with a Pearson correlation coefficient of 0.885 and a $p<10^{-5}$ significance level. The correlation between the contamination-subtracted equivalent widths is also strong, with a correlation coefficient of 0.86, corresponding to a $p<10^{-5}$. This correlation coefficient is lower than the 0.96 and 0.94 reported by \citet{friedman_studies_2011} and \citet{kos_properties_2013} respectively, but agrees well with the 0.85 reported by \citet{vos_diffuse_2011}. We also plot the resulting $E(B-V)$ values against one another, with a $y=x$ line for reference. Almost all of our measured $E(B-V)$ values fall within $1\sigma$ of the 1-1 correlation. 

Our final $E(B-V)$ values are reported in Table \ref{tab:DIBsAvs}, alongside the uncertainties in the measurements and $A_V$ values for different assumed $R_V$ values. We provide weighted means of the $E(B-V)$ measurements from the two DIBs and take the mean uncertainty of the two values to be the uncertainty on our measurement. 

\begin{deluxetable*}{c|c|c|c|c|c}
	\caption{Fits of the $E(B-V)$-$W_{DIB}$ correlation for the $\lambda$5780 and $\lambda$6614 DIBs derived from the literature. 
	         The format of the presented fit is $E(B-V) = a + b \times  W_\lambda$}\label{tab:DIBsFits}
	\tablewidth{0pt}
	\tablehead{
	    \multicolumn{2}{c}{$\lambda$5780} & \multicolumn{2}{c}{$\lambda$6614} & \colhead{} &\colhead{}\\
		\colhead{$a_{5780}$} & \colhead{$b_{5780}$} & \colhead{$a_{6614}$} 
		& \colhead{$b_{6614}$} & \colhead{Sightline} & \colhead{Source} 
	}
	\startdata
	-0.0058 $\pm$ 0.0071  & 2.165 $\pm$ 0.060 & 0.012 $\pm$ 0.0067 & 5.624 $\pm$ 0.146 & Avg & \citet{vos_diffuse_2011}\tablenotemark{*} \\
	-0.0047 $\pm$ 0.012 & 1.56 $\pm$ 0.106 & $\cdots$ & $\cdots$ & $\sigma$ & \citet{vos_diffuse_2011} \\
	0.0568 $\pm$ 0.0119 & 2.385 $\pm$ 0.078 & $\cdots$ & $\cdots$ & $\zeta$ & \citet{vos_diffuse_2011}  \\
	-0.0084 $\pm$ 0.0035 & 1.98 $\pm$ 0.01 & 0.02 $\pm$ 0.0037 & 4.63 $\pm$ 0.04 & Avg & \citet{friedman_studies_2011}  \\
	0.0028 $\pm$ 0.0062 & 1.977 $\pm$ 0.0197 & -0.0016 $\pm$ 0.0086 & 5.216 $\pm$ 0.0697 & $\sigma$ & \citet{kos_properties_2013}\tablenotemark{*}  \\
	0.0138 $\pm$ 0.0057 & 1.659 $\pm$ 0.0156 & 0.0066 $\pm$ 0.0057 & 3.882 $\pm$ 0.0366 & $\zeta$ & \citet{kos_properties_2013} \\ 
	\hline
	0.035 $\pm$ 0.009 & 1.978 $\pm$ 0.514 & 0.072 $\pm$ 0.022 & 3.846 $\pm$ 1.192 & $\zeta$ & This Work \\
    -0.030 $\pm$ 0.005 & 1.967 $\pm$ 0.295 & -0.041 $\pm$ 0.009 & 4.479 $\pm$ 1.03 & $\sigma$ & This Work \\
    0.005 $\pm$ 0.001 & 1.961 $\pm$ 0.412 & 0.065 $\pm$ 0.018 & 3.943 $\pm$ 1.104 & Avg & This Work
	\enddata
	\tablenotetext{*}{Linear fits in this work were originally presented in the form $W_{DIB}$ = $A + B\times E(B-V)$. The uncertainties have been propagated through the inversion to get the form we present here.}
    
\end{deluxetable*}

\begin{deluxetable*}{c|cccc|cccc|cc}[!htb]
    \caption{Computed $E(B-V)$ and $A_V$ values using the fits from Equation \ref{eq:5780_model} and Equation \ref{eq:6614_model}, and the $R_V$ = 3.1 and $R_V$ = 5 reddening laws}\label{tab:DIBsAvs}
	\tablewidth{0pt}
	\tablehead{
	    \colhead{} & \multicolumn{4}{c}{$\lambda$5780} & \multicolumn{4}{c}{$\lambda$6614} & \multicolumn{2}{c}{Weighted Average}\\
		\colhead{Targets} & \colhead{$E(B-V)$} & \colhead{$\sigma_{E(B-V)}$} 
		& \colhead{$A_V (3.1)$} & \colhead{$A_V (5)$} & \colhead{$E(B-V)$} & \colhead{$\sigma_{E(B-V)}$} 
		& \colhead{$A_V (3.1)$} & \colhead{$A_V (5)$}   & \colhead{$E(B-V)$} & \colhead{$\sigma_{E(B-V)}$}
	}
	
    \startdata
    \multicolumn{11}{c}{Young Star Outbursters } \\
    \hline
    BBW 76 &                 0.48 &                 0.19 &                 1.50 &                 2.41 &                 0.47 &                 0.16 &                 1.46 &                 2.35 &                 0.48&                0.18 \\ 
    FU Ori &                 0.12 &                 0.07 &                 0.38 &                 0.61 &                 0.08 &                 0.14 &                 0.25 &                 0.40 &                 0.11&                0.11 \\ 
    Gaia 17bpi &                 1.01 &                 0.71 &                 3.14 &                 5.07 &                 1.01 &                 0.77 &                 3.12 &                 5.04 &                 1.01&                0.74 \\ 
    Gaia 19ajj &                 0.05 &                 0.45 &                 0.15 &                 0.25 &                 0.20 &                 0.44 &                 0.61 &                 0.99 &                 0.13&                0.45 \\ 
    V899 Mon &                 0.38 &                 0.17 &                 1.18 &                 1.91 &                 0.33 &                 0.13 &                 1.04 &                 1.67 &                 0.35&                0.15 \\ 
    LkH$\alpha$ 225 S &                 0.76 &                 0.40 &                 2.35 &                 3.80 &                 0.71 &                 0.38 &                 2.20 &                 3.55 &                 0.73&                0.39 \\ 
    V960 Mon &                 0.95 &                 0.39 &                 2.95 &                 4.76 &                 0.67 &                 0.22 &                 2.06 &                 3.33 &                 0.74&                0.31 \\ 
    PTF 10qpf &                 0.53 &                 0.23 &                 1.65 &                 2.66 &                 0.28 &                 0.15 &                 0.88 &                 1.42 &                 0.36&                0.19 \\ 
    PTF 15afq &                 0.65 &                 0.37 &                 2.01 &                 3.24 &                 0.34 &                 0.36 &                 1.06 &                 1.71 &                 0.49&                0.36 \\ 
    RNO 1B &                 1.35 &                 1.56 &                 4.19 &                 6.76 &                 1.23 &                 2.86 &                 3.81 &                 6.14 &                 1.32&                2.31 \\ 
    V1057 Cyg &                 0.42 &                 0.18 &                 1.29 &                 2.09 &                 0.46 &                 0.20 &                 1.42 &                 2.30 &                 0.44&                0.19 \\ 
    V1331 Cyg &                 0.21 &                 0.16 &                 0.64 &                 1.04 &                 0.42 &                 0.25 &                 1.31 &                 2.11 &                 0.27&                0.21 \\ 
    V1515 Cyg &                 0.44 &                 0.19 &                 1.36 &                 2.19 &                 0.42 &                 0.20 &                 1.31 &                 2.12 &                 0.43&                0.19 \\ 
    V582 Aur &                 0.78 &                 0.32 &                 2.41 &                 3.89 &                 0.59 &                 0.24 &                 1.83 &                 2.95 &                 0.66&                0.28 \\ 
    V733 Cep &                 1.30 &                 1.56 &                 4.02 &                 6.49 &                 1.19 &                 2.28 &                 3.68 &                 5.93 &                 1.26&                1.96 \\ 
    V900 Mon &                 0.73 &                 1.23 &                 2.27 &                 3.67 &                 0.97 &                 1.61 &                 3.01 &                 4.85 &                 0.82&                1.43 \\ 
    Z CMa &                 0.31 &                 0.12 &                 0.96 &                 1.55 &                 0.31 &                 0.09 &                 0.95 &                 1.53 &                 0.31&                0.11 \\ 
\hline
    \multicolumn{11}{c}{Other Photometric Variables with DIBs} \\
    \hline
    ASASSN 15qi &                 1.20 &                 0.60 &                 3.73 &                 6.01 &                 1.08 &                 0.74 &                 3.35 &                 5.40 &                 1.15&                0.67 \\ 
    PTF 14jg\tablenotemark{*} &                -0.10 &                 0.81 &                -0.31 &                -0.50 &                 0.67 &                 2.33 &                 2.08 &                 3.36 &                -0.02&                1.74 \\ 
    PTFS1821n &                 1.34 &                 0.61 &                 4.16 &                 6.71 &                 1.04 &                 0.51 &                 3.23 &                 5.20 &                 1.16&                0.56 \\ 
    RNO 1A &                 0.86 &                 0.39 &                 2.66 &                 4.30 &                 0.84 &                 0.33 &                 2.61 &                 4.22 &                 0.85&                0.36 \\ 
    VES 263 &                 1.57 &                 0.69 &                 4.85 &                 7.83 &                 1.19 &                 0.52 &                 3.70 &                 5.97 &                 1.33&                0.61 \\ 
    \enddata
\tablenotetext{*}{For PTF 14jg we report the non-physical negative extinction after the stellar contamination correction for transparency. The computed $E(B-V)$ values for this target prior to stellar contamination correction are 1.515 and 1.594 respectively. The non-zero $W_{DIB}(6614)$ measurement indicates we may have overestimated the stellar contamination to the $\lambda$5780 DIB.}
\end{deluxetable*}

\section{ Comparison of DIBs Extinction Results to Previous Extinction Estimates } \label{sec:litcomp}

Having demonstrated consistency between the two DIBs lines both in their strength and in their implied extinction estimates,
we now compare our extinction results with those previously derived for the objects in our sample. We begin with a general comparison between our measurements and those in Table \ref{tab:OutburstersProperties}, and then discuss in more detail the reliability of the values in the table and the methods by which they were derived.  

In general, the $A_V$ values we find using $R_V = 3.1$ are systematically lower than the literature values in Table \ref{tab:OutburstersProperties}. Adopting $R_V=5$ brings many of the values in closer agreement ($1\sigma$ for some), though some significant outliers remain. The dispersion in the $A_V$ values in Table \ref{tab:DIBsAvs}, as measured by the standard deviation of the differences between the $A_V(5780)$ values and the $A_V(6614)$ values, (adopting $R_V = 3.1$) is 0.75 magnitudes. This worsens slightly to 0.8 magnitudes if the contamination corrections are not applied. 
The scatter in the derived extinctions based on which DIB is used, is comparable to the typical standard deviation of the $A_V$ measurements for any given object in the more well-characterized T Tauri sample we discussed in Section \ref{sec:lit_ext}. We interpret this finding as an indication that our DIB-based extinction method is at least as reliable as existing methods -- with added independence from needing to model the continuum origin, and the additional benefit of robustness to large scale continuum changes.

As discussed in Section \ref{sec:intro} and Appendix \ref{appendix:Extincts}, the methods for determining extinction to young stars and YSOs, including outbursters, often rely on fitting template spectra based on an assumed (or simultaneously fitted) spectral type. Most of the literature $A_V$ values presented in Table \ref{tab:OutburstersProperties} are estimated using this method, which can be unreliable for -- in particular -- highly variable objects.  Besides the photometric variability, outbursting sources are known to change their spectral type during an outburst \citep[e.g.,][]{hillenbrand_gaia_2018, hillenbrand_ptf14jg_2019, munari_2018_2019}, varying sometimes from typical low-mass pre-main sequence KM spectra during quiescence to accretion-dominated FG during outburst. Furthermore, in some cases \citep[FU Ori, PTF 14jg and Gaia 17bpi; ][respectively]{herbig_high-resolution_2003,hillenbrand_gaia_2018, hillenbrand_ptf14jg_2019} estimates of the spectral type can vary with wavelength, from G-type in the optical regions of the spectra to M-type in the infrared regions. Given the difficulty in identifying a proper spectral template for these objects, any template-fitting-based method against which to measure their extinctions can give highly suspicious results.  As noted in the comments of Table \ref{tab:OutburstersProperties}, many of the published $A_V$ values towards our sources are rough estimates based on significant and poorly justified assumptions by the authors.  

Another method, deriving
extinctions from X-rays, seem less influenced by outburst variability \citep{kuhn_comparison_2019}. However, xray measurements cannot always be simply converted into $V$ band extinctions. For example, in the case of FU Ori, translating the inferred $N(H)$ from the X-rays to $A_V$ gives $A_V=81$ mag, which would totally obscure the object in the optical. However, FU Ori is not completely obscured, and the X-ray $N(H)$ value is clearly over-estimating the extinction to the optical source, even if it does accurately describe the obscuration to the X-ray continuum. The two other sources in common (V960 Mon and Z CMa) have X-ray derived $A_V$s from \citet{kuhn_comparison_2019} in rough agreement with our DIBs-inferred $A_V$s.

We assert that our DIBs measurements can be translated into extinction values which are independent of assumptions regarding the underlying spectral energy distribution of the source. 
Furthermore, the established DIBs-based methodology produces quantifiable errors, unlike the case for the existing $A_V$ estimates to our sources.
We find that DIBs absorption strength relative to continuum is unaffected by outbursting or quiescent phases for YSOs. The principal uncertainty in the DIBs-based extinction measurements then, is that which plagues most extinction calculations: what is the source of the continuum to which we are measuring the extinction?  We expand upon this point in the discussion below.

\section{Discussion} \label{sec:discussion}

A main result of our study is the prevalence of DIB absorption towards outbursting young stellar objects.
In none of these sources was there a high-dispersion spectrum taken before the outburst, only after the outburst peak,
so it is impossible to say whether the DIB features were present in the pre-outburst state or not.
However, the existence of post-outburst DIB absorption has raised several possibilities.

First, can the DIBs be used to estimate line-of-sight extinction to outbursting sources, which are otherwise challenging 
to derive extinctions for?   We advocate that the answer to this question is yes, albeit with some caution
as demonstrated in our analysis above.   Second, where do the DIB carriers reside along the line of sight,
and can the presence of DIB absorption inform us about circumstellar environments, or only about the interstellar column?


\subsection{Extinction to What?}
A crucial part of interpreting extinction measurements is understanding ``to what" the extinction is being measured, given the diagnostic being used. Extinction measurements that rely on template fitting and comparison of observed emission to template properties require a good understanding of the underlying source, as we have discussed above. Measurements of spectral absorption features against a continuum, such as we present in this work, measure the extinction to the dominant continuum emission source.

In early-type main sequence stars, due to their relatively featureless spectra, there is some assurance that the source to which extinction measurements are made is the stellar photosphere. In the targets we have studied, however, this assumption is no longer as reliable. Many YSOs have complex, dynamic circumstellar environments with dust and gas which can contribute to the SED at a range of wavelengths. In targets where hot inner disk emission or accretion veiling contribute significantly to the infrared and/or optical continuum, we can not be sure that the extinction measurement extends all the way through that environment to the stellar photosphere. \citet{mcjunkin_direct_2014} present an excellent illustration of the challenges, in their discussion of the various line-of-sight emission and absorption sources for Ly$\alpha$. 

Can we know how deeply the DIBs probe? That the $\lambda$5780 and $\lambda$6614 DIBs correlate well with $N(H)$ \citep{friedman_studies_2011}, indicates that they likely measure well much of the interstellar extinction. 
However, the rapid fall-off of the $W_{DIB}/E(B-V)$ relation in environments with high molecular hydrogen fractions indicates that DIBs extinction measurements may be less sensitive to {\it circumstellar} extinction.

Also, in outburst events, such as those recently undergone by the YSOs in our sample, 
both the source luminosity and at least some aspects of the source temperature or radial temperature distribution, are expected to increase. 
Spectral absorption features which originate deep in the circumstellar environment, nearer to the stellar photosphere, 
could change during the event, either in the spectral feature carriers, or from spectral veiling changes due to the continuum contribution of the outburst. Based on several objects in our sample with multiple spectra over many years, we find that the DIBs are remarkably consistent through significant variability in source brightness, and hence the optical continuum level against which the DIBs are measured. This would seem to indicate that the DIBs carriers reside outside the region where the photometric variability arises in outbursting young stars.


\subsection{Limitations of DIB-Based Extinction Values in Star-Forming Environments}
The difficulty in describing the formation environment of DIBs limits the certainty with which the $A_V$ measurements that we present can be interpreted. The work by \citet{fan_behavior_2017} demonstrates the sensitivity of various DIBs to their molecular environment, as discussed earlier. We estimate, based on the $W_{DIB}-f_{H_2}$
relationships presented in \citet{fan_behavior_2017}, that the $f_{H_2}$ may decrease the $W_{DIB}(5780)/E(B-V)$ slope and the $W_{DIB}(6614)/E(B-V)$ slope by a factor of 3.16\footnote{This comes from comparing the apparent peaks of the lambda curves in \citet{fan_behavior_2017} Figure 4. The peak/min for 5780 is 2.75/2.25 and for 6614 is 2.5/2 in log space.} if the local value of $f_{H_2}$ is closer to 0.7 than 0.2. This effect may cause our measurements to be underestimates of the local extinction.

Another item to consider is the range of $W_{DIB}$ values over which the original literature correlations we used were determined. As we discussed in Section \ref{sec:Extincs}, the extrapolated models of \citet{vos_diffuse_2011} and \citet{kos_properties_2013} are in good agreement with the \citet{friedman_studies_2011} models, which cover almost our entire $W_{DIB}$ measurement range. However, the behavior of DIBs in denser environments is not well understood. The measurements of $W_{DIB}(5780)/E(B-V)$ and $W_{DIB}(6614)/E(B-V)$ presented by \citet{lan_exploring_2015} seem to indicate that the DIBs saturate at values of $E(B-V)$ greater than $\sim 1$. This phenomenon may be due to increased $f_{H_2}$ fractions in the denser, more extincted environments, as discussed by \citet{fan_behavior_2017}. In either case, our extinction measurements may be underestimates of the full extinction column. 

\subsection{Reliability of DIB-Based Extinction Values}

We investigated and found that DIBs-based extinction measurements for typical low-mass T Tauri type stars 
are unreliable due to the high level of contamination from photospheric lines. 
For both higher mass young stars, and for many YSO outbursters, however, the DIBs-based extinction method appears promising. 
In these targets, stellar contamination is low due to the increased continuum above any regular stellar photosphere
caused by the hotter temperatures of the $\tau=1$ surface due to enhanced accretion.  

Furthermore, the DIBs measurements converted to $A_V$ values seem more reliable for rapidly accreting YSOs
than other existing methods of estimating $A_V$. 
Literature estimates of extinction to the targets in our sample are highly uncertain due to the challenges we have presented in Section \ref{sec:litcomp}. 
As our DIBs measurements  do not rely on many of the assumptions traditionally made to compute extinction to YSOs, we believe them to be more robust for these enigmatic sources. 
This is especially true in systems with large continuum variability or large continuum contributions from accretion, which do not impact DIBs extinction measurements.

Additional study of the use of DIBs to estimate extinction towards outbursting and other highly variable objects is warranted. We limited our primary study to two bands, the 5780 \AA\ and 6614 \AA\ DIBs, because they appeared strong in all of the outburster spectra we investigated. The $\lambda$5797 DIB also appears strongly in many of our spectra, but it is on the edge of a spectral order. We discuss our measurements of the $\lambda$5797 DIB and reasons for excluding it from our primary analysis in Appendix \ref{sec:dib5797}. Many of the spectra show the $\lambda$6270 as well, but it does not appear consistently in all of our sources and is often heavily telluric-contaminated.

\subsection{Uncertainties in the Interpretation of DIB Strength}
Although we believe DIBs measurements to be reasonably reliable, especially compared to existing methods for estimating the extinction to YSOs, there are numerous sources of uncertainty that are intrinsic to the method. Most of these are described above, and incorporated into the error bars shown in our figures, and presented in Tables \ref{tab:DIBsEqWs} and \ref{tab:DIBsAvs}. There are, however, additional uncertainties that are important to consider when assessing the utility of the method for YSOs.

The first of these sources of uncertainty is astrophysical, and perhaps the most basic: we do not know much about the interstellar sightlines along which these objects lie. For many of the sources in our sample, they are first detected in wide-field transient surveys as they enter an outburst phase. Consequently, the environments around them may not contain many well-studied sources.  
Study and characterization of targets using our proposed $A_V$ measurement method requires assuming the target lies along a $\sigma$-type or $\zeta$-type sightline.  
We present our derived $A_V$s under the assumption that the targets lie along $\zeta$-type sightlines, as described in Section \ref{sec:DIBsLit} and justified in Appendix \ref{sec:dib5797}. 
However, for new targets, this assumption may not be valid and those which lie along $\sigma$-type sightlines, or some combination of the two, will require a different model, presented also in Appendix \ref{sec:dib5797}.

Another source of uncertainties is methodological, and concerns the stellar contamination correction. To calculate the potential contamination from photospheric features to the DIBs spectral region without assuming spectral types for them, we relied on measurements of neighboring line complexes. This introduces two possible sources of uncertainty. 

First, the act of performing an equivalent width subtraction introduces an additional uncertainty term in the stellar-contamination-corrected equivalent widths. This is signal-to-noise dependent, and approximately the same magnitude as the original uncertainty, so the resulting total uncertainty on the corrected measurements can be obtained by multiplying our reported values by $\sqrt{2}$. 
Second, in applying the contamination corrections, we have taken the mean of equivalent width ratios that appear to be sensitive to spectral type. 
This uncertainty is more difficult to estimate, but we believe it is relatively inconsequential. The ratios of the equivalent width of the contaminating stellar line complexes to the neighboring stellar line complexes increases towards earlier spectral types. This is important because the actual quantity of contamination decreases towards earlier spectral type, so that averaging the ratios over the standards as we have done will serve to slightly over-estimate the contamination in later spectral types. Another consideration is that the majority of our targets are believed to be G or F spectral types, so the actual degree of stellar contamination is expected to be low. 

We reiterate that although the DIBs-derived extinction measurements are uncertain, they provide extinction measurements which are independent of the source of the optical continuum and of any continuum variations. The extinctions also do not require a precise determination of the stellar line contributions to the equivalent width measurements. In the cases where such corrections may be necessary, we provide a method by which to make the correction which relies minimally on assumptions about the spectral type of the target.  A benefit of our method compared to popular gas-line extinction measurements is that our species do not appear in the underlying stellar spectrum.

\subsection{Considerations When Using DIBs to Measure Extinction}
As discussed above, the DIBs provide a means to measure extinction to objects for which traditional SED or spectral template fitting are challenging. This is especially the case in highly variable targets such as the outbursters presented in this work. Where traditional methods fail, the DIB equivalent width conversions we present may provide a starting point for estimating the extinction to a target, along with some sense of the uncertainty in that measurement.

Those who would use this method should consider two important assumptions required to convert the $W_{DIB}$ value to an $A_V$ value. The first is the conversion from the $E(B-V)$ to $A_V$ via an assumed $R_V$ value. As we've discussed above, there is significant evidence that the canonical $R_V = 3.1$ value typically assumed may be wrong. \citet{ramirez-tannus-HAeBeDIBs_2018} demonstrate in their study of sightlines toward M17 that there can be significant variability in $R_V$, finding values ranging from 3.3-4.8. Notably, they do not report any $R_V$ values close to or less than 3.1.

Another assumption is that of sightlines. We have provided a new calibrations for $\zeta$-type sightlines, along with calibrations for $\sigma$-type sightlines or an average of both (see Appendix \ref{sec:dib5797}). The reader should consider which sightline is most relevant to their objects when using the calibrations provided. The calibration we ultimately selected seemed most relevant for our targets and may be most applicable for targets in similar environments such as other YSOs.  

\section{Conclusion} \label{sec:conclusion}
We have presented a promising technique for measuring extinctions to outbursting YSOs by using equivalent width measurements of identifiable DIBs (diffuse interstellar bands) in a sample of 23 sources (22 outbursting YSOs plus RNO 1a) observed using the Keck Observatory's HIRES instrument. We focused our analysis on the strong 5780 and 6614 \AA\ lines, both of which appeared in all of our spectra, though other weaker DIBs complexes are also present in most sources. 

We carefully measured the equivalent widths of the DIBs in the continuum normalized spectra. 
To account for the possibility of photospheric line contamination within the DIBs integration ranges, 
we computed the expected contribution to the DIBs absorption profiles using line strength ratios 
derived for  main sequence standard stars.  
We used the ratios of total equivalent widths of known stellar line complexes near each DIB 
to those within the DIBs integration regions,
calibrated by the ratios in standard stars, to estimate the stellar contamination to the measured DIBs features in the outburster spectra. 

With DIBs band equivalent width measurements $W_{DIB}$,
we then computed the best-fit linear model to $E(B-V)$ and $W_{DIB}$ values assembled from three separate published surveys to determine the $E(B-V)$ towards each outburster spectrum. We considered both the measured $W_\lambda$ and the stellar contamination corrected $W_{DIB}$. Although the small corrections shift some of the reddening values, the overall correlation between the extinction values for each of the two DIBs remains strong, with $p < 10^{-5}$. We therefore conclude that -- for these particular sources -- the stellar line contamination within the DIBs features does not significantly affect the strength of the $E(B-V)$ -- $W_{DIB}$ correlation in these targets.   Finally, we converted the $E(B-V)$ reddening values to visual extinction $A_V$ values using total-to-selective reddening ratios of $R_V = 3.1$ and $R_V = 5$.  With the inferred $E(B-V)$ values and an $R_V$ of 3.1, we measure $A_V$ values up to 6 mag for our outburster sample.

We have demonstrated that DIBs equivalent widths measurements can be used to estimate the extinction to outbursting YSOs. We have also shown that the $A_V$ values computed from the DIBs measurements show only slightly greater dispersion at $\sigma(A_V)\sim 0.8$ mag than is present in $A_V$ values computed for much less extincted, and better understood, T Tauri star where $\sigma(A_V)\sim 0.5$ mag is common among different measurements.

Consequently,  we recommend searching for DIBs in spectra of outbursting variables and utilizing them as a starting point for extinction estimates. 
The DIBs appear more reliable than other extinction estimation prospects for these enigmatic sources. 
Furthermore, the DIBs are apparently robust to the sources of variability in the spectra, remaining constant as the continuum level changes
as YSOs enter and exit from an outburst or other variable phase. 
As more is understood about the origin of DIBs, they may also prove a powerful probe of the circumstellar environment around even embedded variables.

\acknowledgements{ We thank Greg Herczeg and Michael Kuhn for discussions and Chris Johns-Krull and Lisa Prato for access to several standard star spectra obtained at McDonald Observatory that were used in this analysis.  
}

\bibliography{Outbursters_DIBs_Paper}{}
\bibliographystyle{aasjournal}

\appendix 

\restartappendixnumbering

\section{Extinction Methods Historically Used for Young Stars} \label{appendix:Extincts}

In this Appendix, we review in detail the literature containing the most robustly derived extinction estimates for young stars in the Taurus region. The subsections are organized following the column order in Table \ref{tab:TTSExts}. The order is based on the general attention to the scientific details underlying the extinction determination method.
Both dust-based and gas-based diagnostics have been employed. 

\subsection{Dust Extinction Measurements}\label{sec:DustExt}

\subsubsection{Valenti 1993}
\citet{valenti_tts_1993} use low resolution spectra with WTTS standards as templates against which to measure the excess spectra of their CTTS program stars. 
The WTTSs first serve as templates for a veiling correction, and then after determining the veiling continuum, they are used to measure the reddening excess. 

Veiling is measured in three regions: 4226 \AA\ (CaI resonance line), 4400-4600 \AA\ (V and Fe complexes at 4380, 4405, and 4460 \AA\, and CH contribution at bluer wavelengths), and 3580-3590 \AA\ (Fe complexes). The WTTS spectra are normalized and the veiling corrections applied manually to best match the CTTS spectra. 

To account for color excess in the CTTS system caused by accreting material, \citet{valenti_tts_1993} use a model of a hydrogen slab appearing between the observer and the stellar photosphere. They then perform a simultaneous fit of the slab and stellar contributions, and the interstellar extinction to the veiling-corrected WTTS template, assuming the slab suffers the same extinction as the stellar photosphere. In cases where they are unable to find a good fit for the $A_V$ after fitting the slab, they allow the slab and $A_V$ parameters to vary simultaneously. When they were unable to identify a good fit when varying $A_V$, they adopted an existing literature value.

\subsubsection{Gullbring 1998}
\citet{gullbring_disk_1998} perform extinction measurements by simultaneous fitting of the veiling factor $r$ and the visual extinction $A_V$ to a template spectrum (taken from a known WTTS). 

The function for the observed flux $F_i^o$ in wavelength bin $i$, which they seek to fit, is given as
\begin{equation} \label{eq:flux_ext}
    F_i^o = C_1 F_i^t 10^{0.4(A_i^t - A_i^o)}(1+r_i).
\end{equation}
Here $C_1$ is a constant scaling factor accounting for the different angular sizes of template and CTTS, and $F_i^t$ is the flux of the template in wavelength bin $i$.
The veiling coefficient in wavelength bin $i$, $r_i$, is given by 
\begin{equation}
    r = \frac{F_l^o/F_c^o - F_l^*/F_c^*}{1-F_l^o/F_c^o}.
\end{equation}
$A_i^t$ is the visual extinction of the template, and $A_i^o$ is the visual extinction of the observed star. 

\subsubsection{Hartigan and Kenyon 2003}
We did not include their measurements in Table \ref{tab:TTSExts}, but it is important to describe the work of \citet{hartigan_kenyon_2003} in any discussion of TTS extinction measurement methods. This sub-arcsecond binary study captures both WTTS and CTTS using STIS spectra. The authors determine $A_V$ values for the WTTS using the $m_{5400}-m_{7035}$ color, computing the continuum magnitudes in 30 \AA\ bands around the two wavelengths. 

For the CTTSs, they first determine the spectral type and then fit the veiling continuum using the best matching WTTS template. The veiling fit is based on similar processes in \citet{valenti_tts_1993} and \citet{gullbring_disk_1998}. The critical assumption is that the observed continuum is an arithmetic mean of a photometric spectrum and a veiling spectrum, so that by use of non-accreting template spectra, the excess can be identified and subtracted.

Reference wavelengths of 6100 \AA\ and 8115 \AA\ are used along with the assumption for the medium-resolution data of a linear model for the veiling between those wavelengths, and for the low resolution spectra a quadratic model. The authors then fit their veiling continuum models to the template WTTS spectra. These veiling fitted spectra are then used to determine the reddening, at that which results in a veiling-fitted template with a flat residual spectrum in the wavelength range 6000 \AA\ to 8500 \AA.  

\subsubsection{Fischer 2011}
\citet{fischer_characterizing_2011} use IRTF/SpeX low resolution spectra to compute line veilings and then extinction for CTTS targets. Veiling is computed in regions around 0.82, 0.91, 0.97, 1.05, 1.18, 1.31, 1.98, 2.11, 2.20, and 2.26 $\mu$m by comparison to spectral templates and three WTTS spectra. The $A_V$ values are computed by fitting the same reddening law given in Equation \ref{eq:flux_ext} above, but in the rearranged form
\begin{equation}
    2.5 \log \left[ T_\lambda (1+r_\lambda)/O_\lambda \right] = k_\lambda (A_{V,O} - A_{V,T}) - 2.5 \log C,
\end{equation}
where $O$ denotes quantities of the observed spectrum, $T$ denotes quantities of the template spectrum, 
$k_\lambda = A_\lambda/A_V$, and the other parameters are defined as above. 

The best-fit line to the $2.5 \log \left[ T_\lambda (1+r_\lambda)/O_\lambda \right]$ vs $k_\lambda$ relation provides the value of $A_V$ for the target star. 

\subsubsection{McClure 2013}
\citet{mcclure_characterizing_2013} use new observations of WTTS plus templates from the IRTF spectral library \citep{Rayner_IRTF_2009} 
to determine the veiling and extinction in SpeX spectra of CTTS through a similar procedure to \citet{fischer_characterizing_2011}. The sample has little overlap with our selected stars, and as a consequence we do not report their measurements in  Table \ref{tab:TTSExts}, but we summarize their procedure for completeness. 
The authors target 14 lines ranging from 0.8 to 2.3 $\mu$m for their veiling measurements, using an equivalent width relation to determine $r_\lambda$
\begin{equation}
    \frac{W_\lambda^{phot}}{W_\lambda} = 1+r_\lambda.
\end{equation}
Using the equivalent widths of selected lines to determine veiling using this relation is a more robust way to determine the veiling contribution than when computing it directly for many lines simultaneously, as demonstrated in \citet{basri_veiling_1990}. This is in large part because with individual measurements, lines with anomalous profiles due to surface gravity sensitivity or differential veiling can be ignored.

After determining the veiling continuum, the authors use a version of Equation \ref{eq:flux_ext} to compute the extinctions to the target CTTSs. 
Using the fit, they approximate the 3$\sigma$ uncertainties on their measurements.

\subsubsection{Herczeg and Hillenbrand 2014}
\citet{herczeg_survey_2014} use Keck LRIS low resolution optical spectra in a large survey of stars in the Taurus-Auriga clouds, 
presenting extinction measurements for many of the most-studied targets in the region. 
They perform an analysis similar to that of \citet{hartigan_kenyon_2003}. 

The authors first demonstrate that the accretion continuum in CTTSs can be treated as a flat spectrum, given that veiling measurements using WTTS templates are approximately flat. We note this assumption for the continuum differs from that adopted by \citet{hartigan_kenyon_2003}. 
\citeauthor{herczeg_survey_2014} then use the veiling fit and estimated spectral type to compute the extinction necessary to match the template to the observation. 

Using the values from this first pass as the starting values for a second pass, \citeauthor{herczeg_survey_2014} then simultaneously fit the spectral type, veiling, and extinction. The veiling fit is performed by studying the excess emission in the broad and deep \ion{Ca}{1} 4227\AA\ absorption line and visually confirming the quality of the fit in this region while simultaneously fitting the entire spectrum for the template spectral type and the extinction.

\subsubsection{Furlan 2011}
\citet{furlan_spitzer_2011} use the 2MASS survey magnitudes to compare $J-H$ and $H-K$ colors for their T Tauri targets to assumed photosphere colors from \citet{luhman_disk_2010} and compute color excesses. For objects with little long wavelength excess emission, and thus little $H$-band excess, they use the extinction derive from the $J-H$ calculation. For objects where these $J-H$ derived extinctions fall below similarly derived $H-K$ extinctions, they use $J-K$. If an excess at $H$-band was inferred based on the source colors at longer wavelengths, the authors use either the slopes of the SpeX spectra around 1$\mu$m (compared to those expected from their template spectra), 
or the intrinsic $H-K$ colors for late K and M CTTSs \citep{meyer_intrinsic_1997}. We convert the values reported by \citeauthor{furlan_spitzer_2011} from $A_J$ to $A_V$, using the \citet{Mathis_Rv_1990} $R_V=3.1$ conversion $A_V = 3.55 A_J$.

\subsubsection{Kenyon and Hartmann 1995}
\citet{kenyon_hartmann_1995} use a simple color excess method to determine $A_V$. They compare the optical and IR colors 
observed in their program stars (listed in their Appendix A) with the colors of Main Sequence standards of similar spectral type, and take the difference to be the color excess. They then convert the measured excesses to $A_V$ and $A_J$ values by the extinction law in \citet{BessellBrettRv_1988}. 

This color excess method, although popular for a simple estimation of the color excess in a target, and also used by the next several authors we discuss, can be misleading. Young stars (especially CTTSs) are known to have many possible sources of excess emission in their spectra, from flares to spots to accretion to disk emission. Even determining the spectral type of a CTTS for comparison to an assumed standard color is nontrivial. 

\subsubsection{Luhman 2017}
\citet{luhman_survey_2017} estimate the extinction in $A_J$ to their program stars through a variety of methods. Most of the stars in Table \ref{tab:TTSExts} have their $A_J$s determined by computing the slope of the SED around 1$\mu$m and comparing to observed WTTS standards. Some extinctions are computed by comparison of $J-H$ from \citet{furlan_spitzer_2011} and some by assuming the $J-H$ intrinsic colors for CTTSs given in \citet{meyer_intrinsic_1997}.

\subsubsection{Grankin 2017}
\citet{Grankin_reliable_2017} use photometric color excesses to measure extinction to their target CTTSs. They use long-term photometry of young stars over 22 years (1984-2006) to study the sources of excess emission in the spectra and determine the degree of variability. They identify stars whose variations are caused by the presence of dark spots or hots spots or accretion variability. For WTTS stars, \citeauthor{Grankin_reliable_2017} identify the $V$ band observations they believe to be least affected the presence of dark or bright spots. In the case of a dark-spot-covered star, that would be the brightest. If the star is covered by hot spots, the optimal observation would be the dimmest. They assume this dimmest $V$ magnitude, $V_{ph}$, as the most representative of the photosphere of the target star. 

In the case of CTTSs, where accretion dominates the variability, the authors look to the $E(B-V)$ observed in the star versus that expected for the spectral type of the target. They then choose the $V$ magnitude, $V_{ph}$, where $E(B-V)=0$, so that the blue excess due to accretion disappears. The corresponding $E(V-R)$ is calculated according to the $V-R$ color relation and used to compute $A_V = 3.7 E(V-R)$. 

\subsubsection{Rebull 2020}
\citet{rebull_rotation_2020} use $V$ magnitudes (either measured or interpolated from Gaia G magnitudes) and 2MASS $K_s$-band photometry to compute the $V-K_s$ colors for their program stars. They then use the spectral type of the target to determine the expected $V-K$ color and thus compute the excess $E(V-K_s)$.  For some stars the 
$E(V-K_s)$ values come from de-reddening in the $(J-H)$ vs $(H-K_s)$ diagram, or from SED fitting.  We convert the reported color excesses to $A_V$ values in Table \ref{tab:TTSExts} using the relation $A_V = 1.1 E(V-K_s)$.

An advantage of the $V-K_s$ color is the long wavelength baseline, which allows relatively small extinction values to be detected.  
Also, infrared photometry is less susceptible to the bright bluer accretion excess that can appear in optical photometry of these targets.
However, a disadvantage is the possible confusion between reddening excess and disk excess, as mentioned above.

\subsection{Gas Column Measurements}\label{sec:GasExt}
\subsubsection{McJunkin 2014}
\citet{mcjunkin_direct_2014} use careful modeling of the Ly$\alpha$ emission expected from a young star to predict the original profile for the geocoronoal contamination near line center. They then employ a detailed radiative transfer model to account for the line of sight absorption in Ly$\alpha$ to estimate the neutral hydrogen column $N(H)$. The \citet{bohlin_NH1_1978} relation: $N(H)$ = $4.8\times 10^{21}$ $E(B-V)$ is used to compute $E(B-V)$. The authors also probe how different assumed reddening laws, $R_V$, can affect the measured $A_V$. 

The procedure relies on the ability to accurately predict the Ly$\alpha$ emission at the source and estimate the degree of absorption. 

\subsubsection{McJunkin 2016}
\citet{mcjunkin_empirically_2016} employ a slightly different method from \citet{mcjunkin_direct_2014}. They take the modeled Ly$\alpha$ profiles and measure the amount of H$_2$ absorption in the line of sight to the target. This method relies less on the ability to model the Ly$\alpha$ profile accurately, as the absorption measurements are generally taken from the much broader profile, which can be normalized in the narrow H$_2$ line ranges. 

The challenge to this method is that there is an unknown amount of H$_2$ self-absorption between the emitting source and the observer. The method is also generally unable to distinguish between H$_2$ in the circumstellar environment and that in the interstellar environment. Thus it is not clear how deeply the H$_2$ measurement probes. Furthermore, the conversion from N(H$_2$) to $A_V$ the authors present results in $A_V$ values which differ significantly from those derived in \citet{mcjunkin_direct_2014} for the same targets.  

The assumption may be that the H$_2$-based measurement probes the circumstellar environment, whereas the Ly$\alpha$-based measurement can only capture interstellar absorption. In either case, the question of what the emitting source to which extinction is being measured is remains consequential to the interpretation of the measurement.

\subsubsection{XMM Newton XEST}
\citet{gudel_xest_2007} perform model fits to X-ray spectra from the XMM Newton satellite using XSPEC \citep{arnaud_xspec_1996}. They fit the $vapec$ thermal collisional equilibrium ionization model to their spectra, using single and double spectral component fits, and a separate, power-law based model. They allow the $N(H)$ out to their targets to vary as a model parameter. We convert this value to an $A_V$ value following the prescription in \citet{mcjunkin_direct_2014, mcjunkin_empirically_2016}, where $A_V = 3.1 \times \frac{4.8\times 10^{21}}{N(H)}$. The $A_V$ values we compute from the $N(H)$ are reported in Table \ref{tab:TTSExts}.

Although this method avoids the problem of optical variability in young star systems, it still may be affected by X-ray variability from extreme flare activity. It also depends on the gas-to-dust conversion assumed along the line of sight to compute a color excess or extinction value.

\section{Line profiles and Integration Limits for Individual DIBs} \label{appendix:EqW}

In this Appendix, we illustrate for each source in our sample the individual DIBs features
and the integration limits used for our equivalent width measurements. 

\begin{figure}[!htb]
    \centering
    \includegraphics[width=0.22\linewidth]{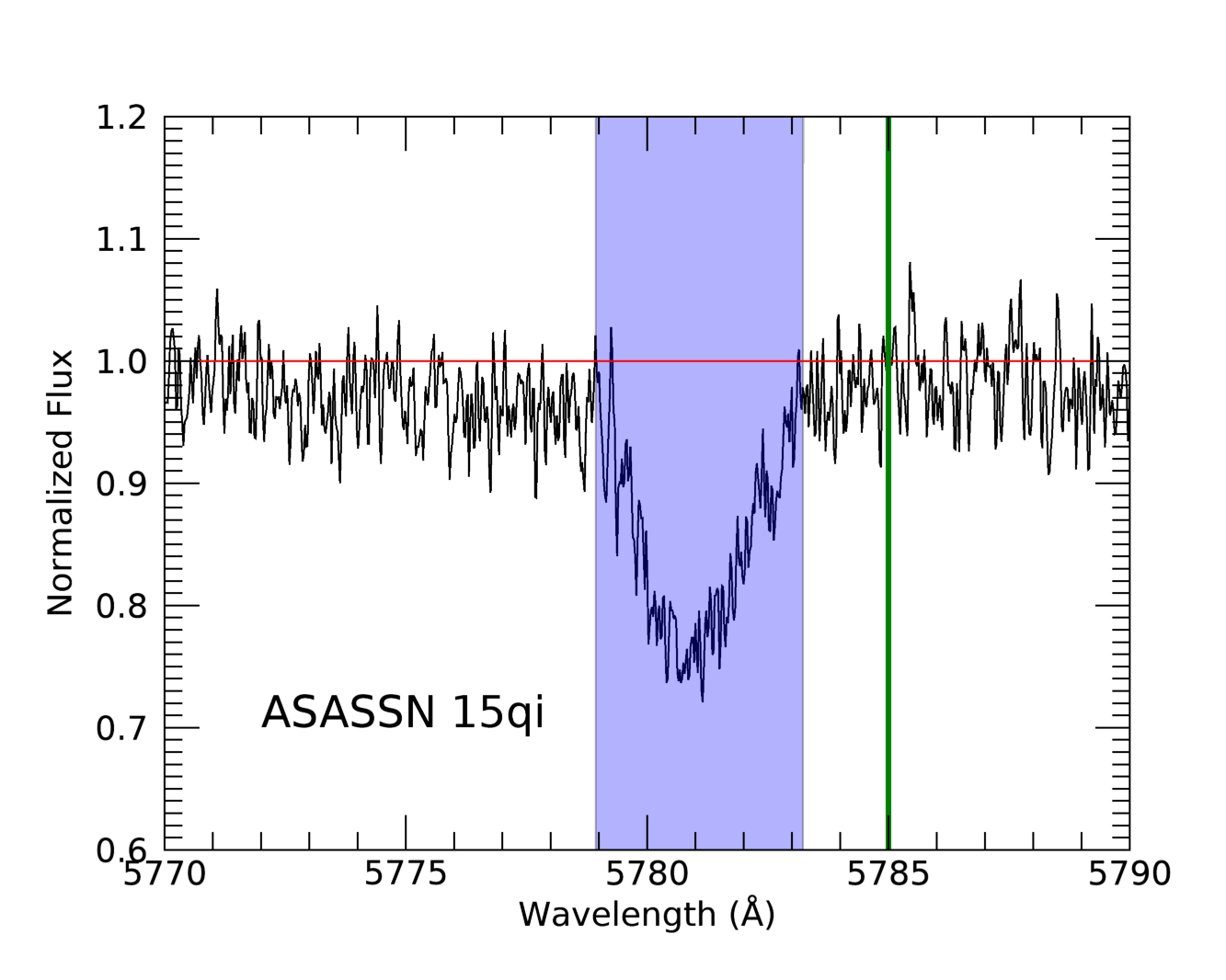}
        \includegraphics[width=0.22\linewidth]{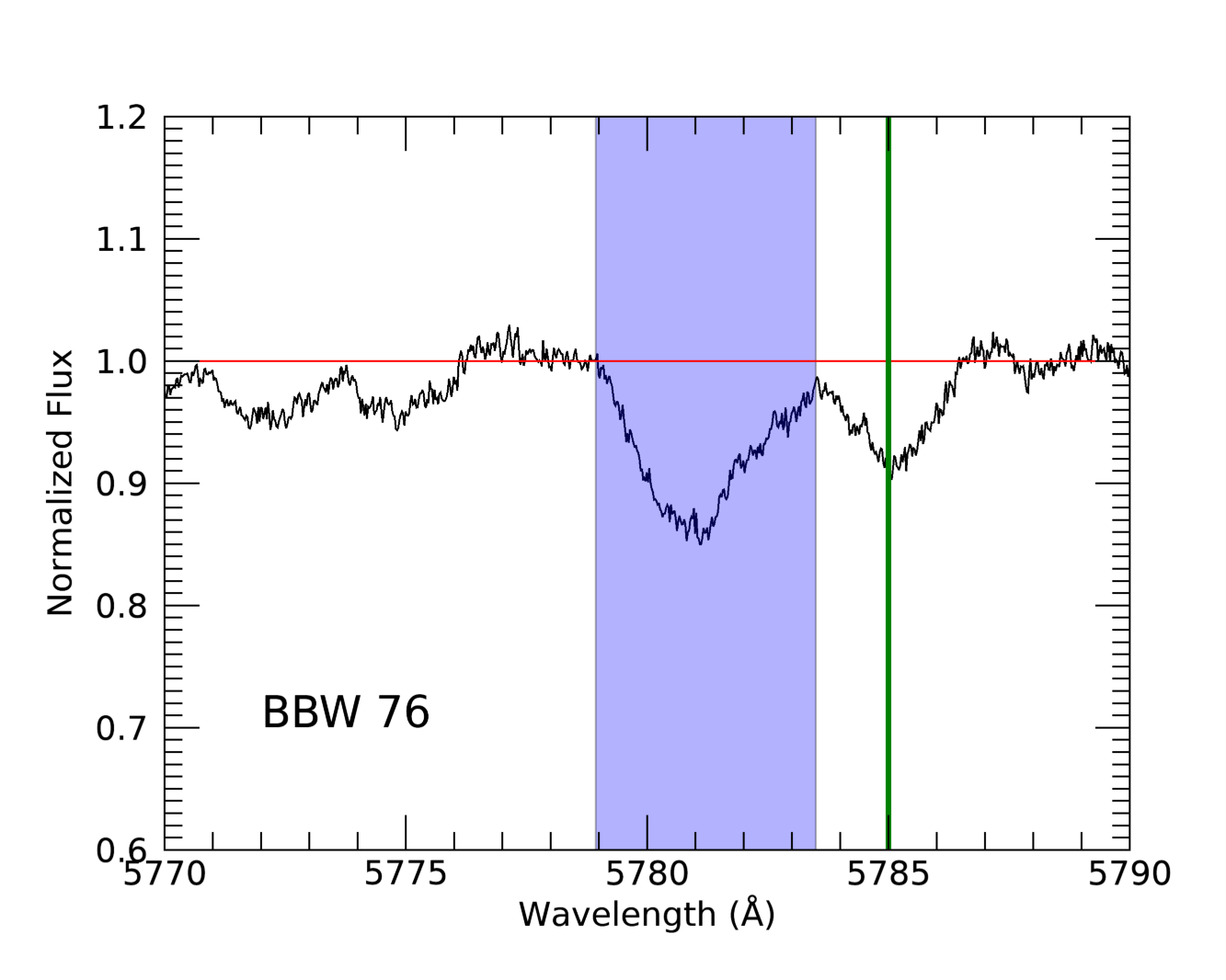}
        \includegraphics[width=0.22\linewidth]{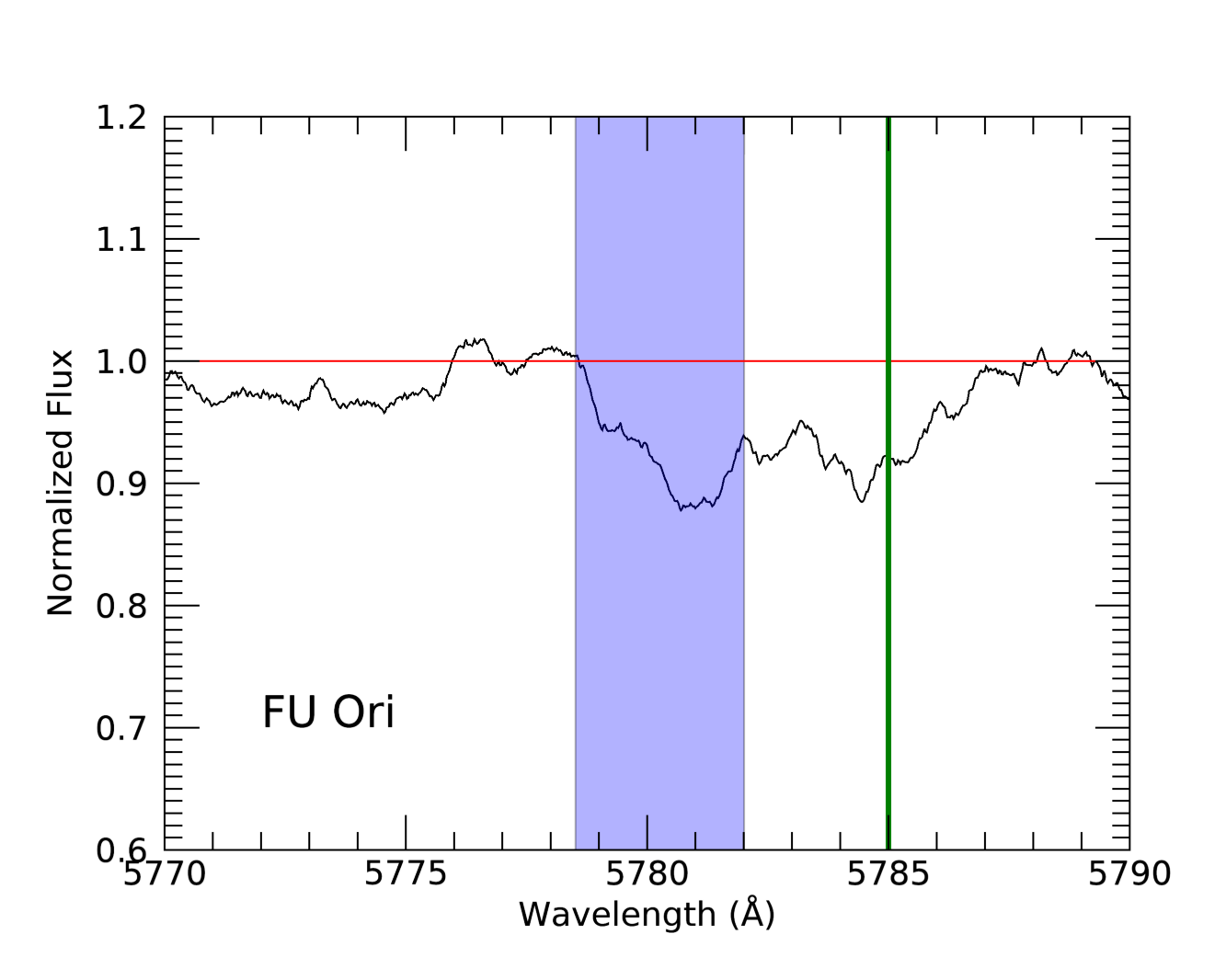}
    \includegraphics[width=0.22\linewidth]{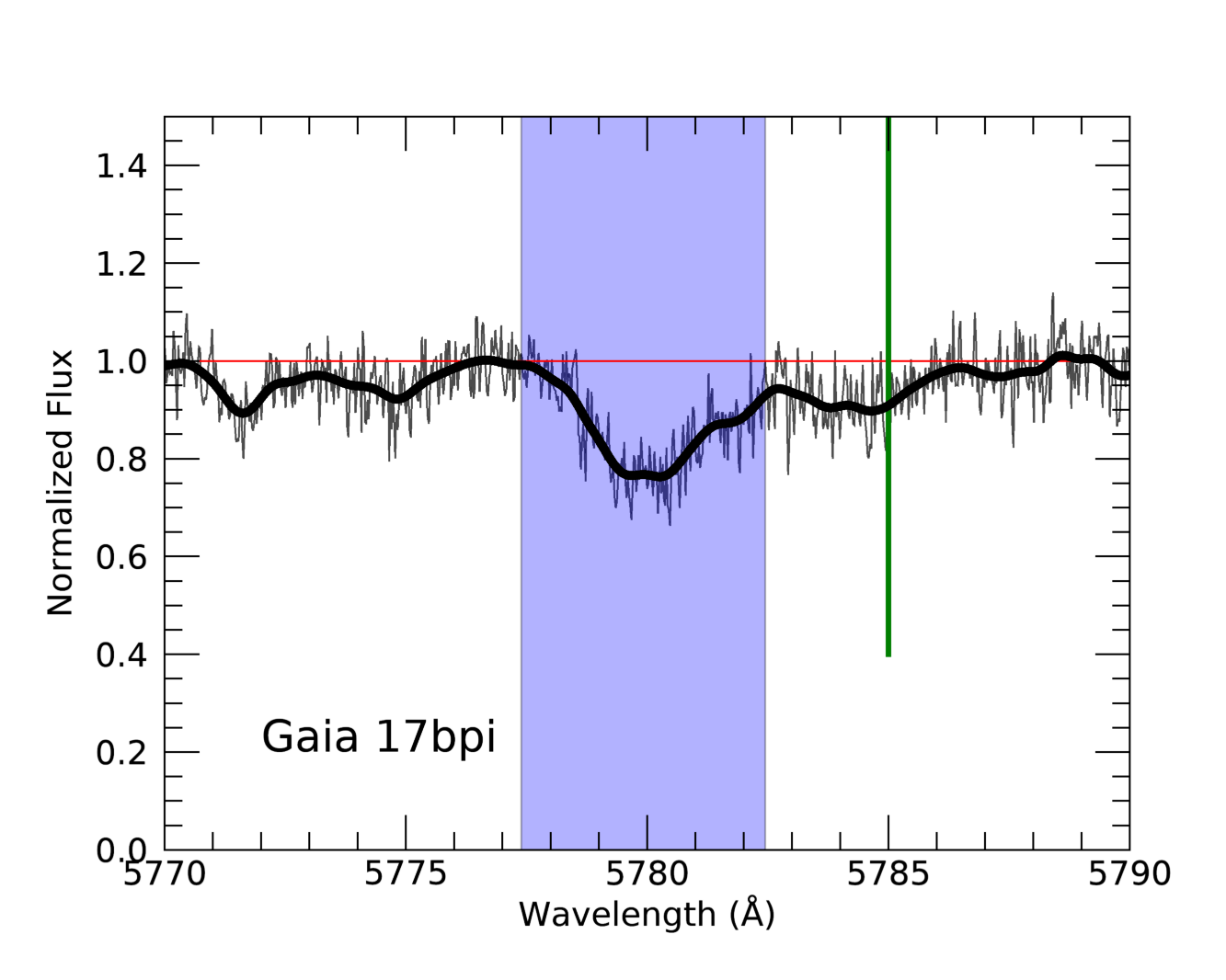}
        \includegraphics[width=0.22\linewidth]{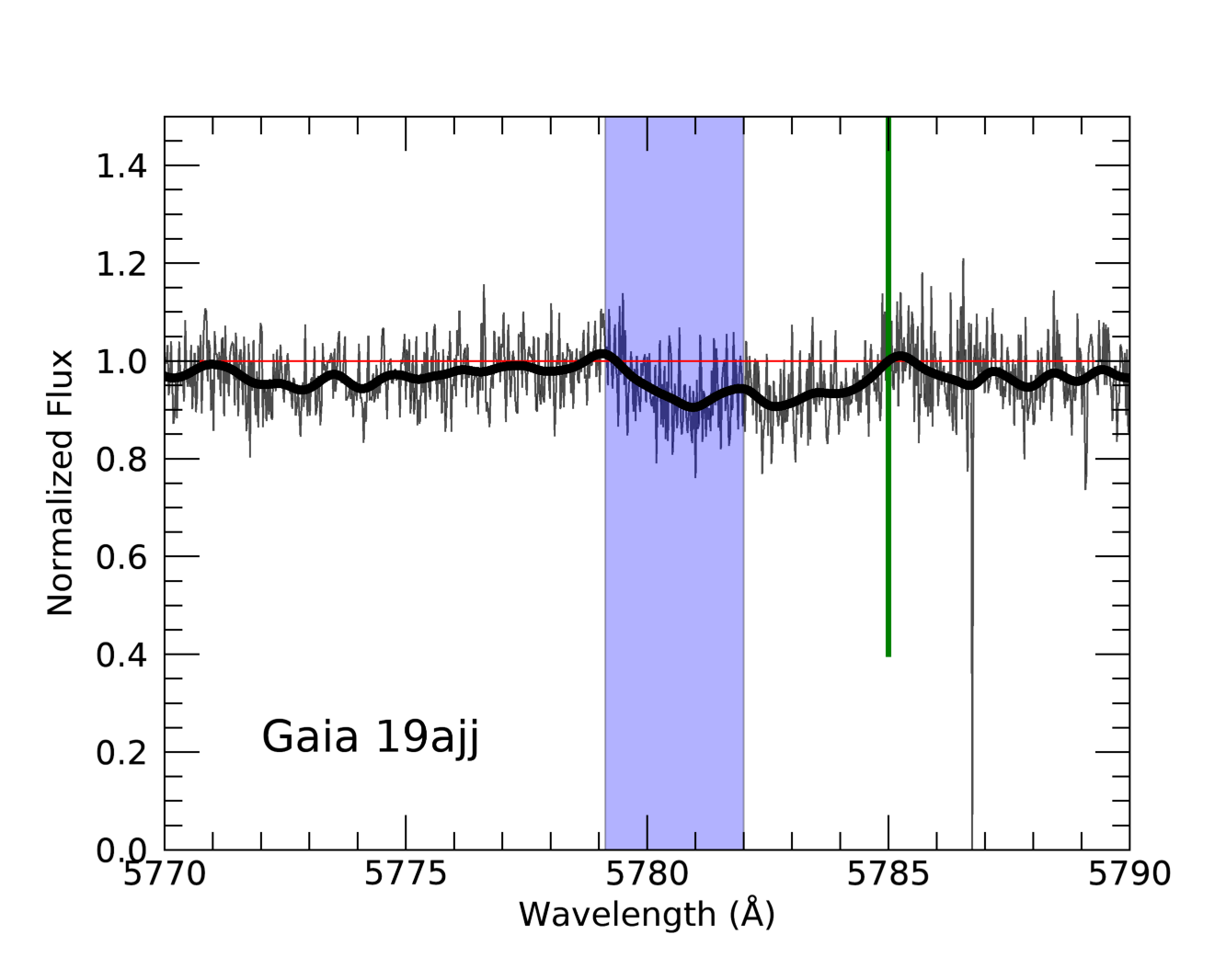}
        \includegraphics[width=0.22\linewidth]{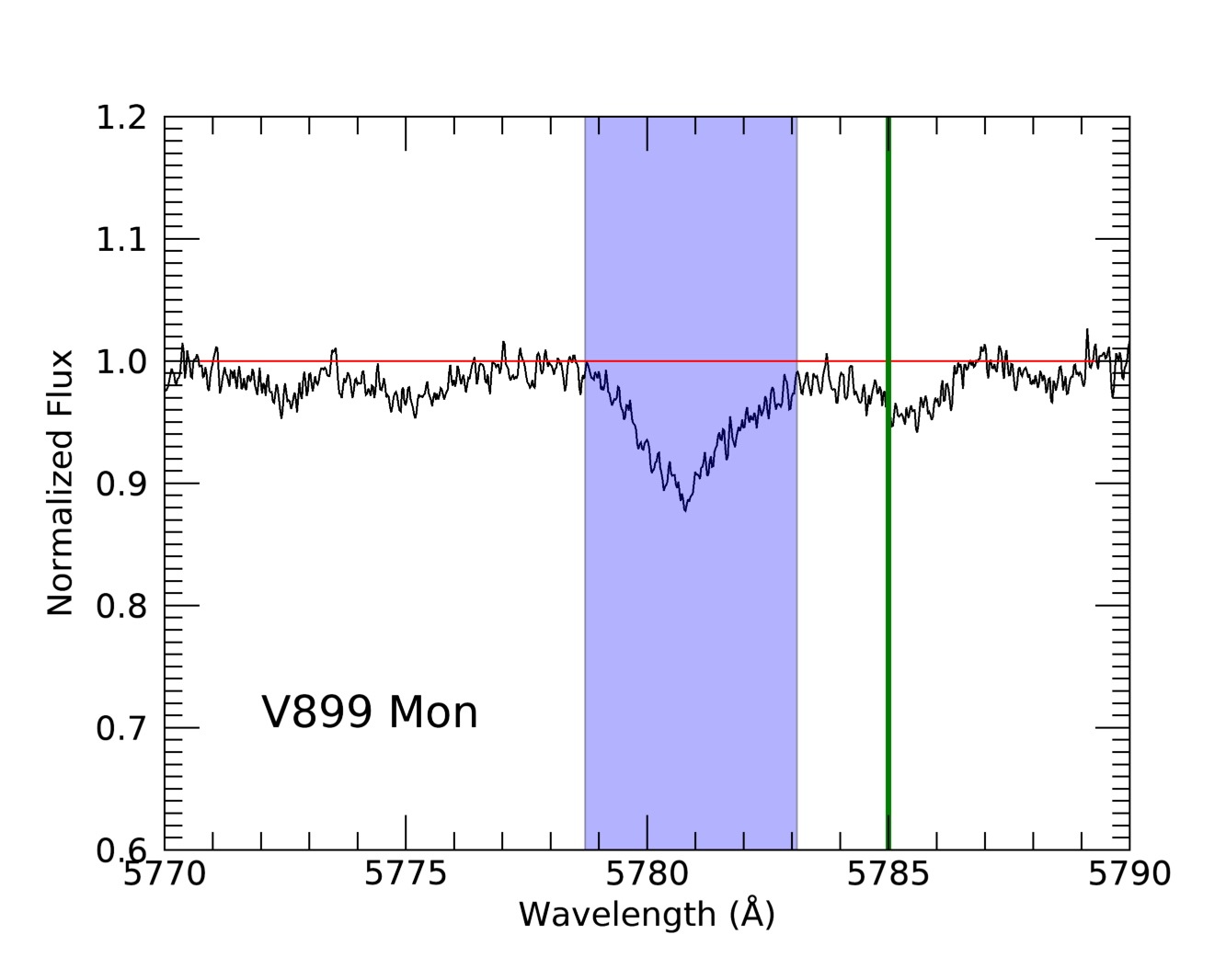}
    \includegraphics[width=0.22\linewidth]{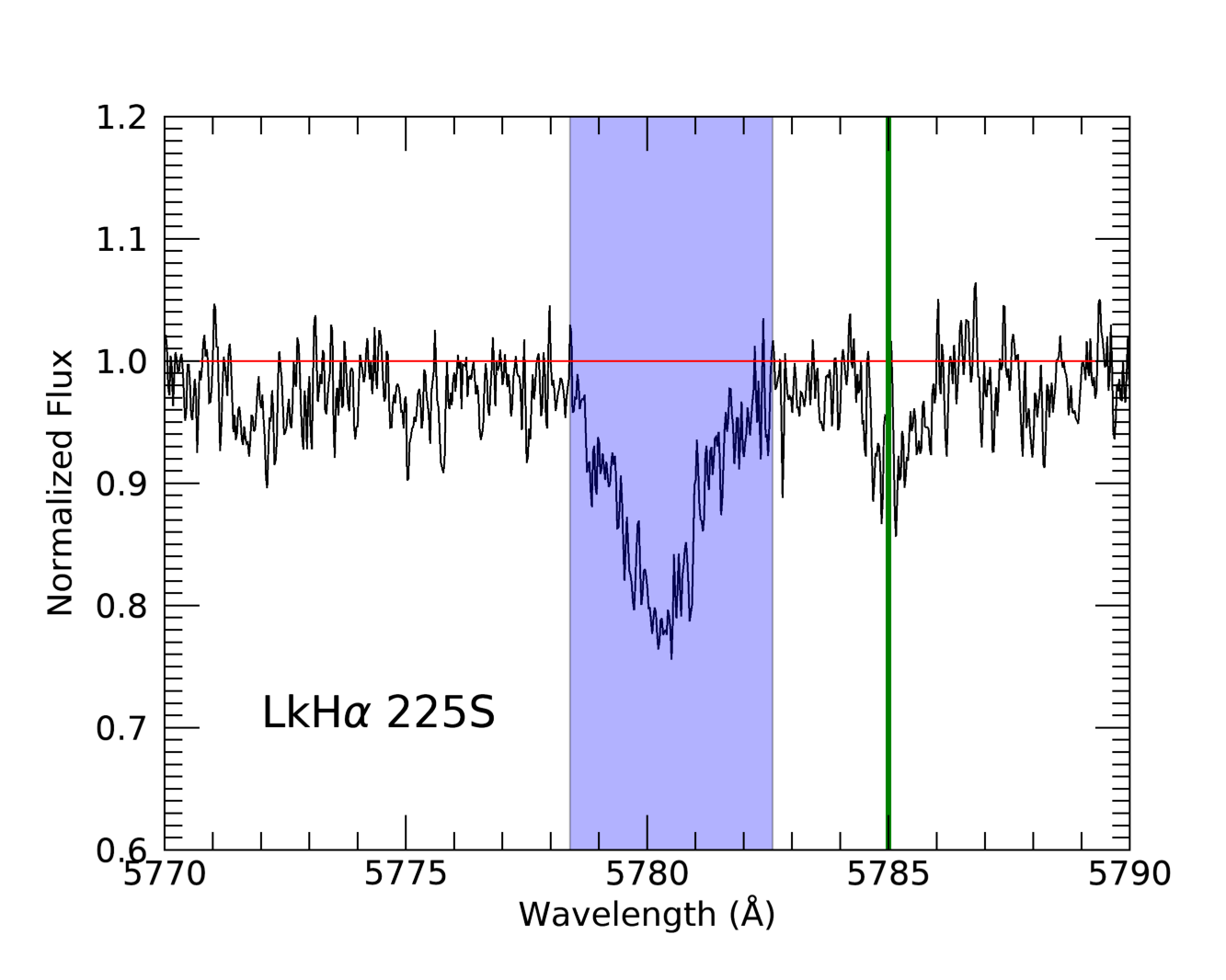}
        \includegraphics[width=0.22\linewidth]{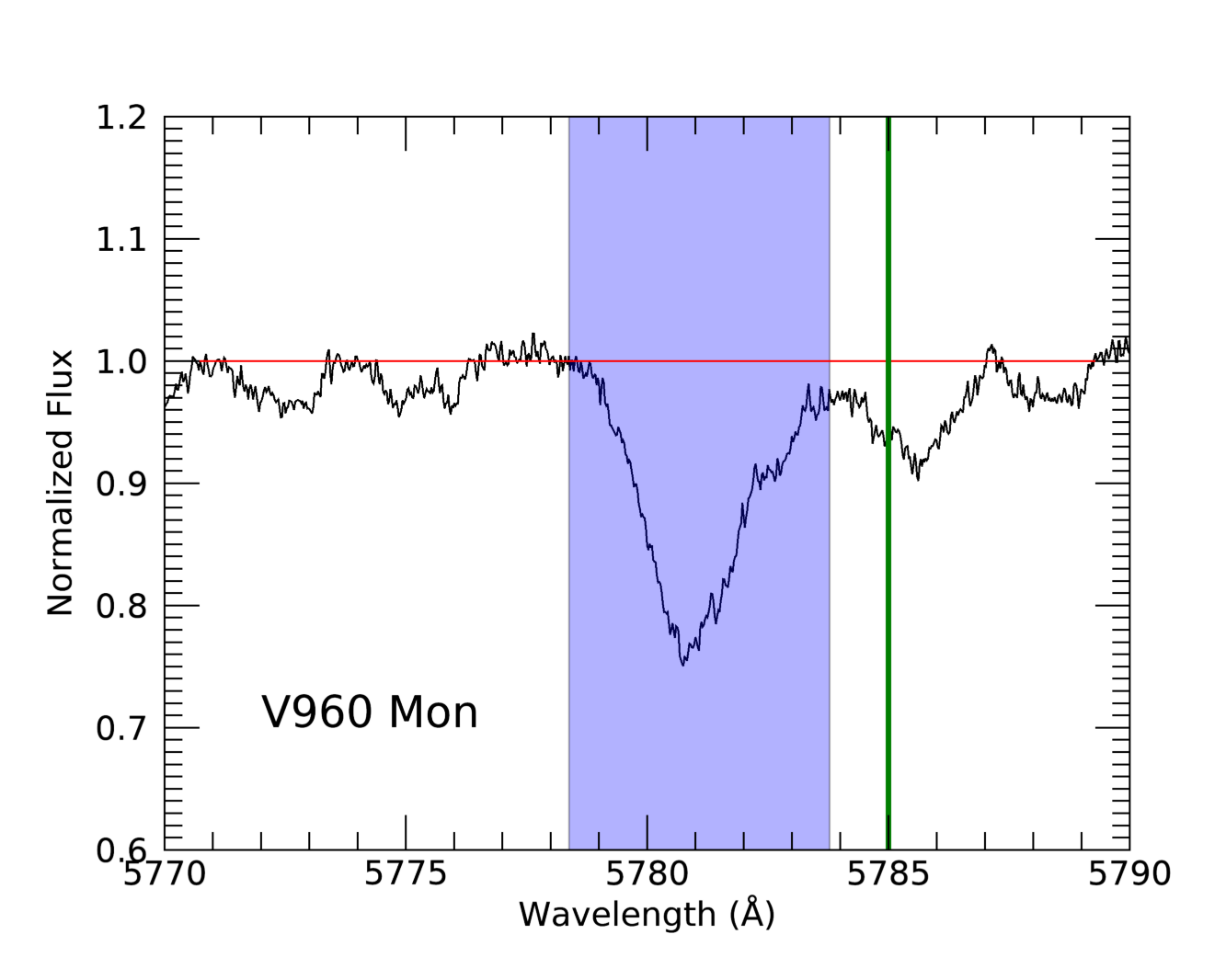}
        \includegraphics[width=0.22\linewidth]{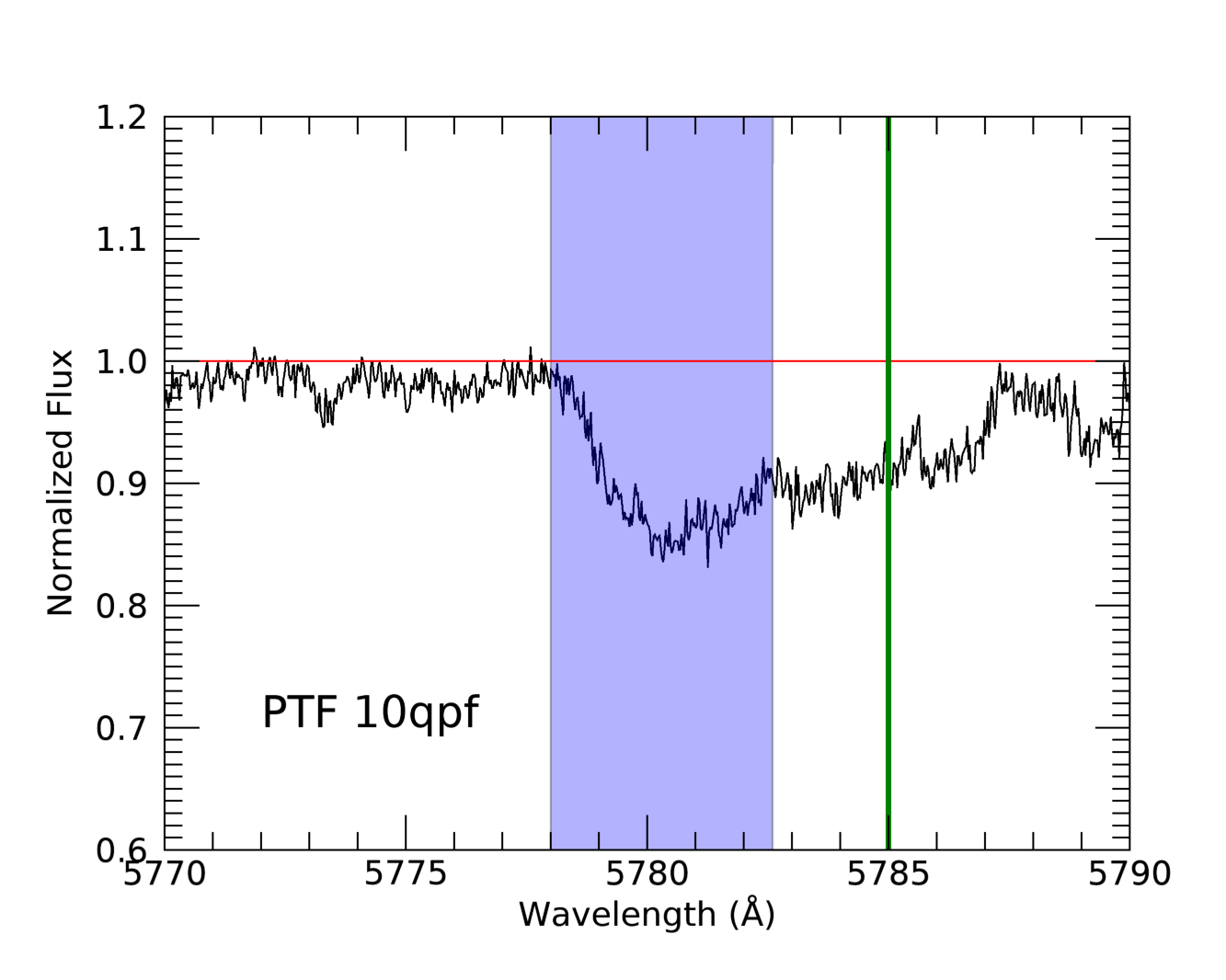}
    \includegraphics[width=0.22\linewidth]{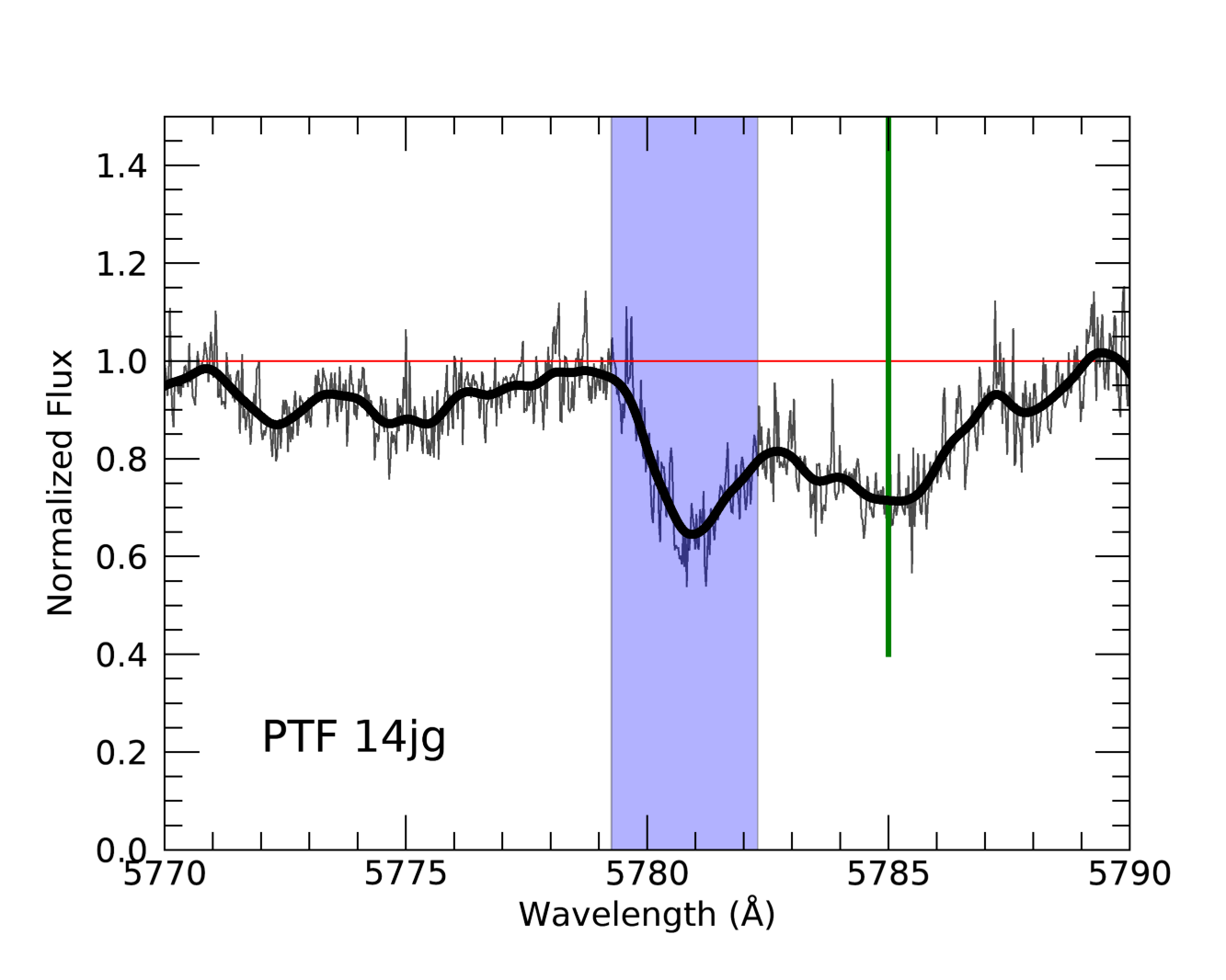}
        \includegraphics[width=0.22\linewidth]{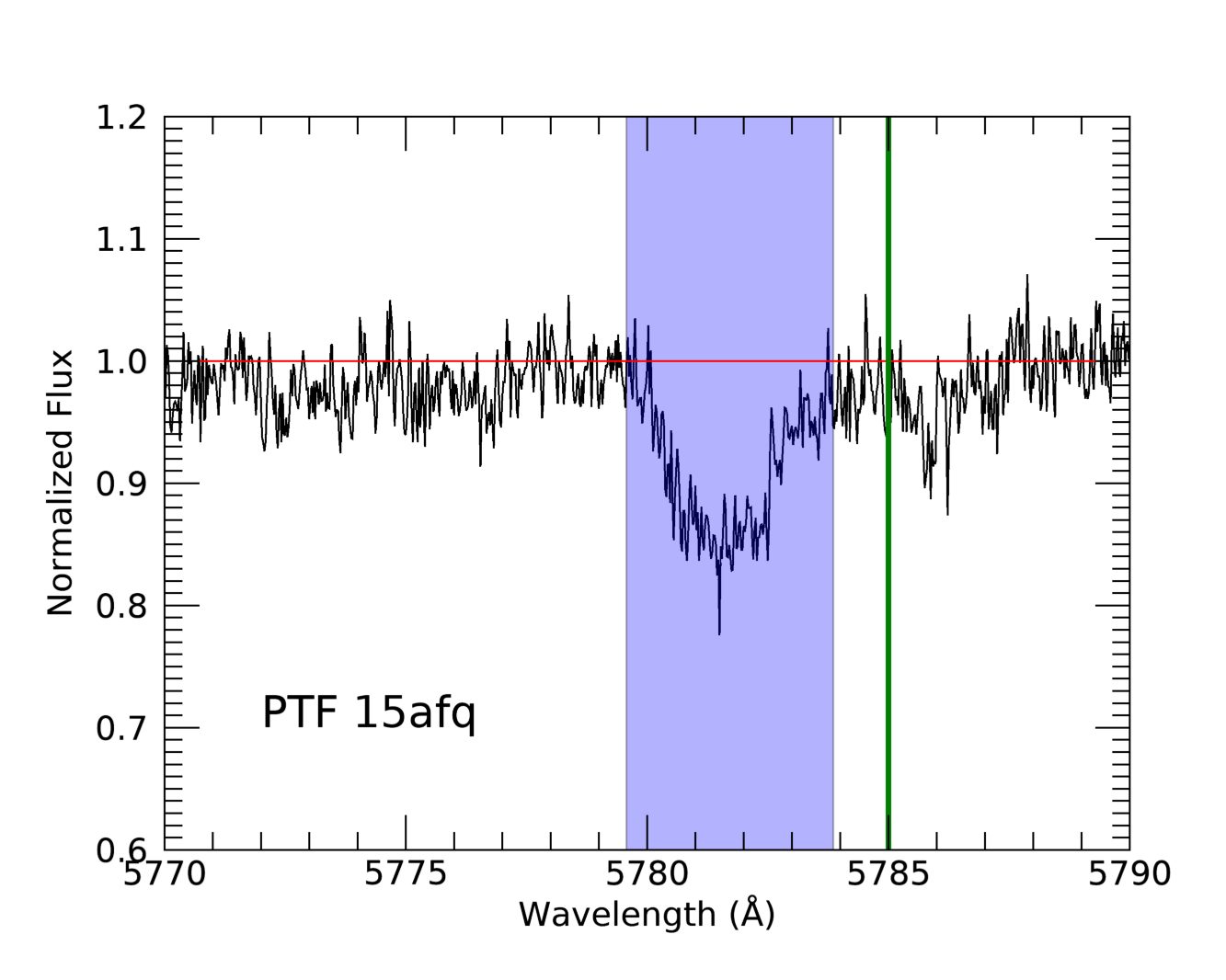}
        \includegraphics[width=0.22\linewidth]{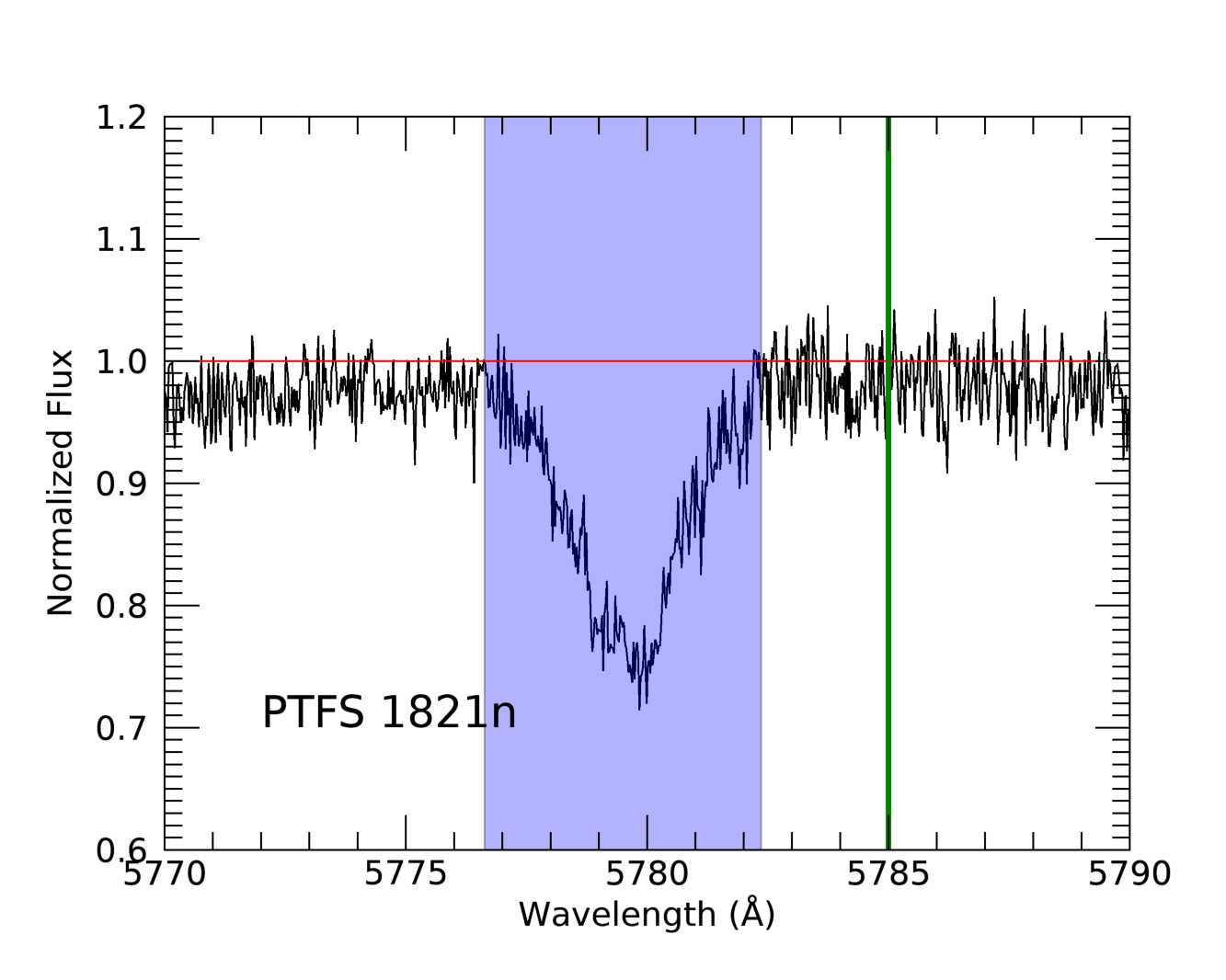}
    \includegraphics[width=0.22\linewidth]{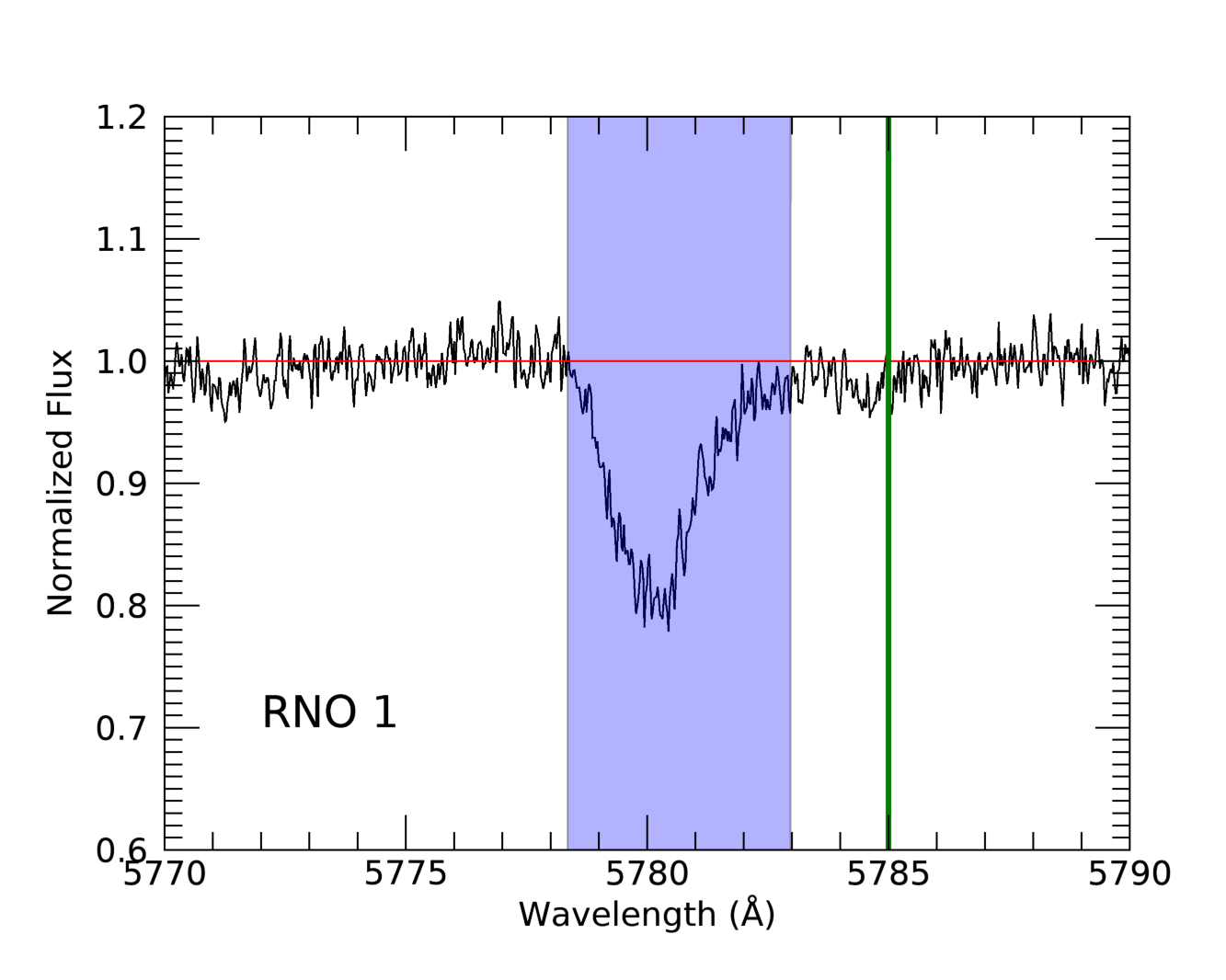}
        \includegraphics[width=0.22\linewidth]{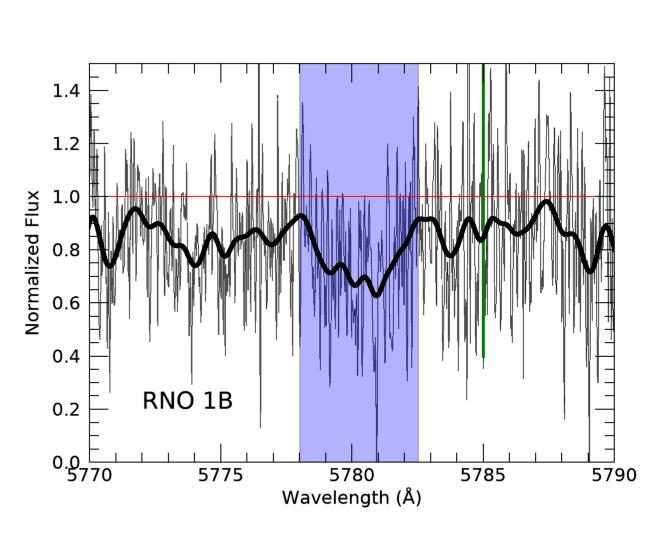}
        \includegraphics[width=0.22\linewidth]{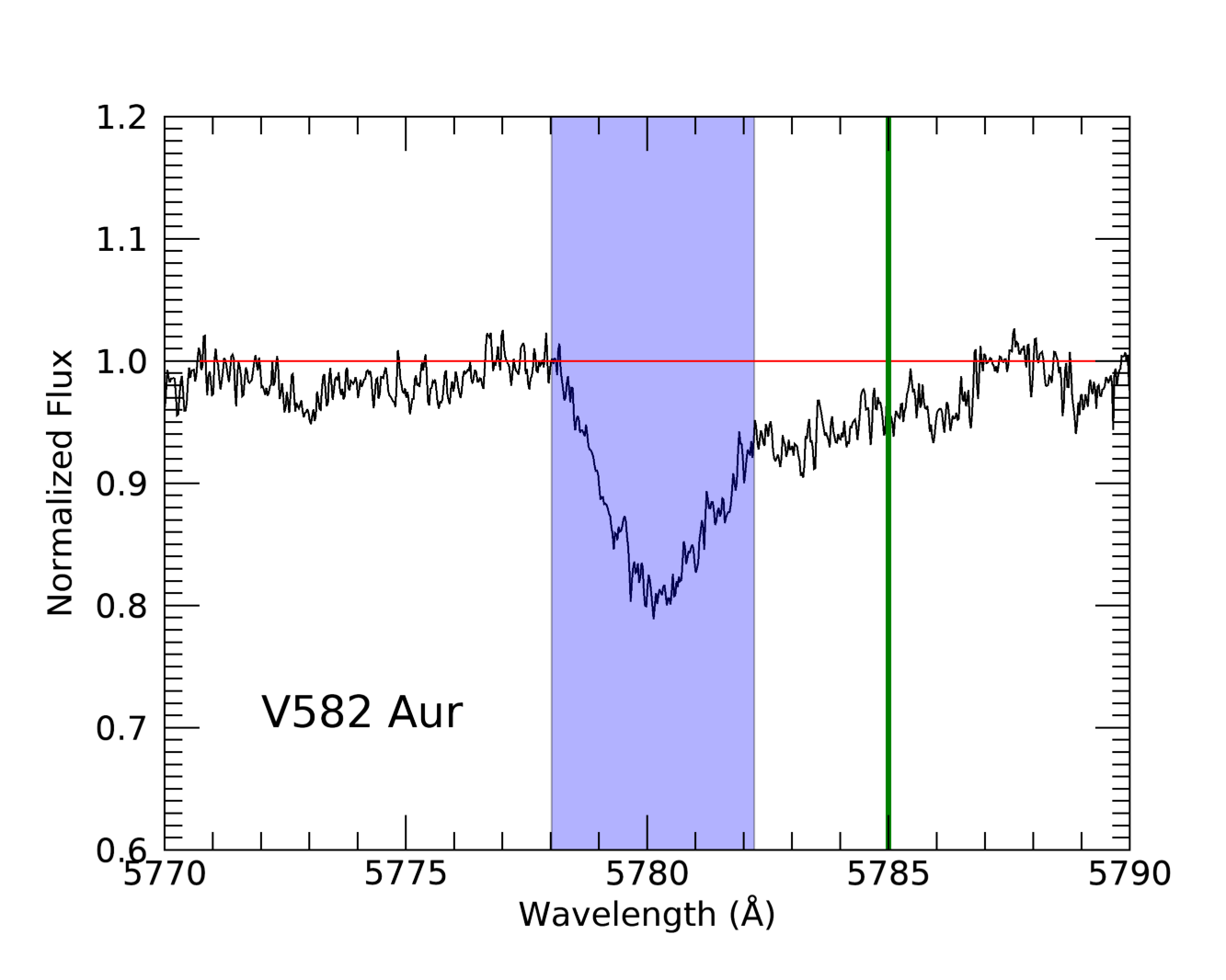}
    \includegraphics[width=0.22\linewidth]{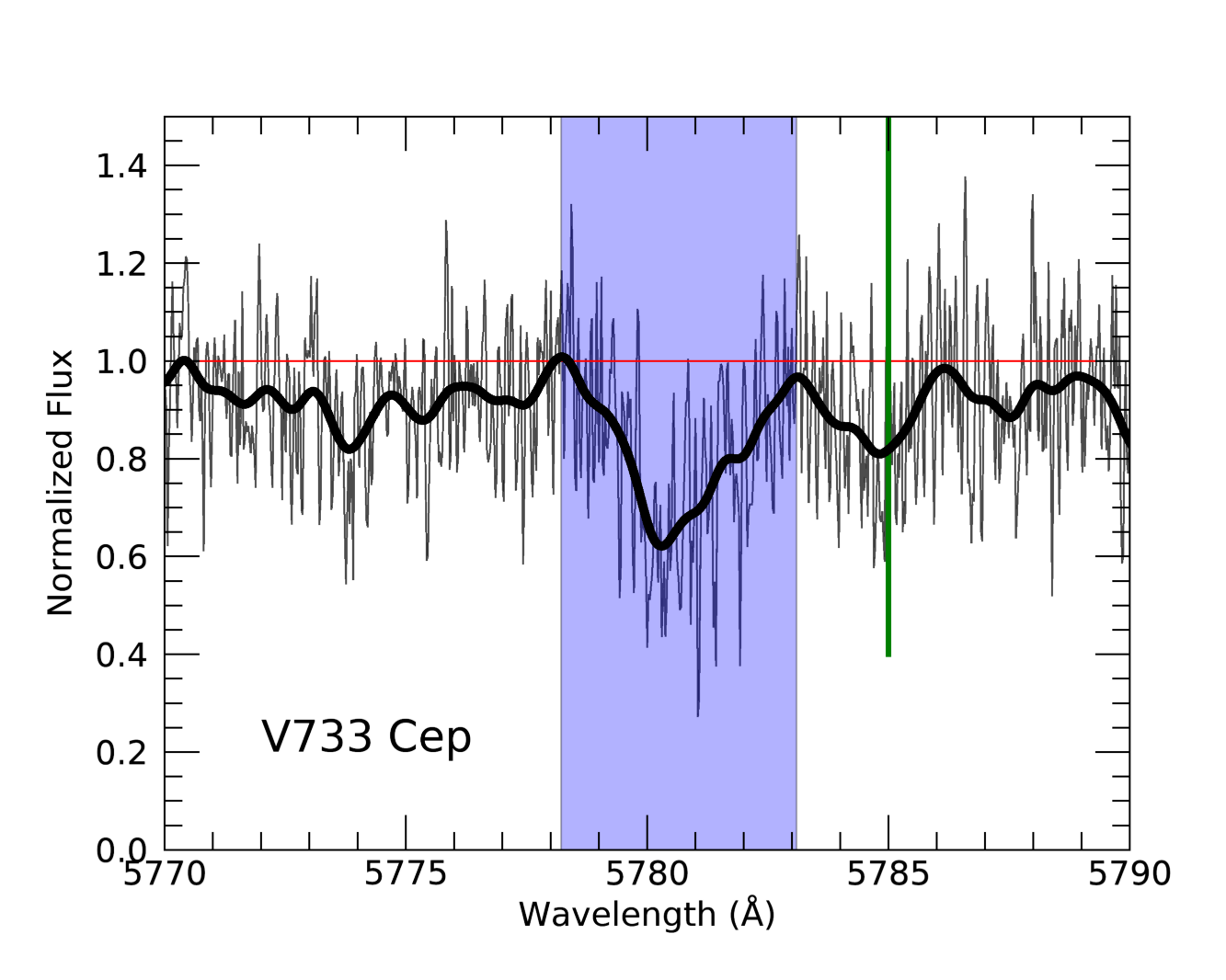}
        \includegraphics[width=0.22\linewidth]{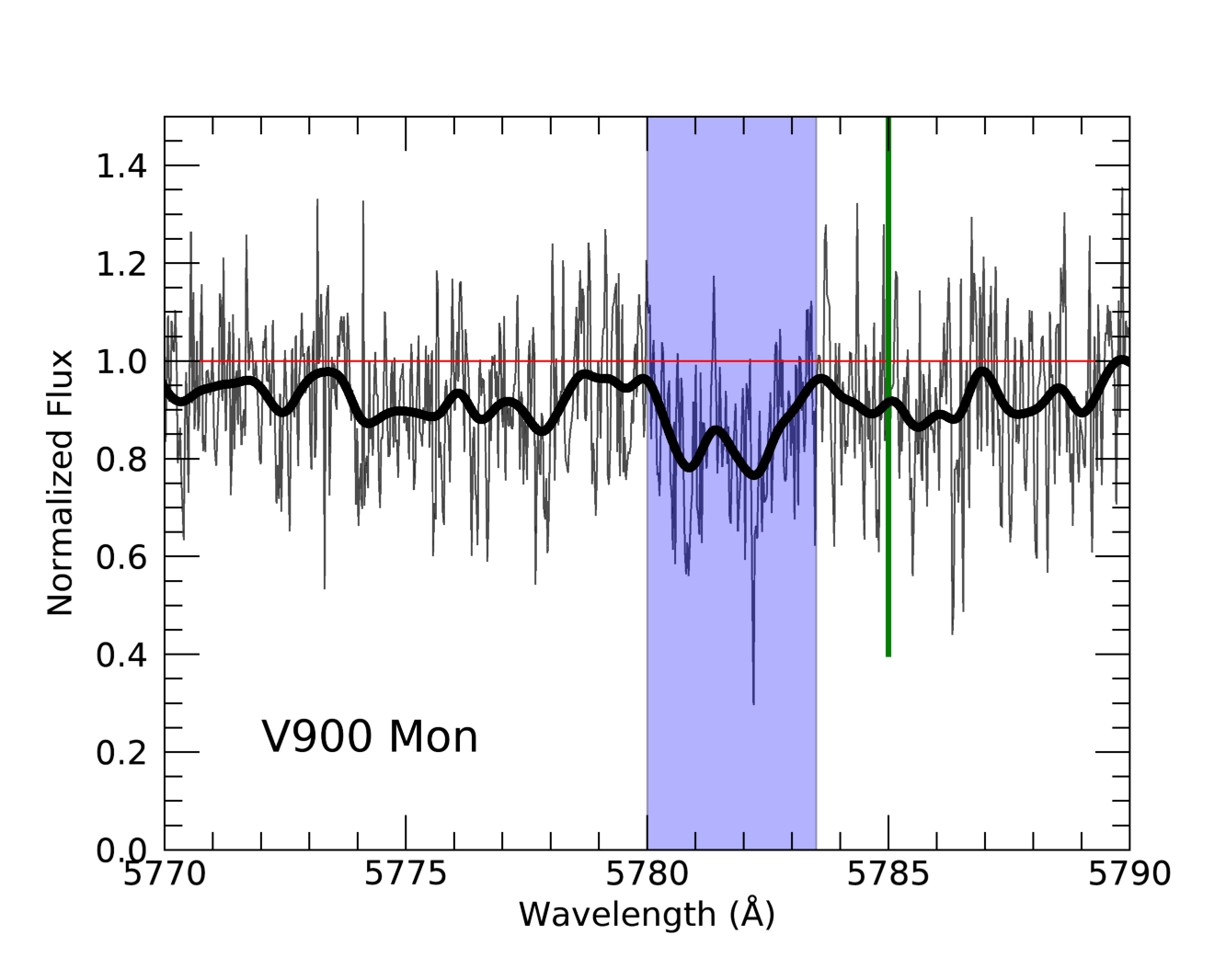}
        \includegraphics[width=0.22\linewidth]{v1057cyg_5780IntegrationSample.png}
    \includegraphics[width=0.22\linewidth]{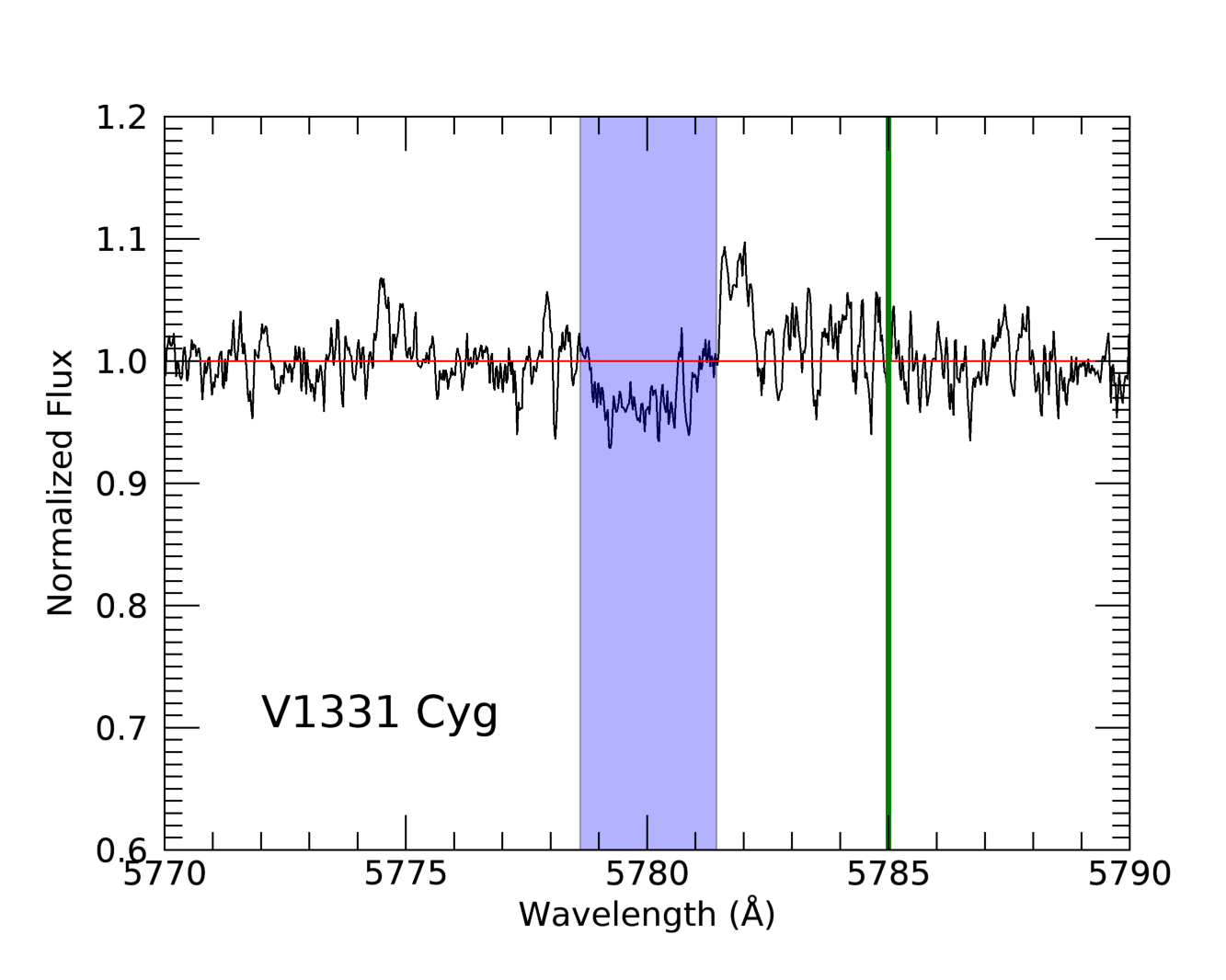}
        \includegraphics[width=0.22\linewidth]{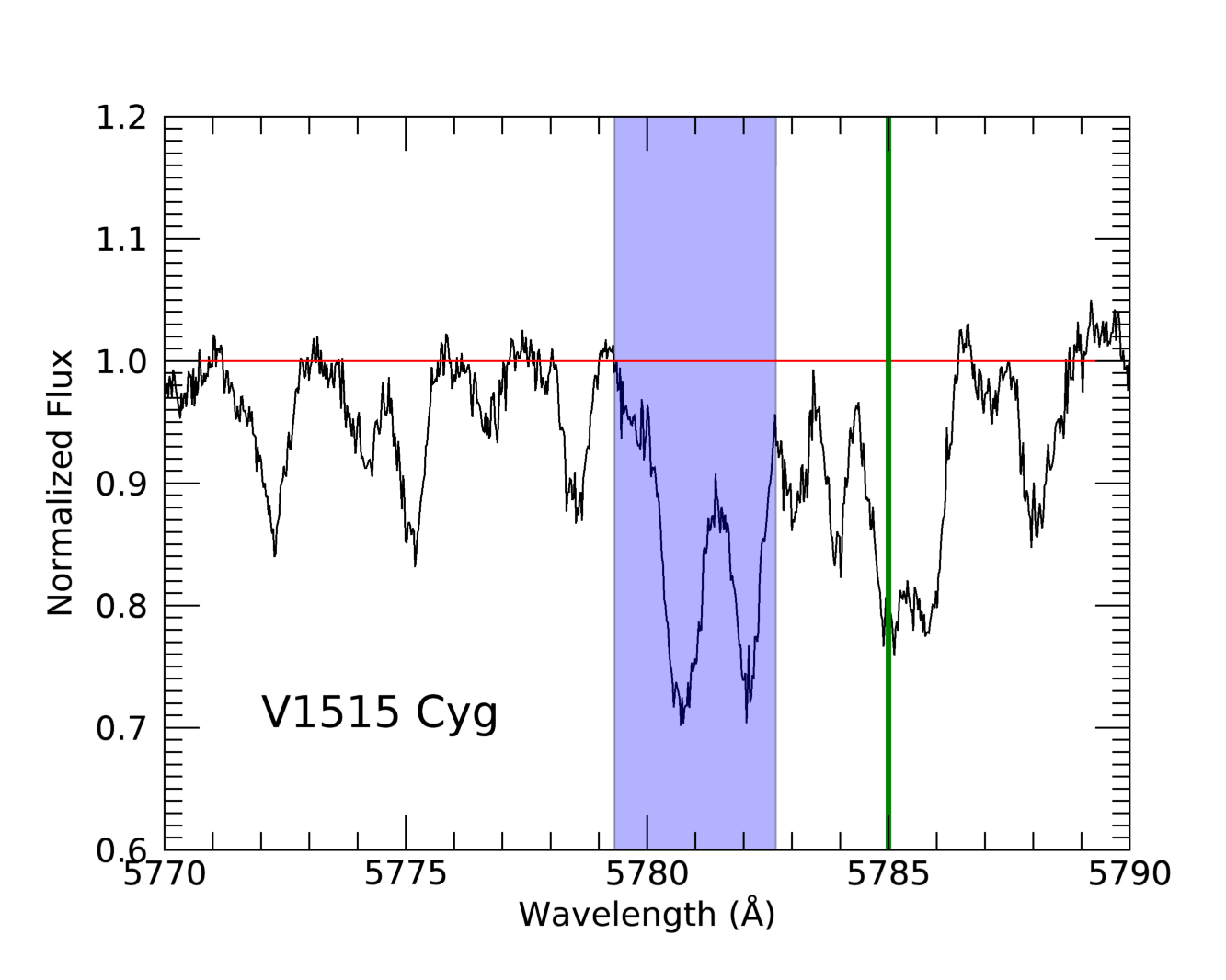}
        \includegraphics[width=0.22\linewidth]{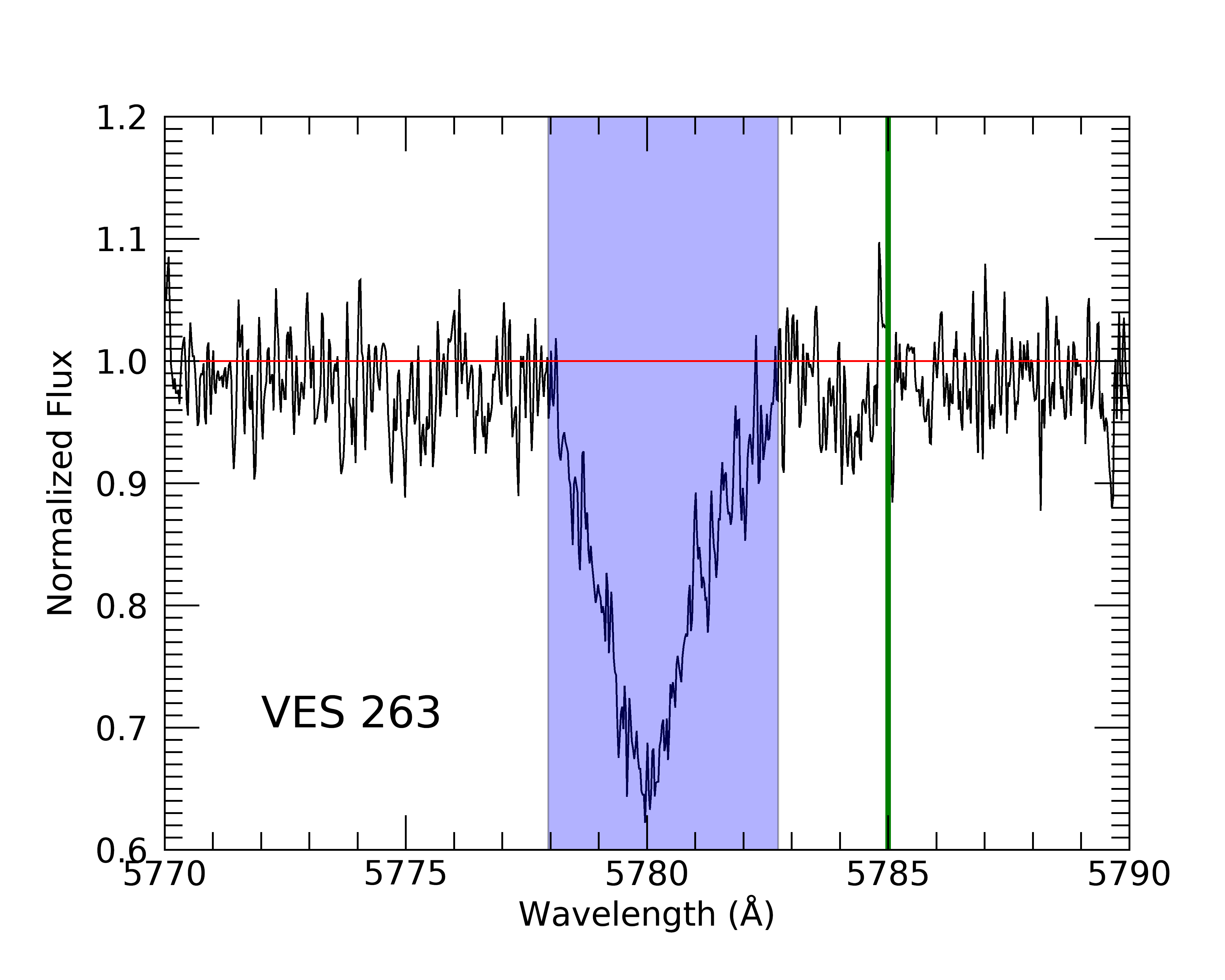}
    \includegraphics[width=0.22\linewidth]{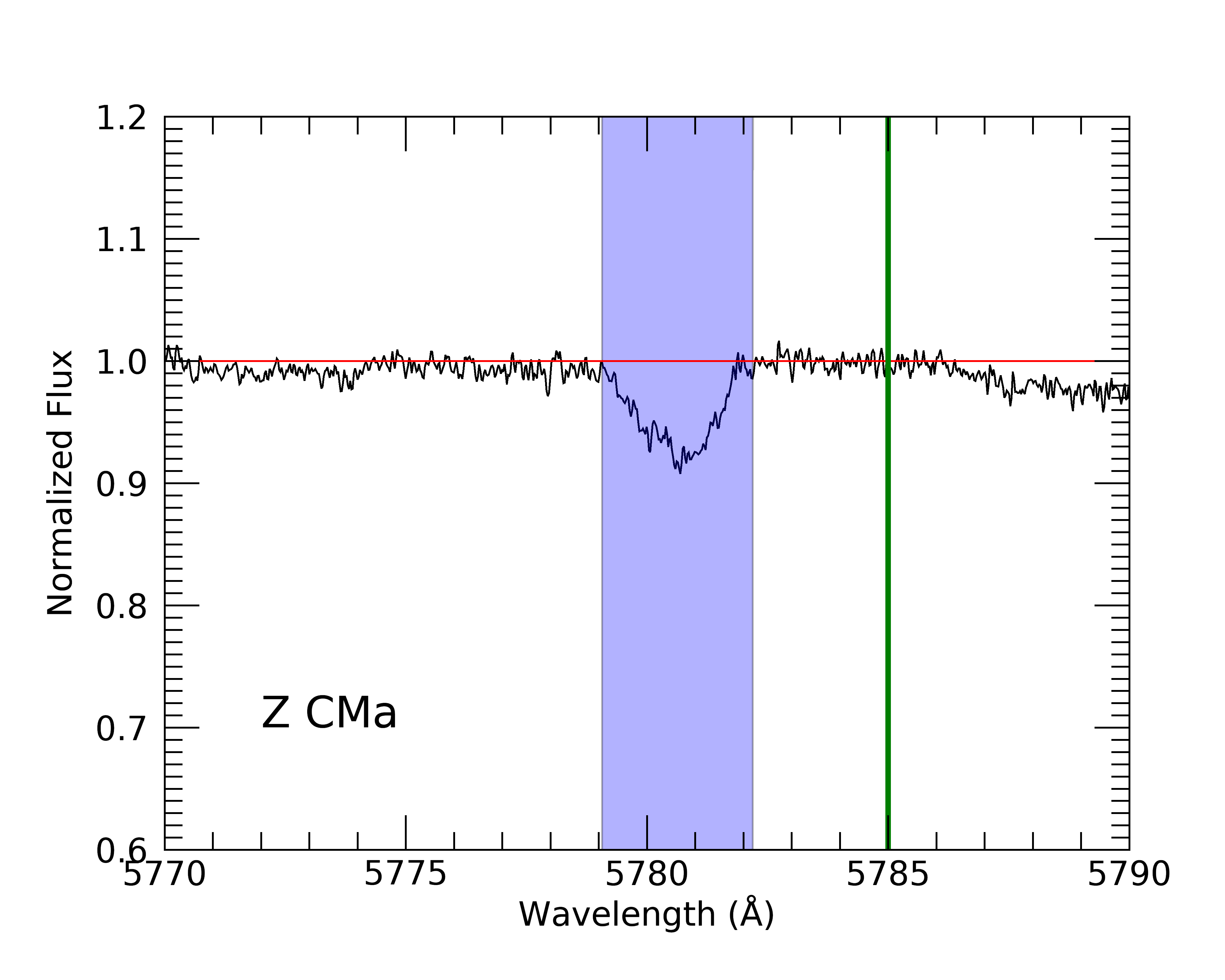}
    \caption{The $\lambda$5780 DIB in our sample. The blue region shows the region of integration for the equivalent width calculation. The red line lies at 1.0 for reference. For the noisiest spectra (Gaia 17bpi, Gaia 19ajj, PTF 14jg, RNO 1B, V733 Cep, V900 Mon) a $\sigma=10$ pixels Gaussian-smoothing of the spectrum is shown in black, with the raw data shown in grey. The green vertical line marks the center of the $\lambda$5785 band used for the stellar contamination correction.}
    \label{fig:eqWIntegration5780s}
\end{figure}
\begin{figure}[!htb]
    \centering
    \includegraphics[width=0.22\linewidth]{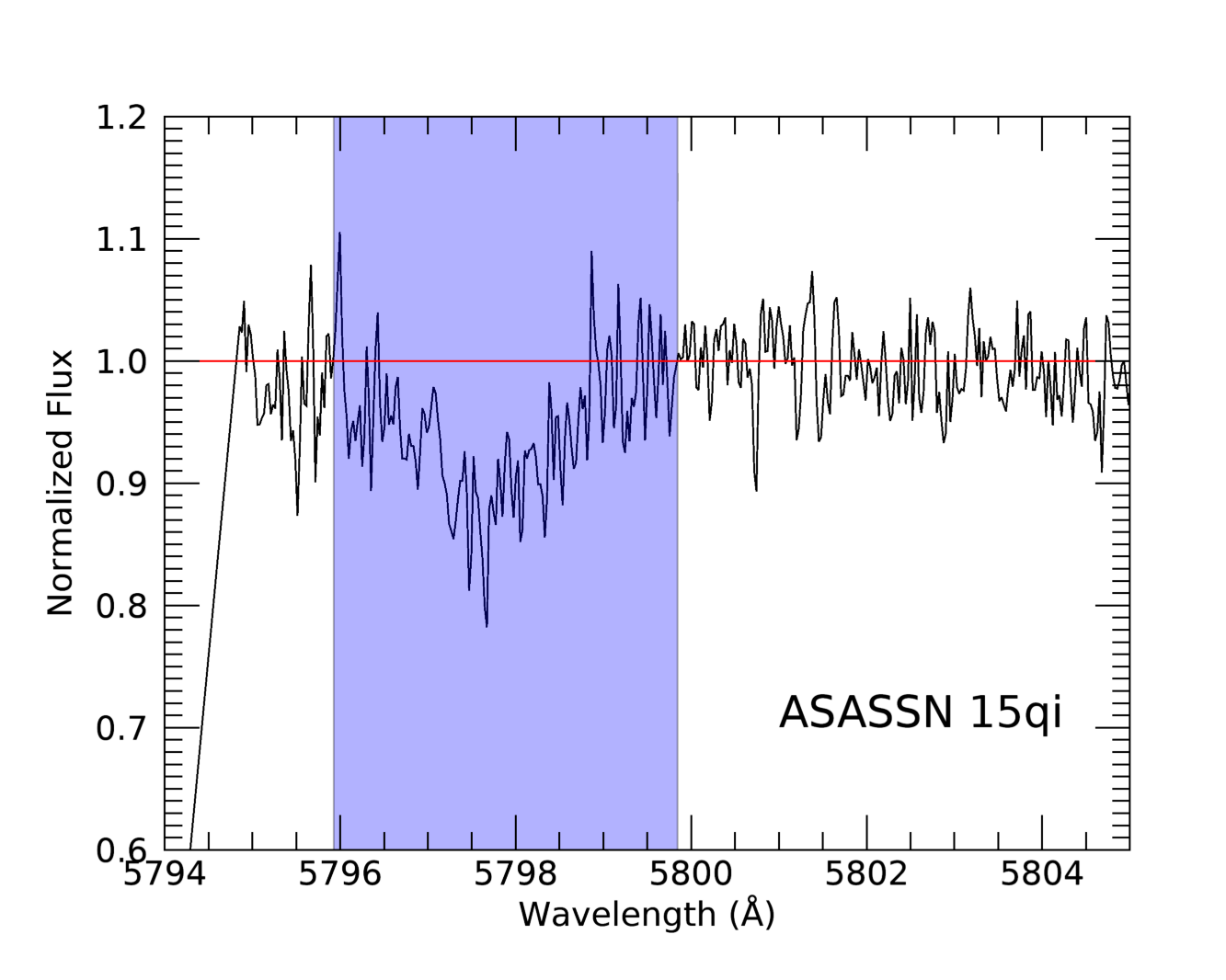}
        \includegraphics[width=0.22\linewidth]{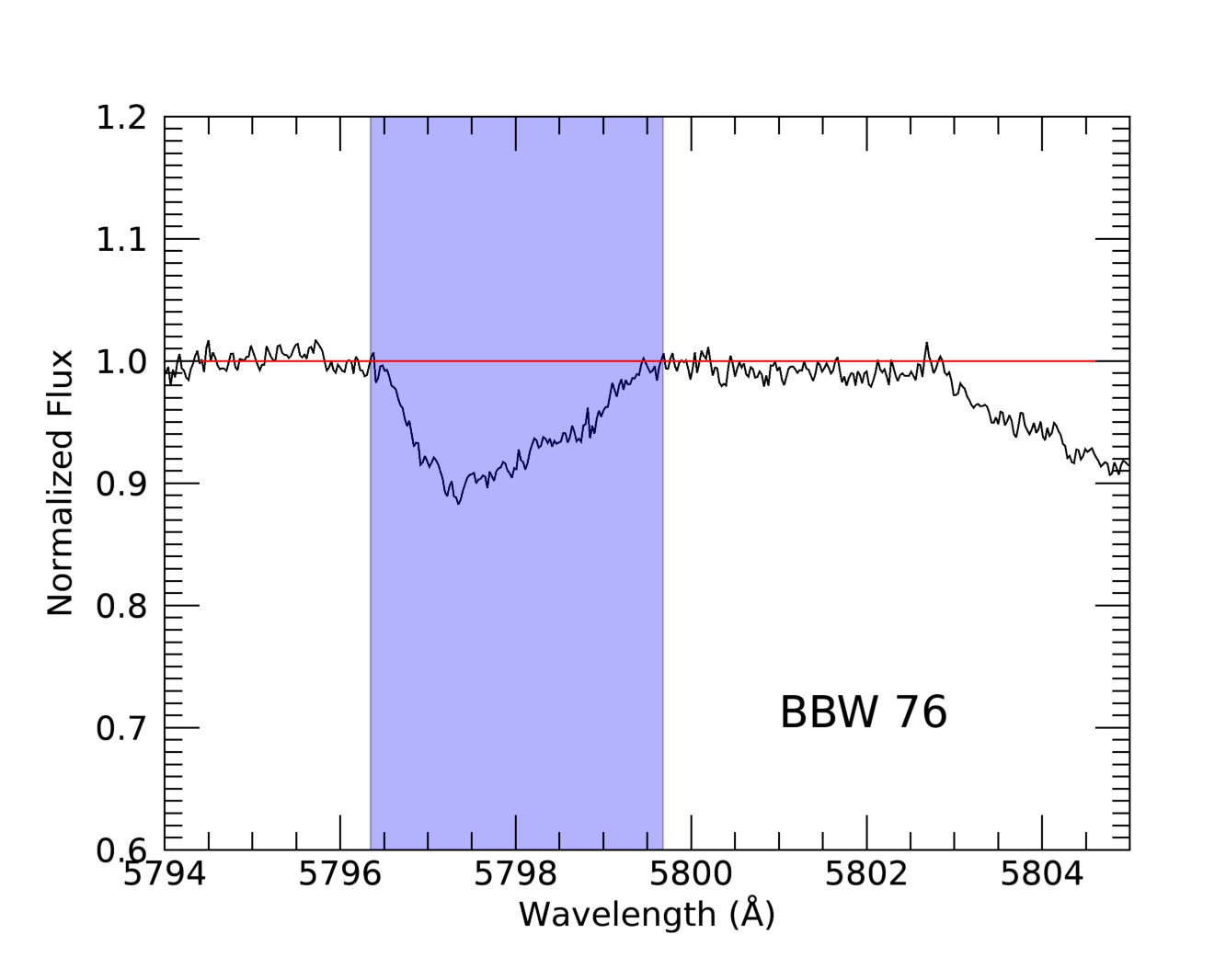}
        \includegraphics[width=0.22\linewidth]{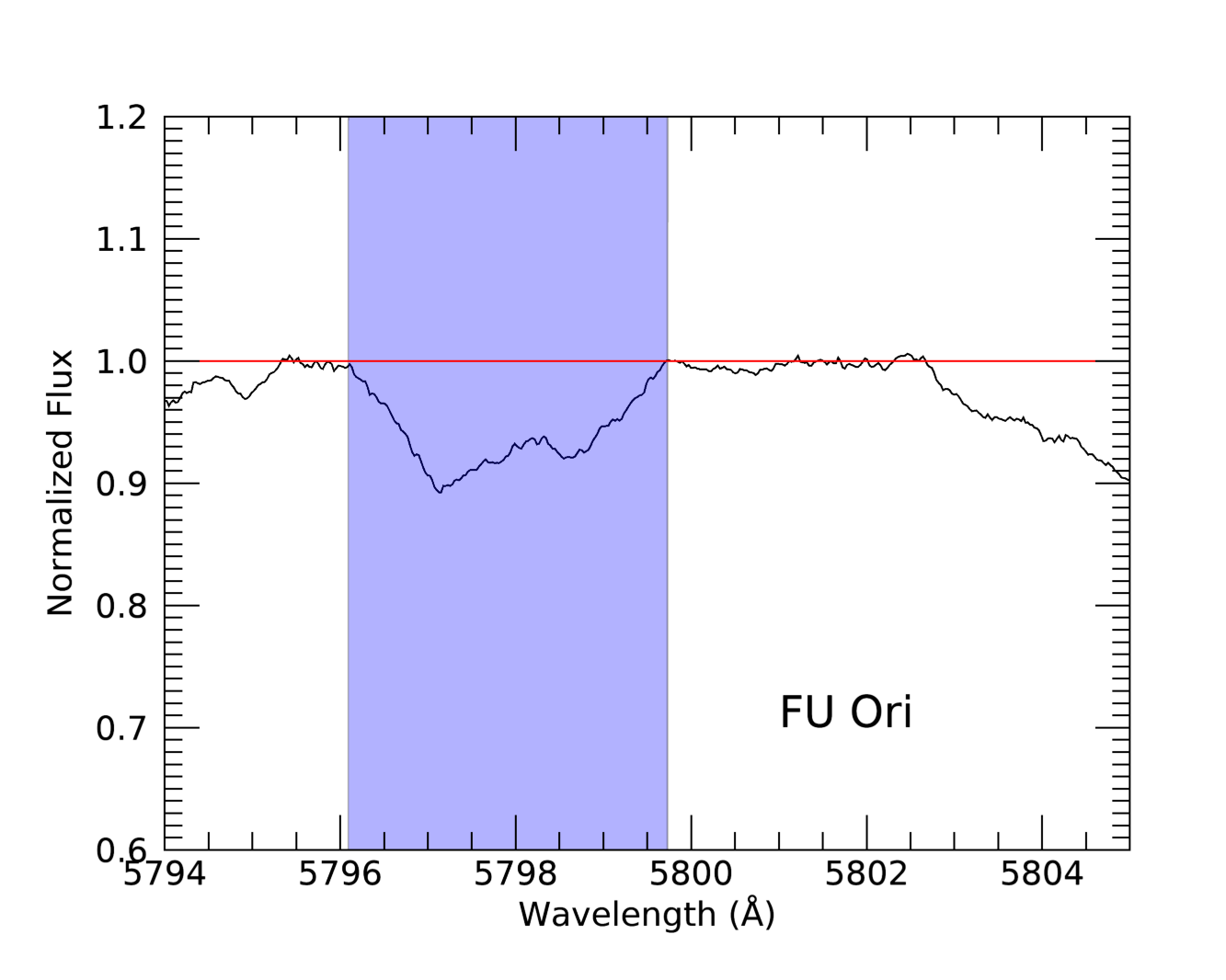}
    \includegraphics[width=0.22\linewidth]{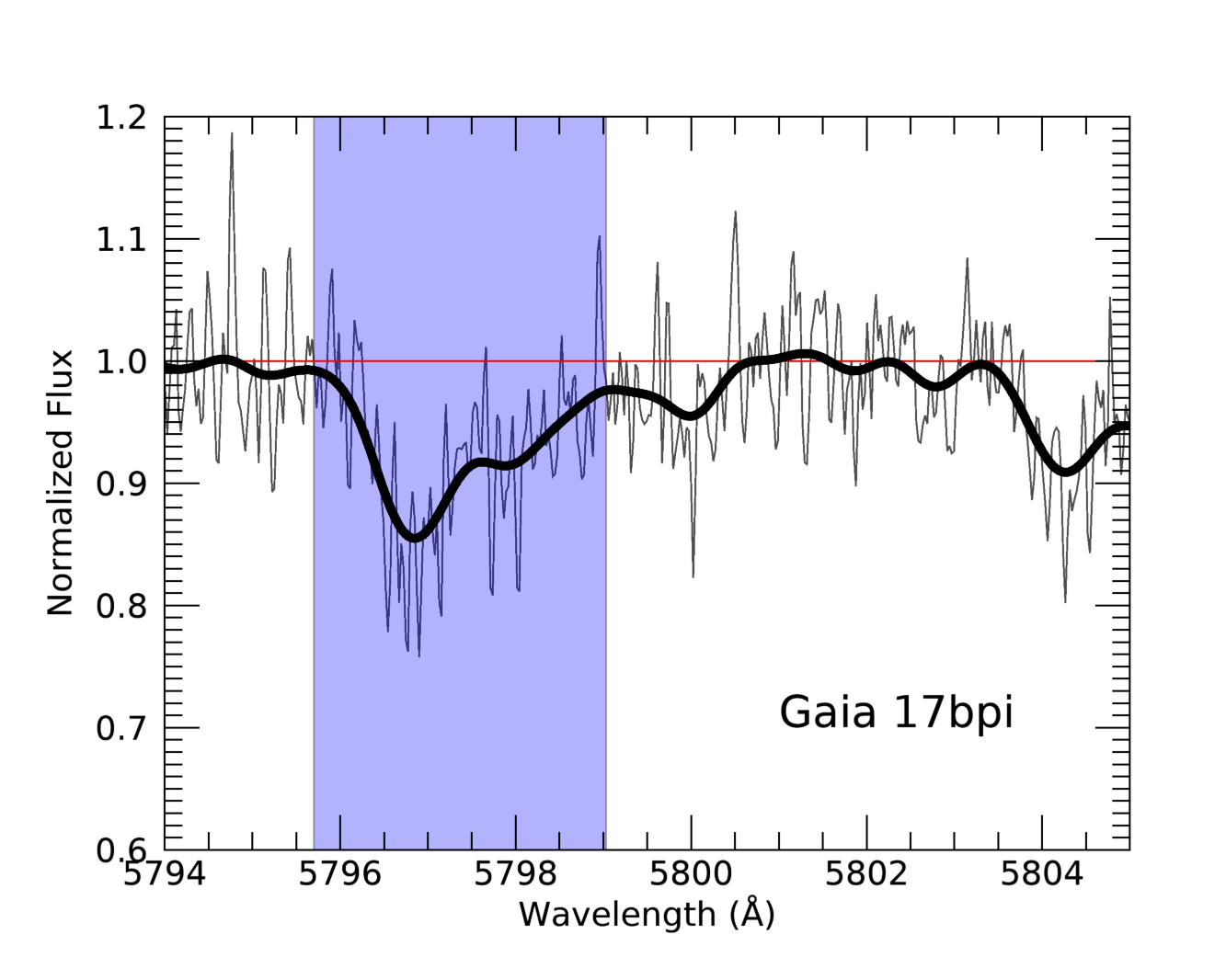}
        \includegraphics[width=0.22\linewidth]{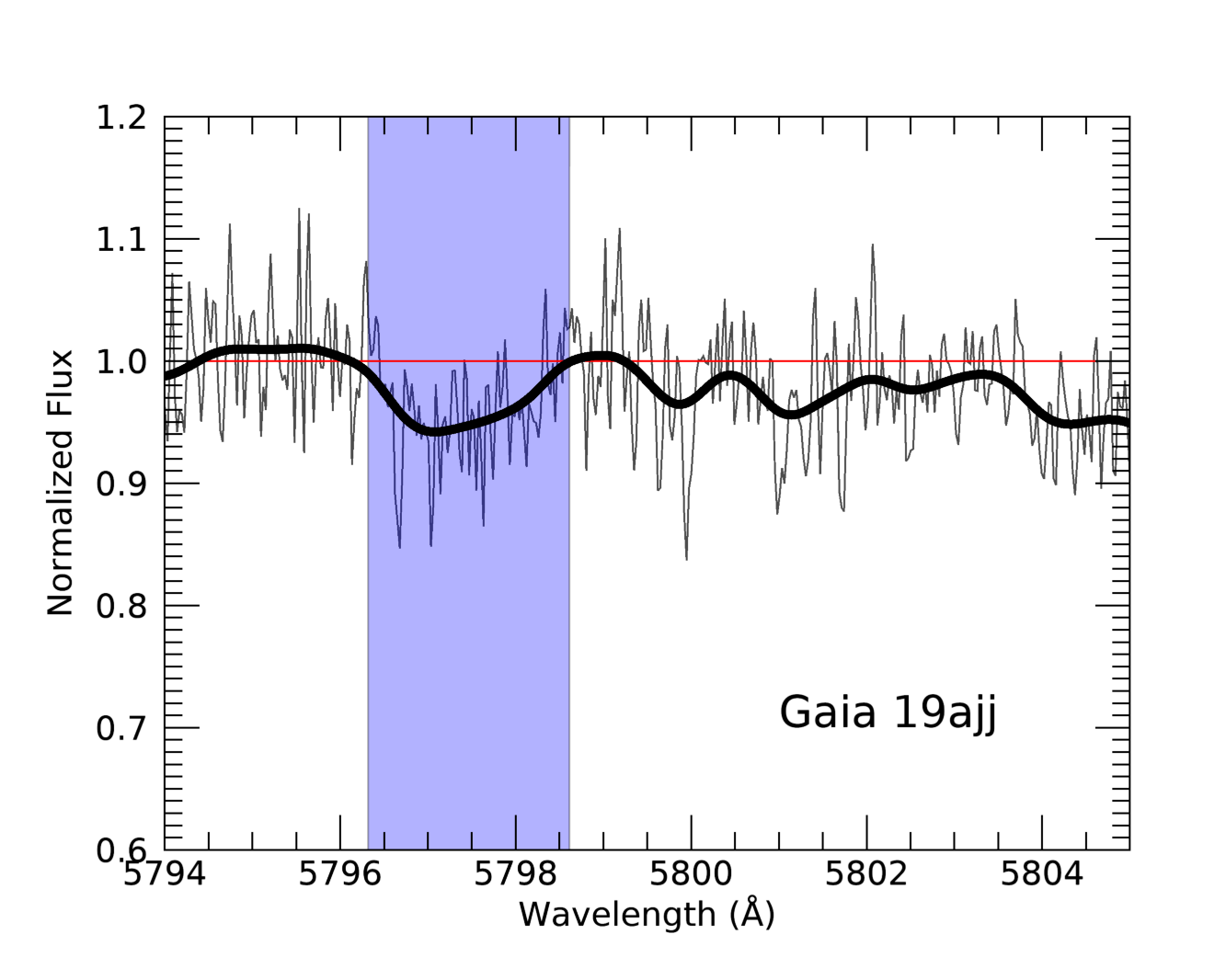}
        \includegraphics[width=0.22\linewidth]{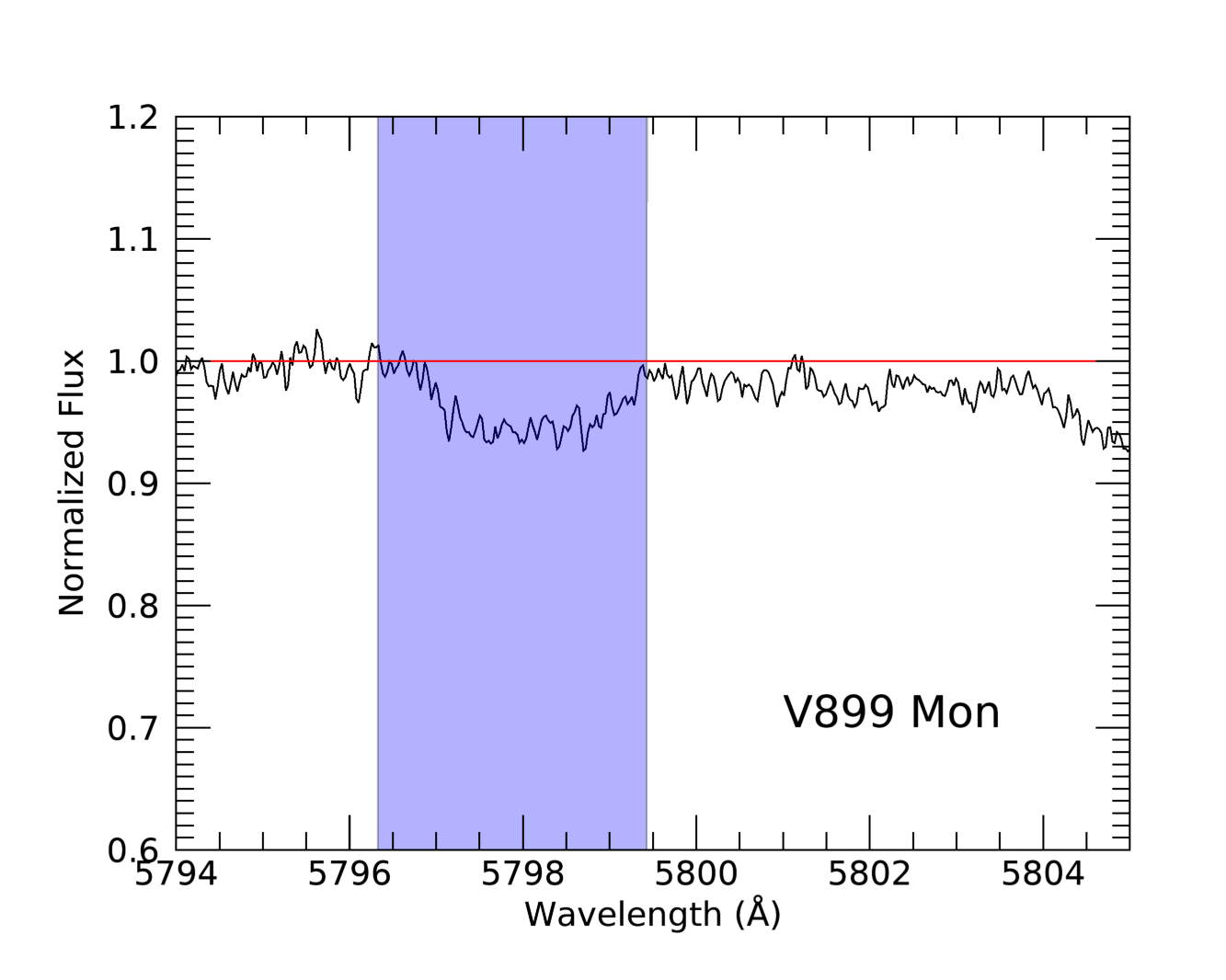}
    \includegraphics[width=0.22\linewidth]{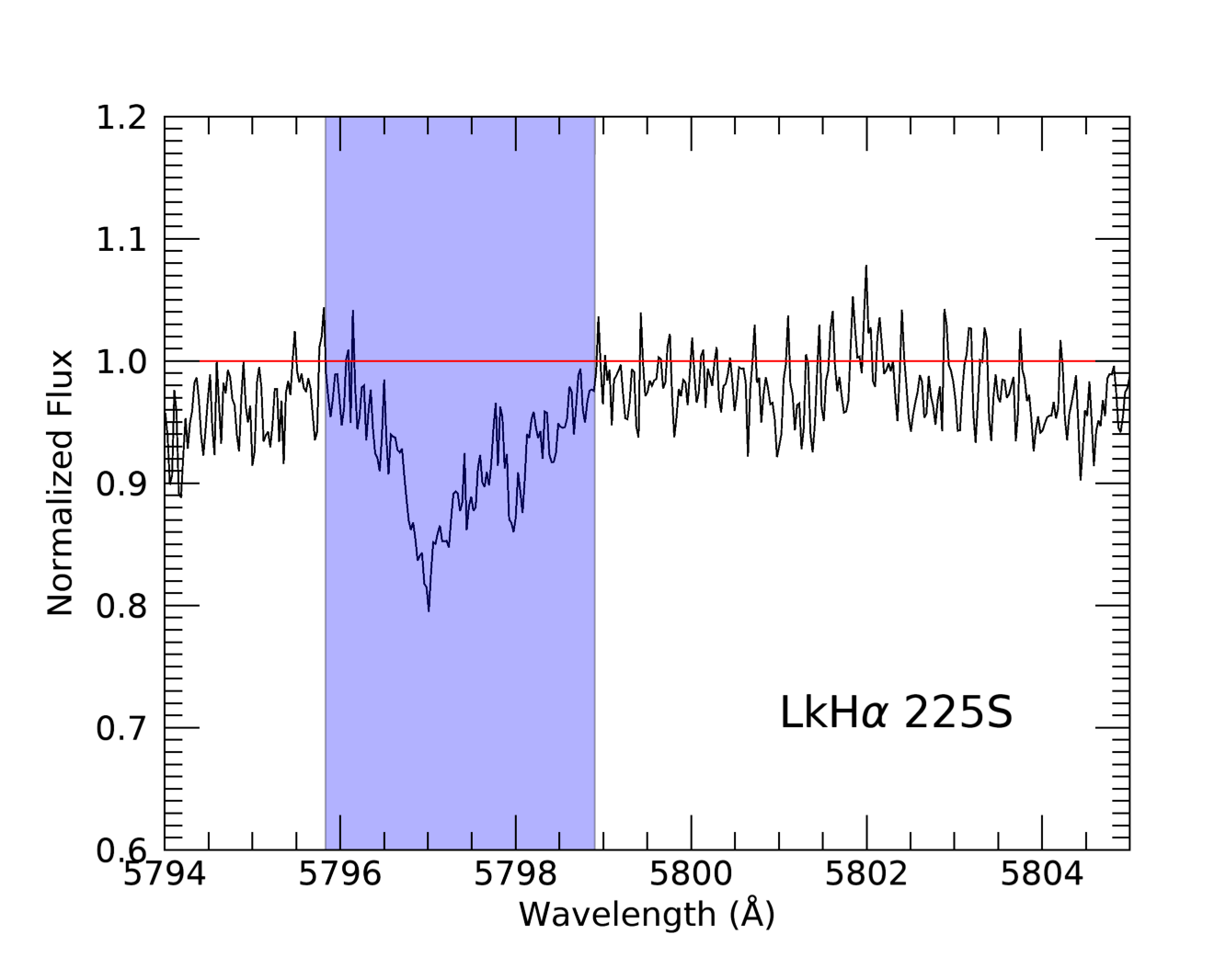}
        \includegraphics[width=0.22\linewidth]{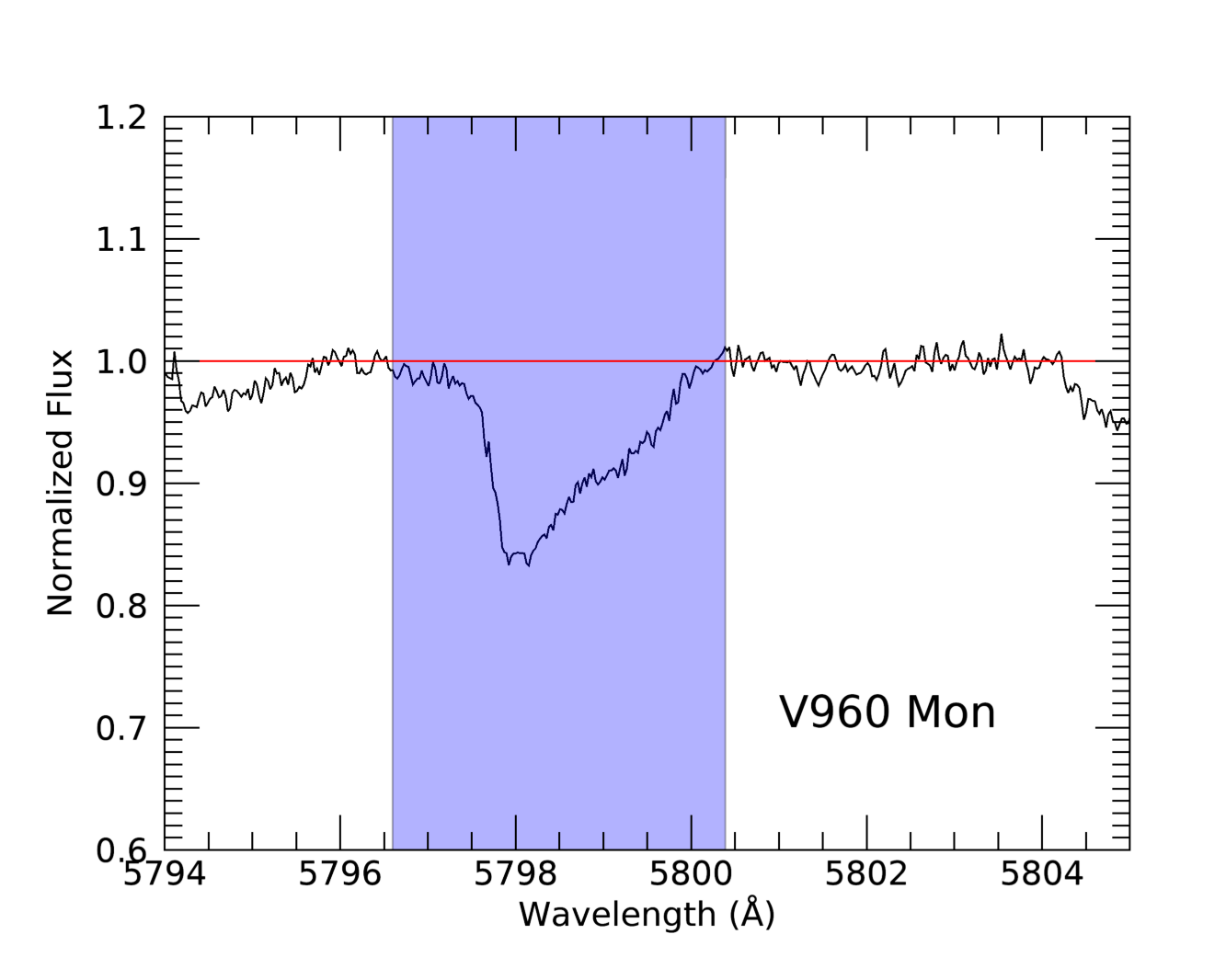}
        \includegraphics[width=0.22\linewidth]{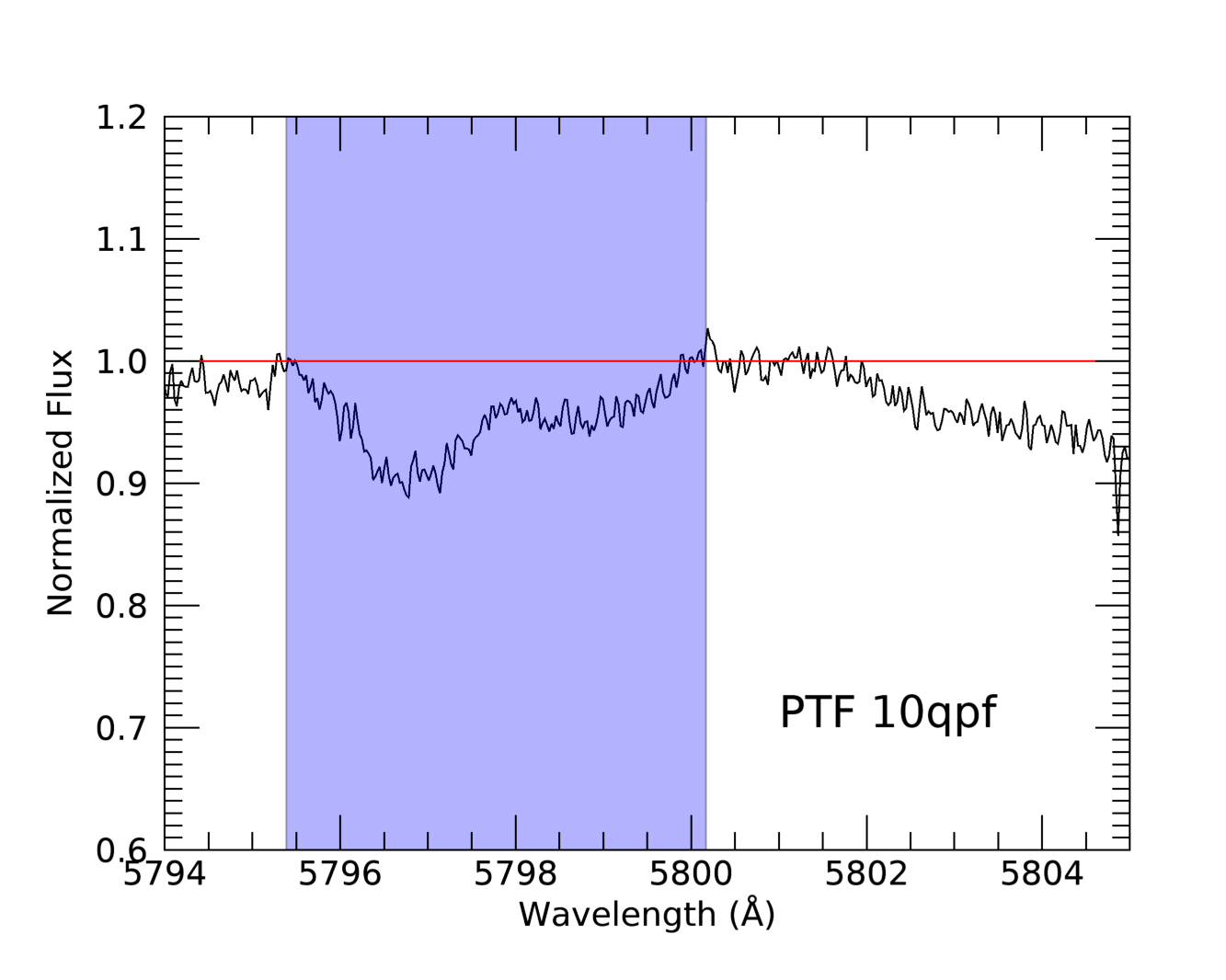}
    \includegraphics[width=0.22\linewidth]{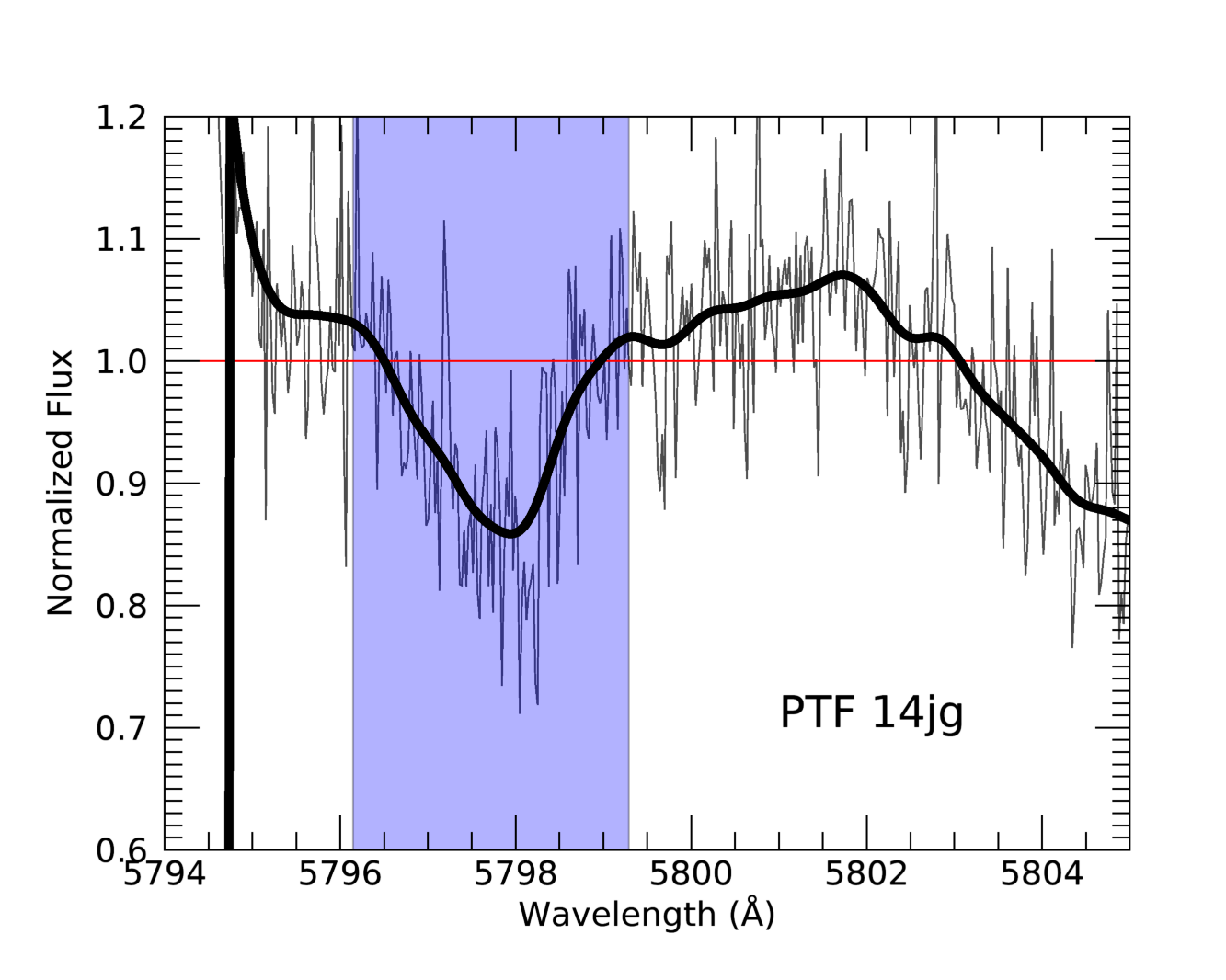}
        \includegraphics[width=0.22\linewidth]{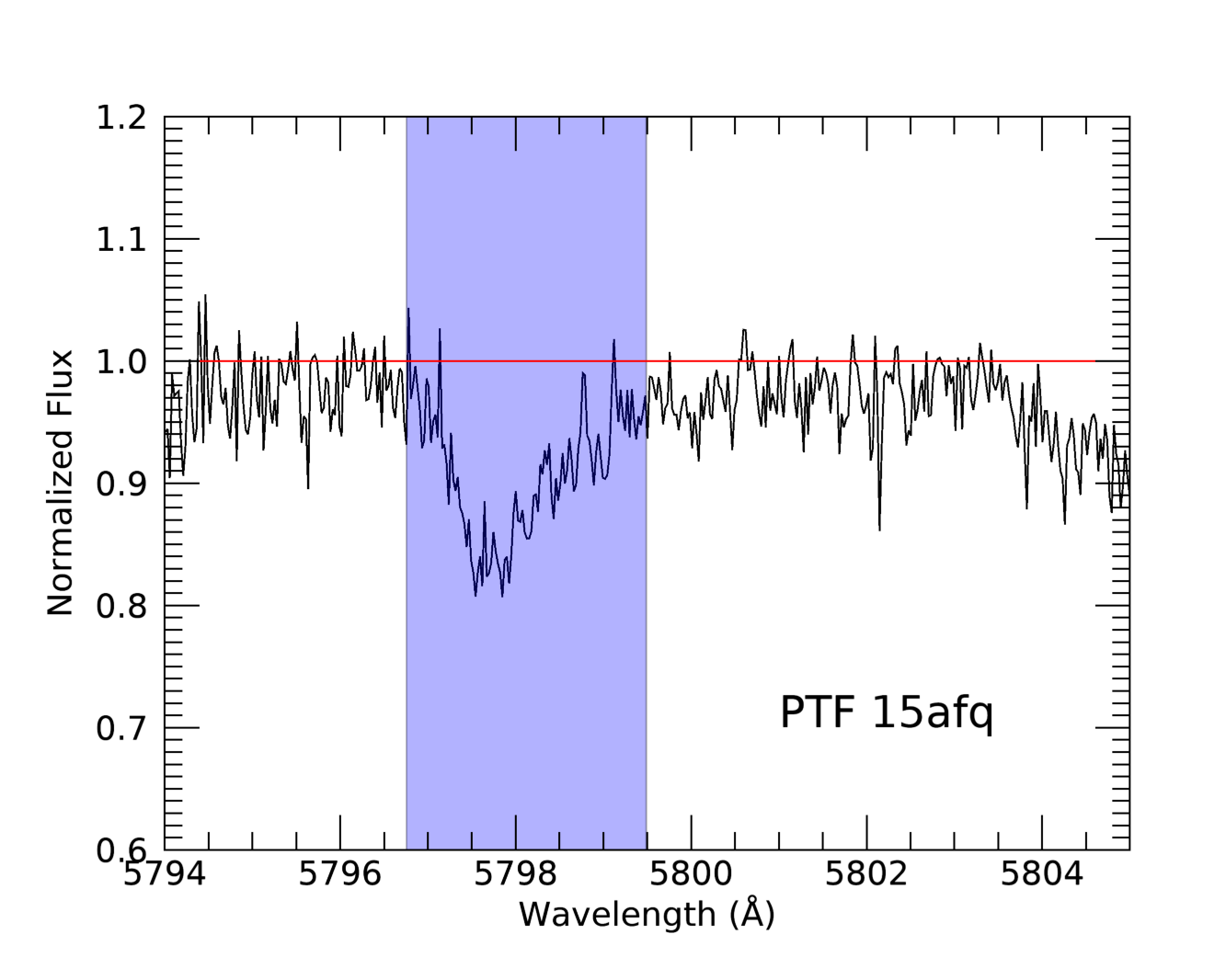}
        \includegraphics[width=0.22\linewidth]{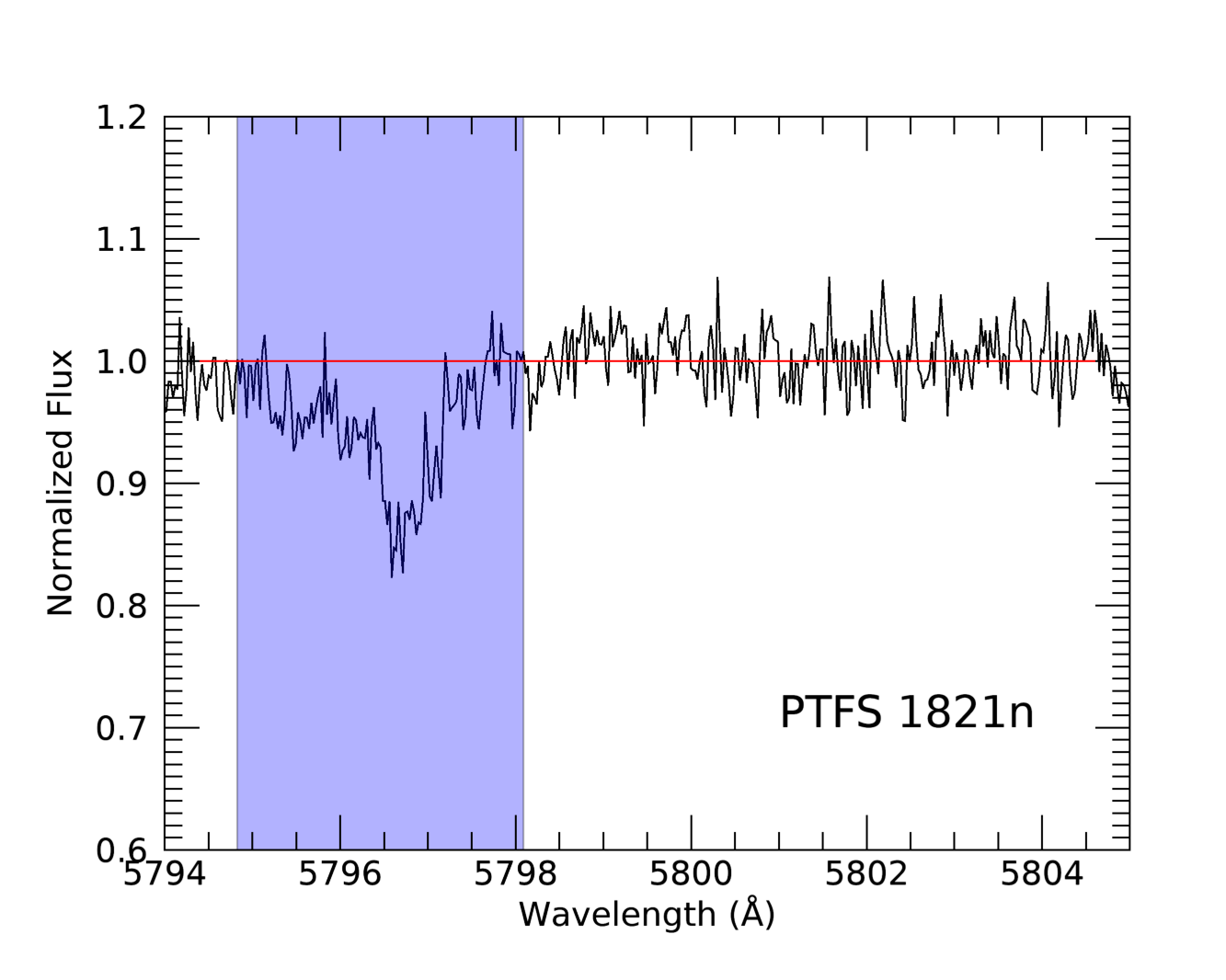}
    \includegraphics[width=0.22\linewidth]{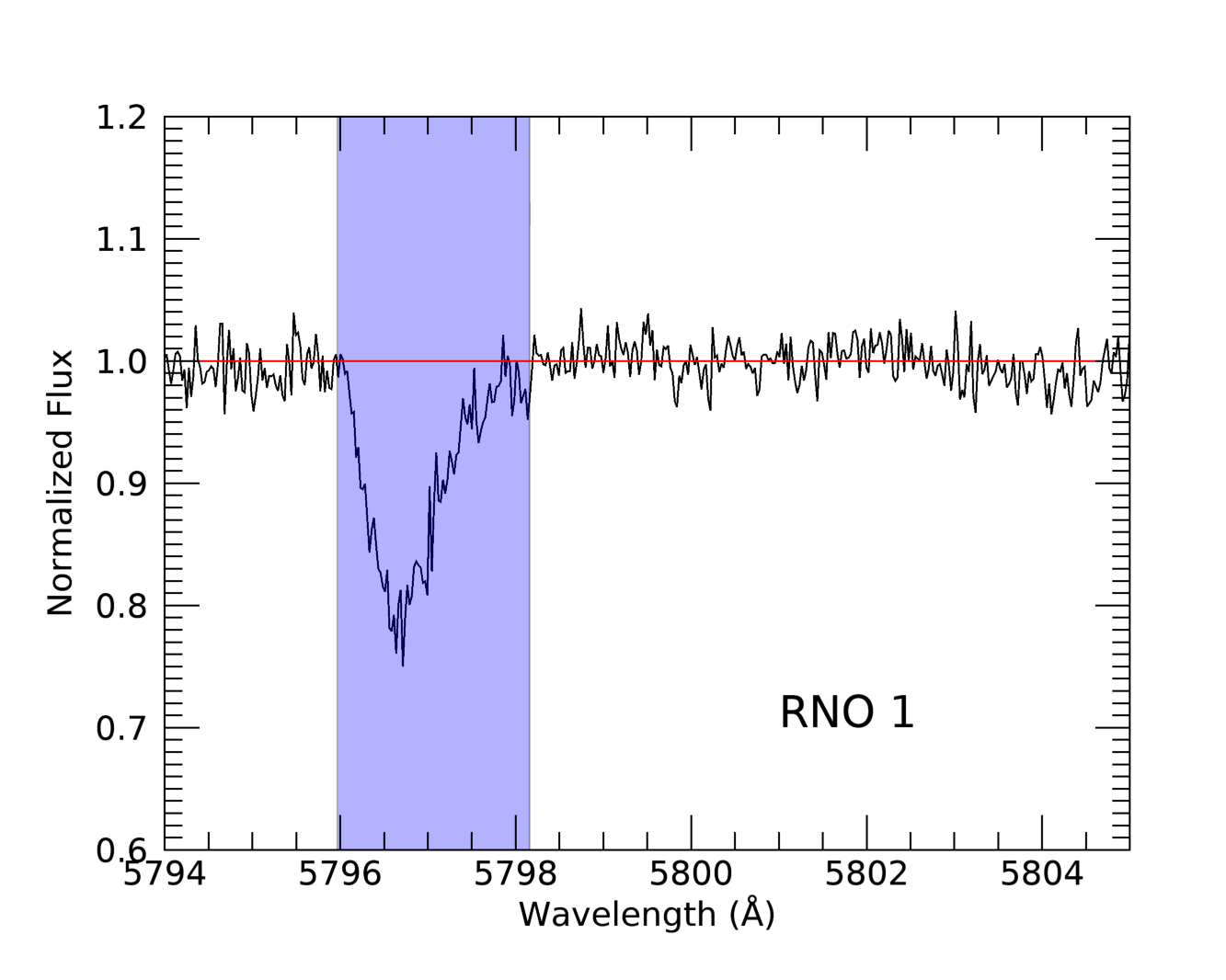}
        \includegraphics[width=0.22\linewidth]{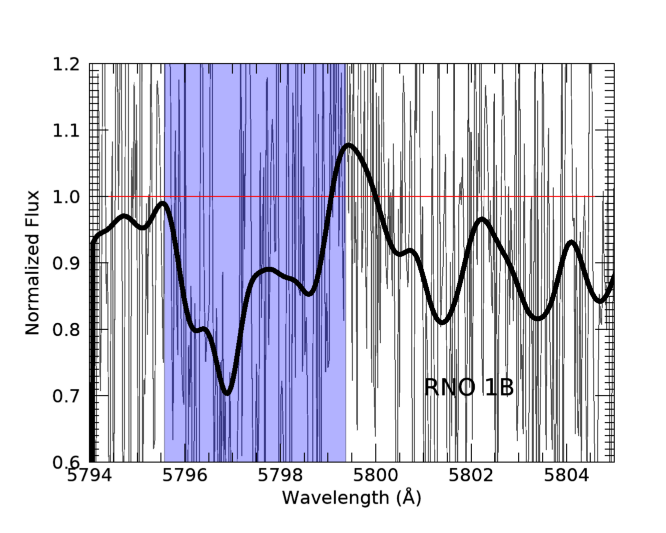}
        \includegraphics[width=0.22\linewidth]{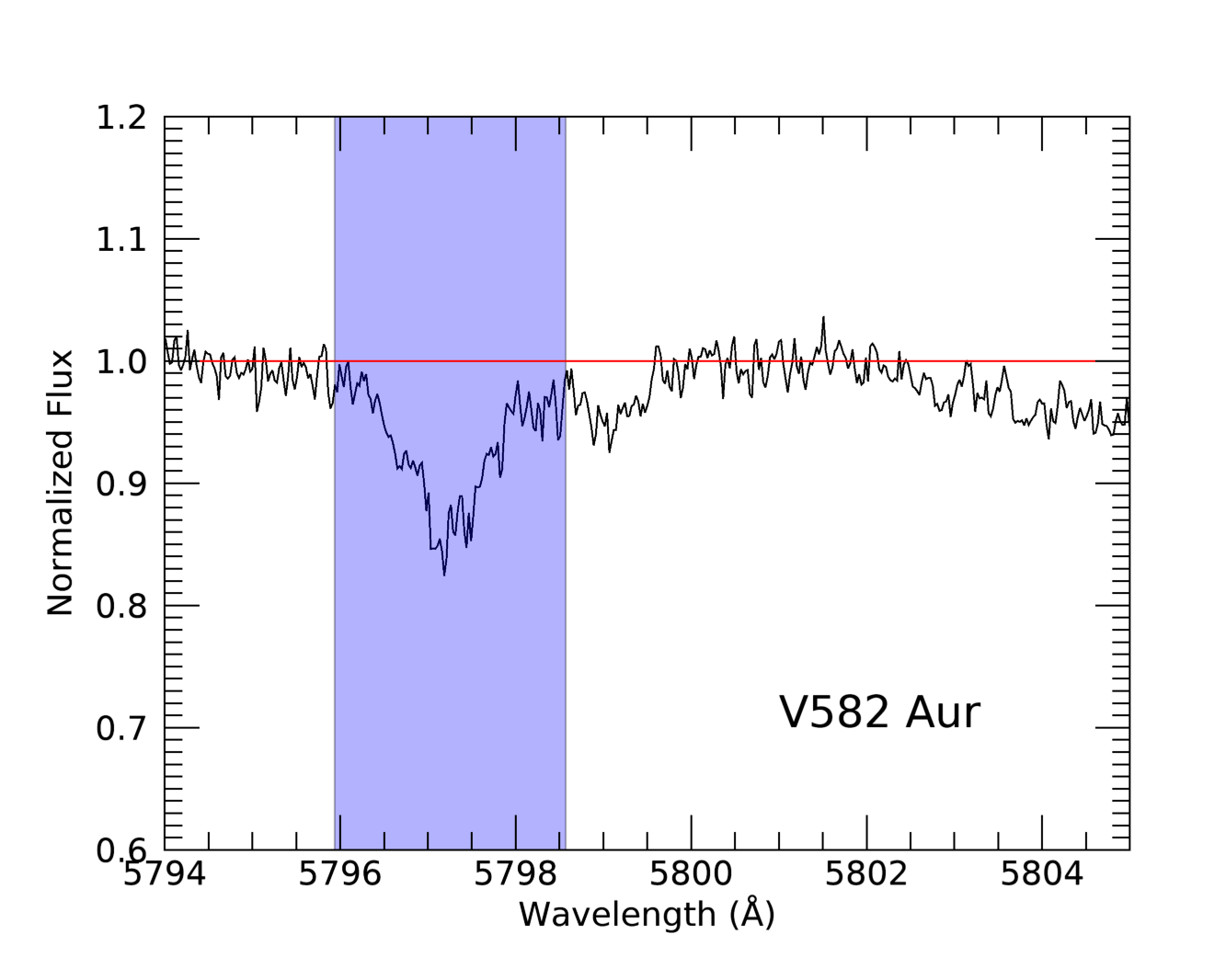}
    \includegraphics[width=0.22\linewidth]{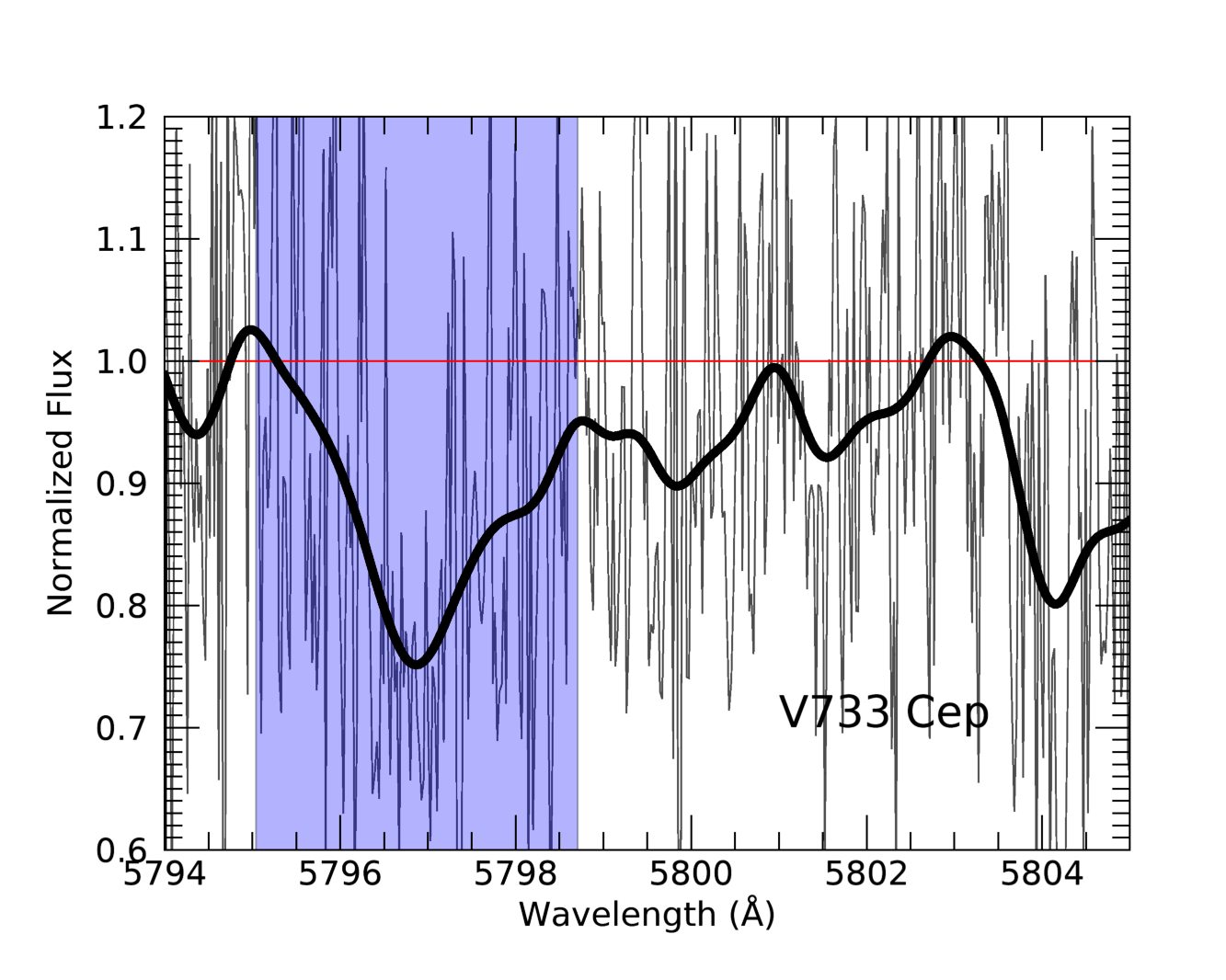}
        \includegraphics[width=0.22\linewidth]{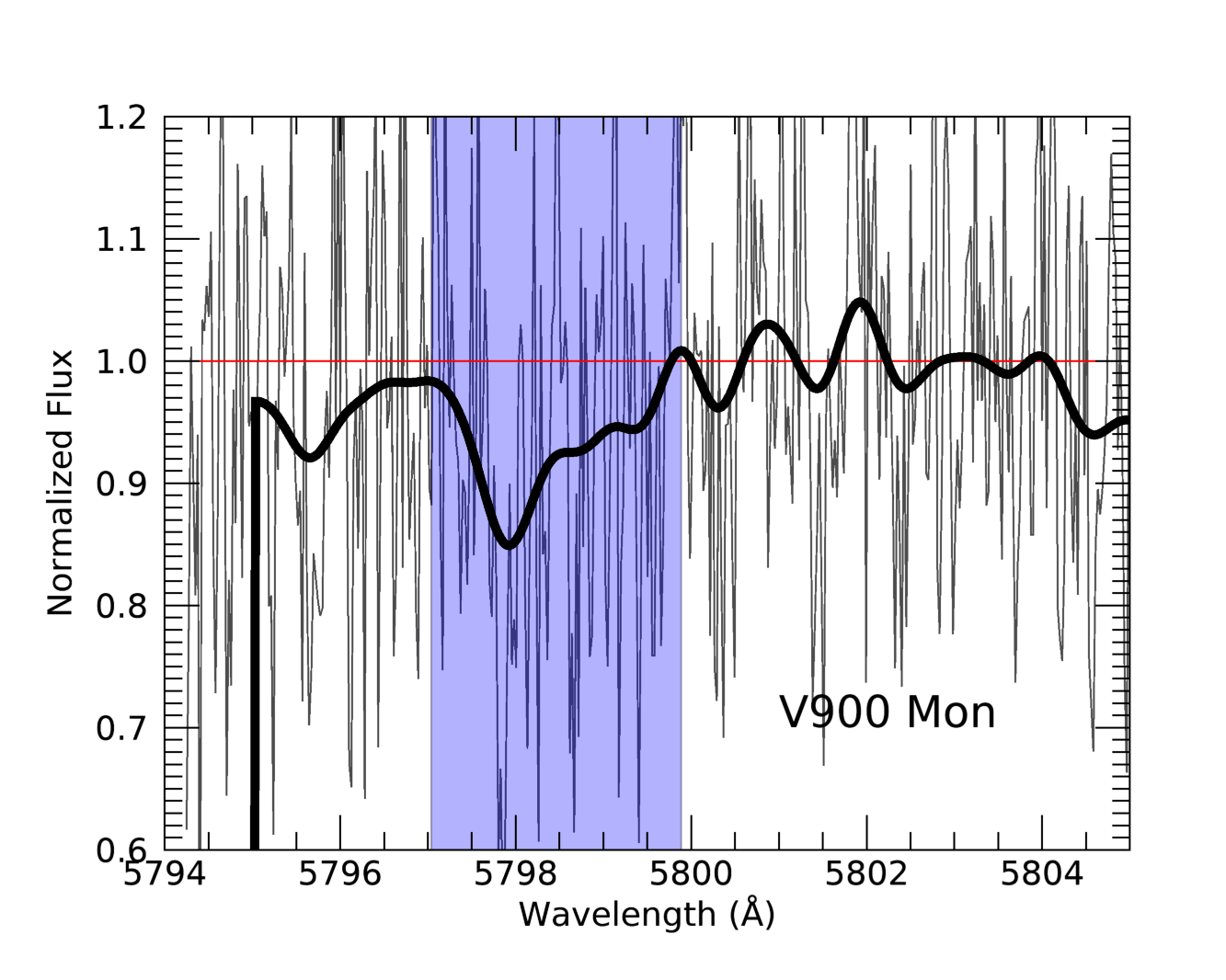}
        \includegraphics[width=0.22\linewidth]{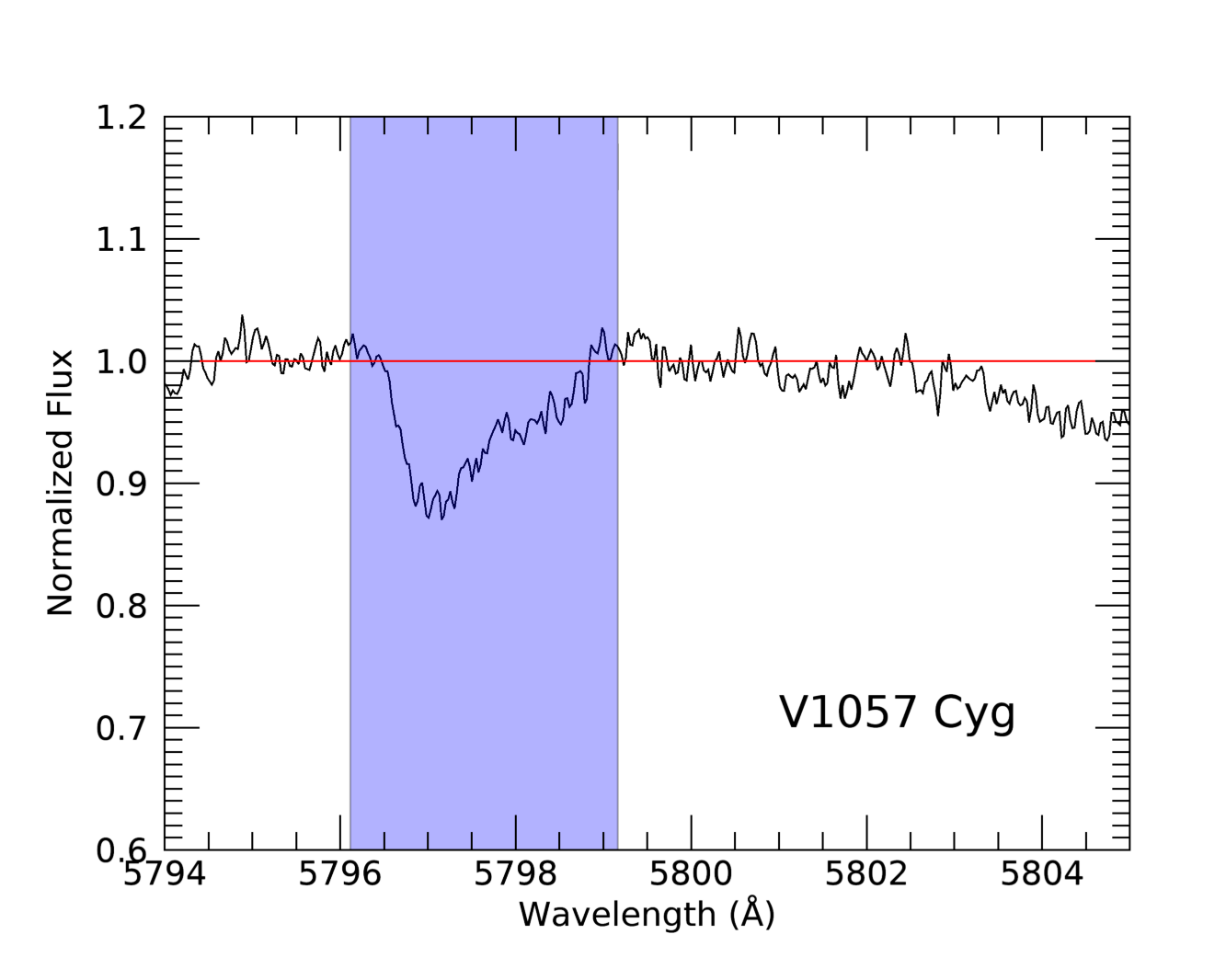}
    \includegraphics[width=0.22\linewidth]{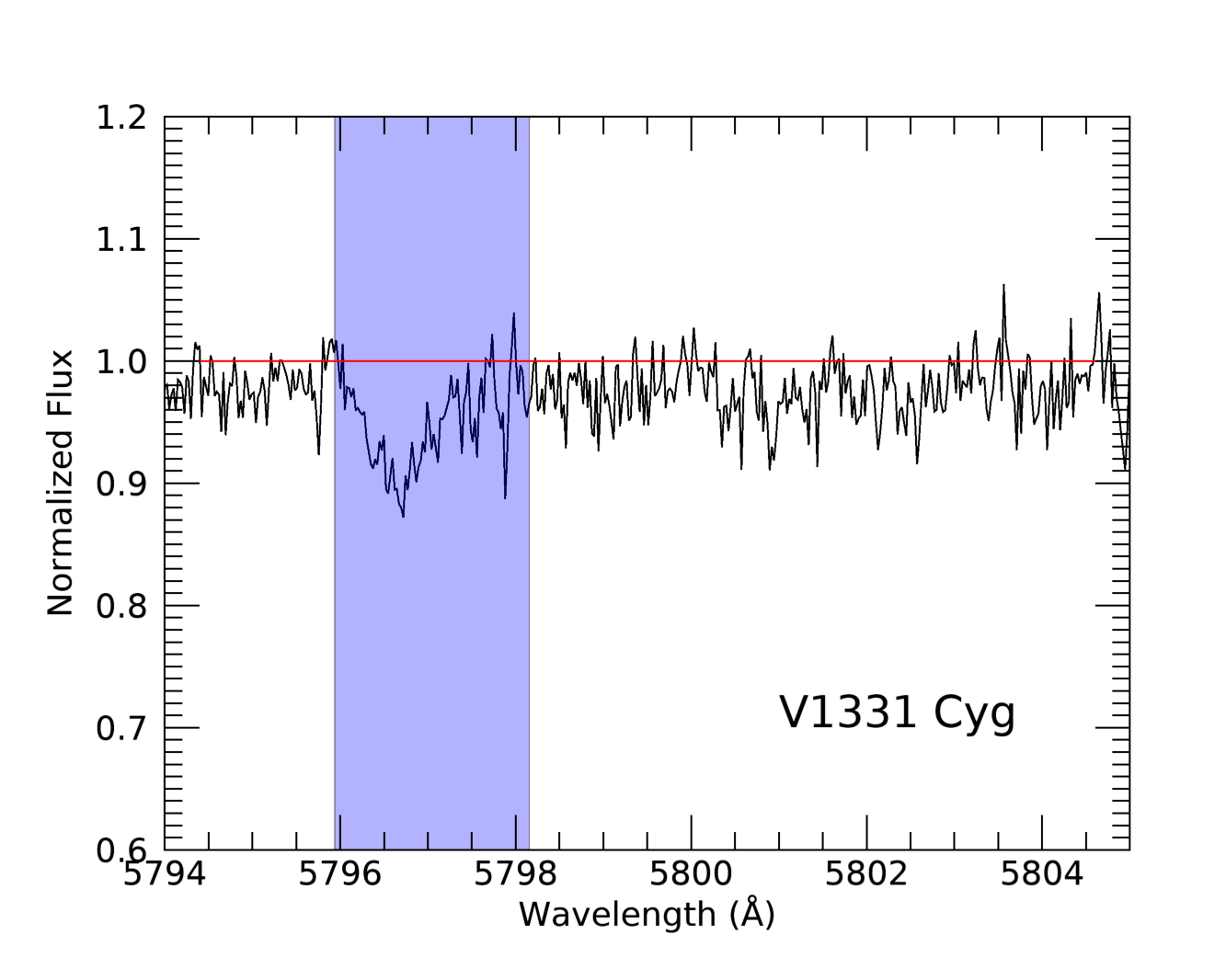}
        \includegraphics[width=0.22\linewidth]{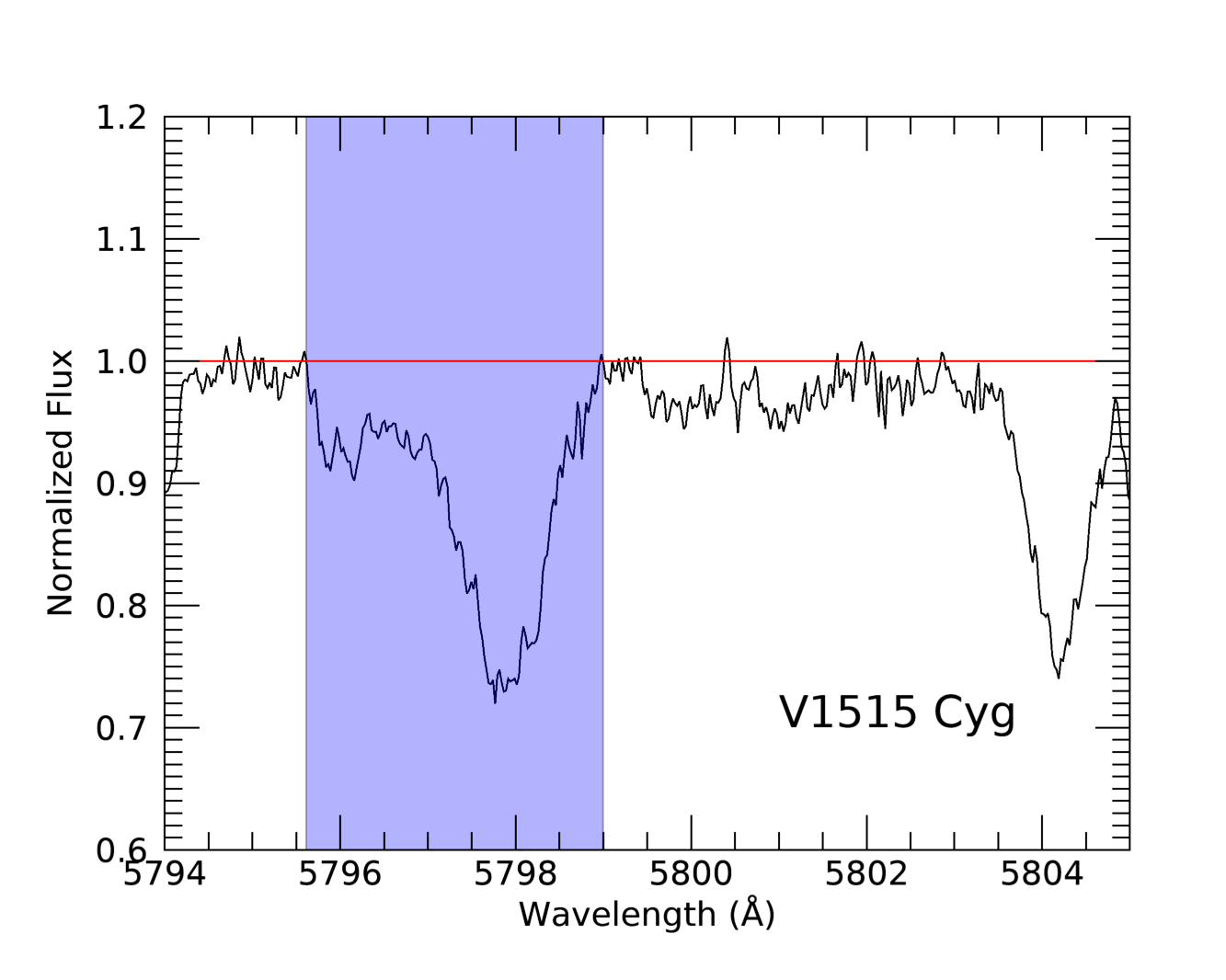}
        \includegraphics[width=0.22\linewidth]{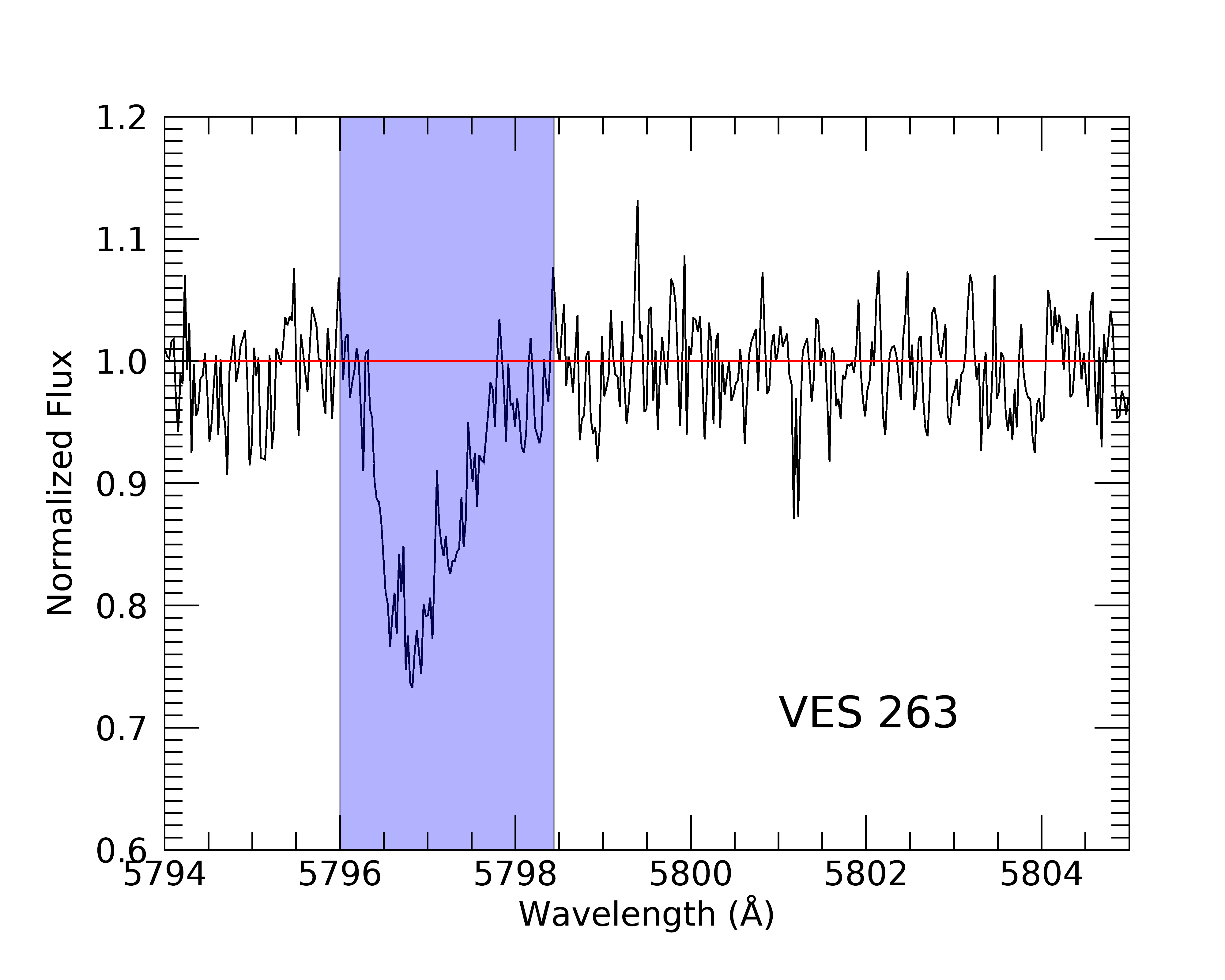}
    \includegraphics[width=0.22\linewidth]{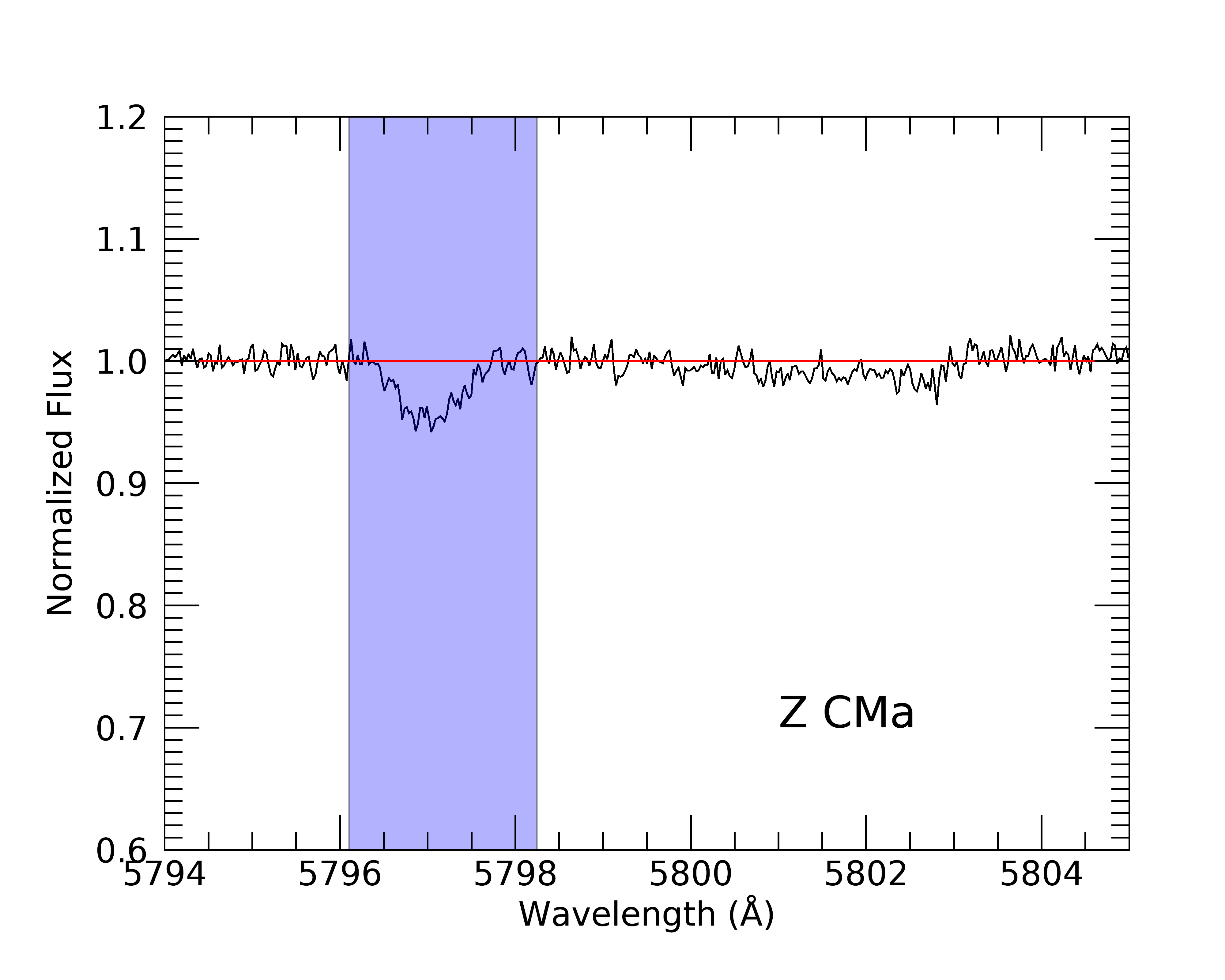}
    \caption{The $\lambda$5797 DIB in our sample. The blue region shows the region of integration for the equivalent width calculation. The red line lies at 1.0 for reference. For the noisiest spectra (Gaia 17bpi, Gaia 19ajj, PTF 14jg, RNO 1B, V733 Cep, V900 Mon) a $\sigma=10$ pixels Gaussian-smoothing of the spectrum is shown in black, with the raw data shown in grey.}
    \label{fig:eqWIntegration5797s}
\end{figure}

\begin{figure}[!htb]
    \centering
    \includegraphics[width=0.22\linewidth]{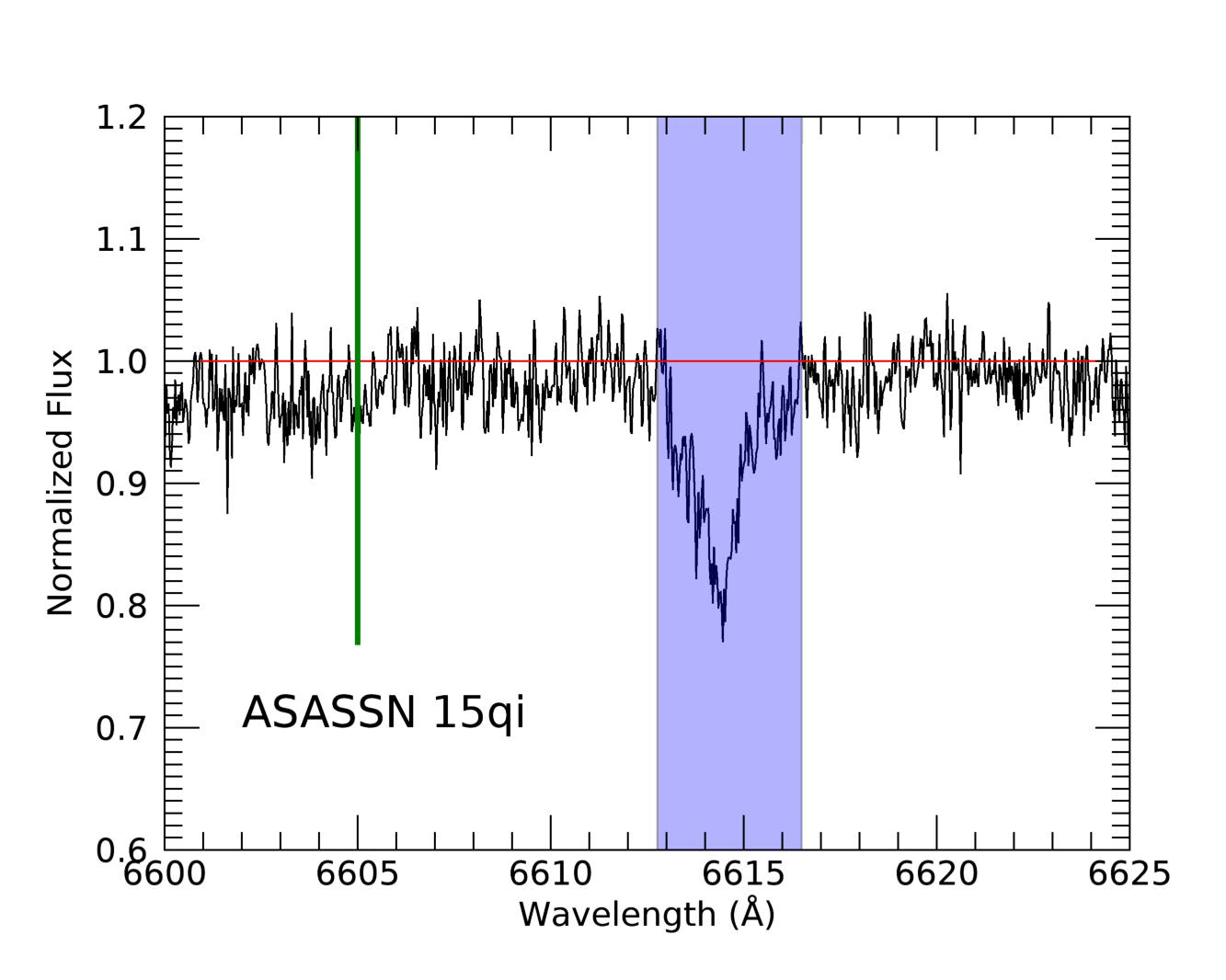}
        \includegraphics[width=0.22\linewidth]{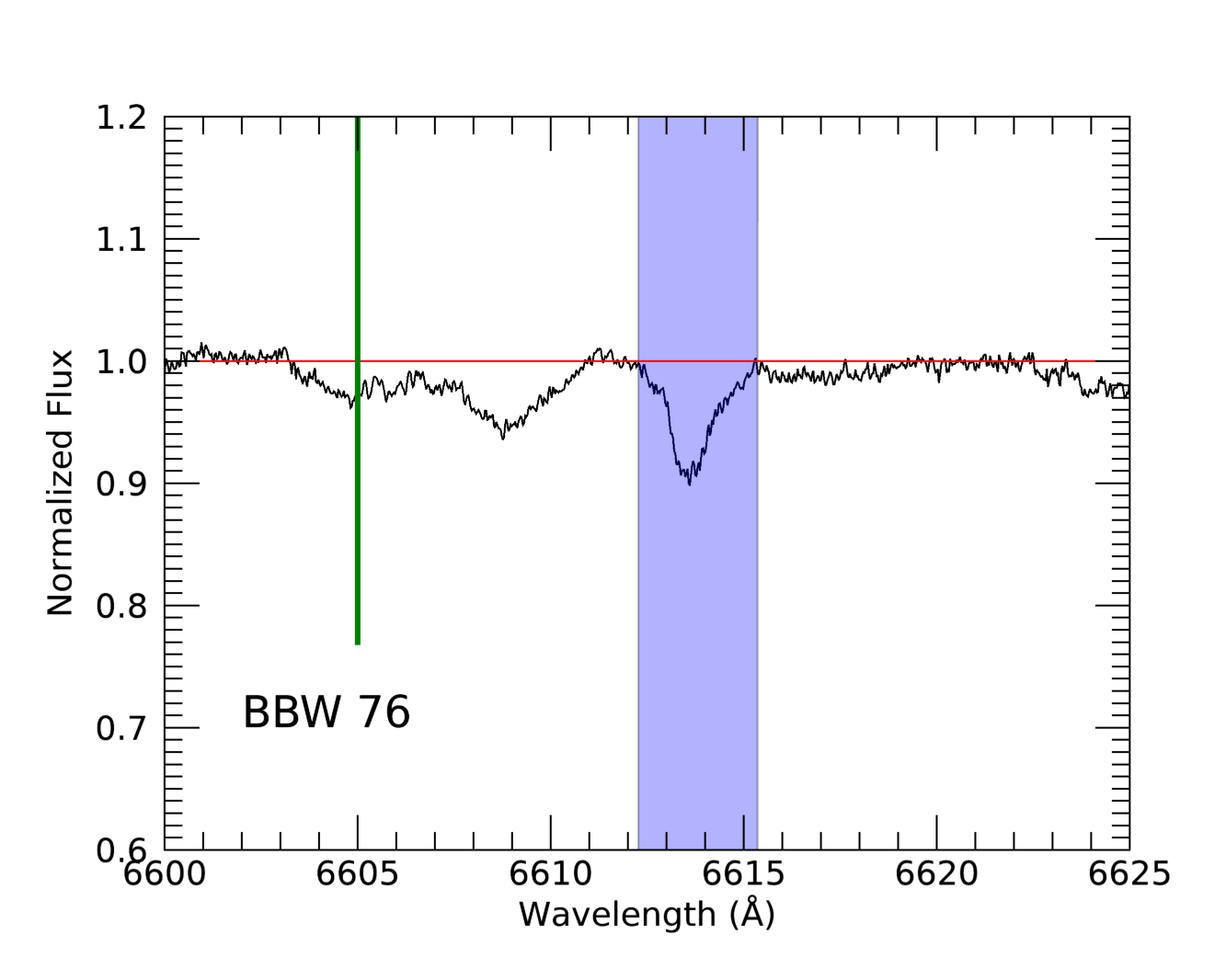}
        \includegraphics[width=0.22\linewidth]{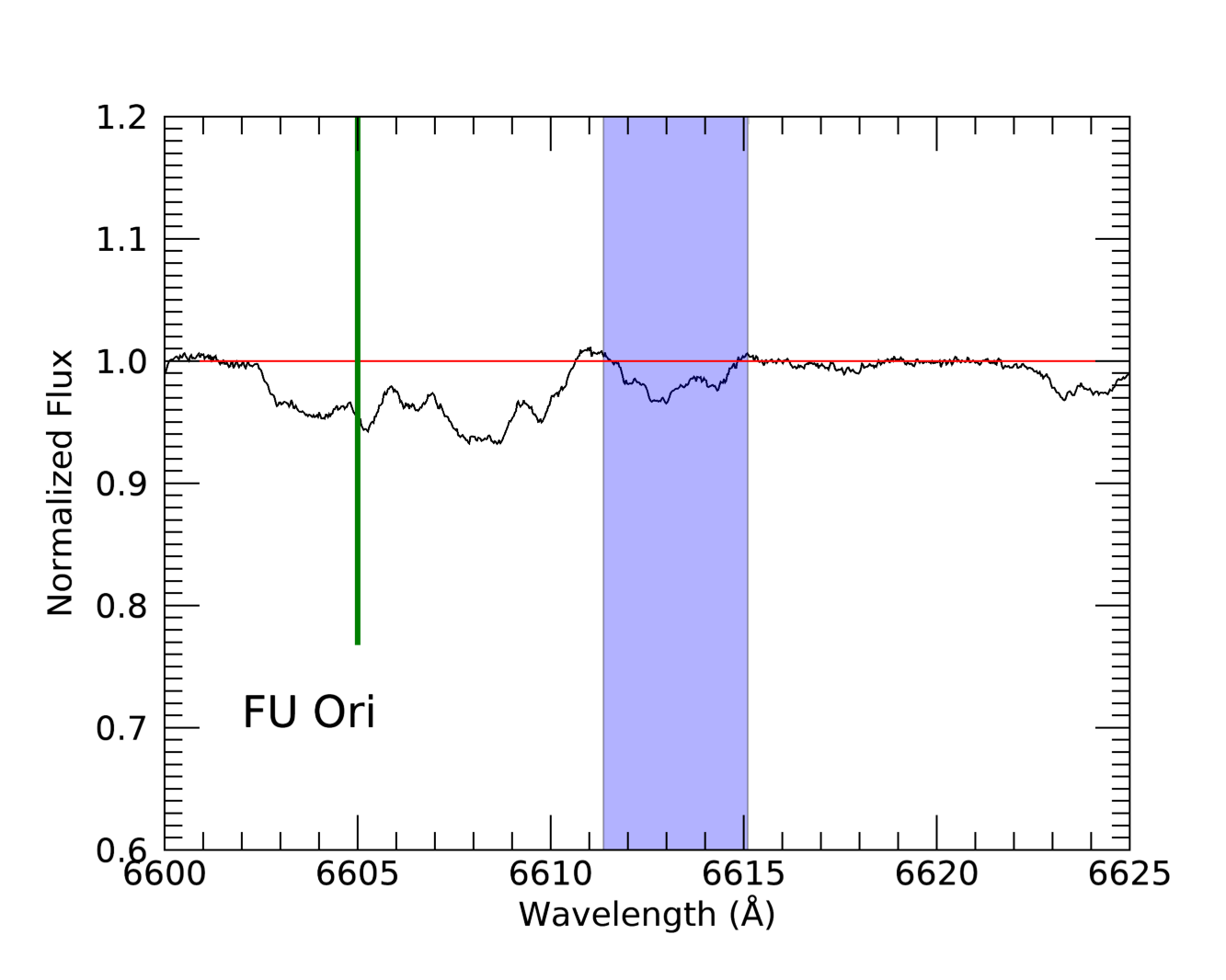}
    \includegraphics[width=0.22\linewidth]{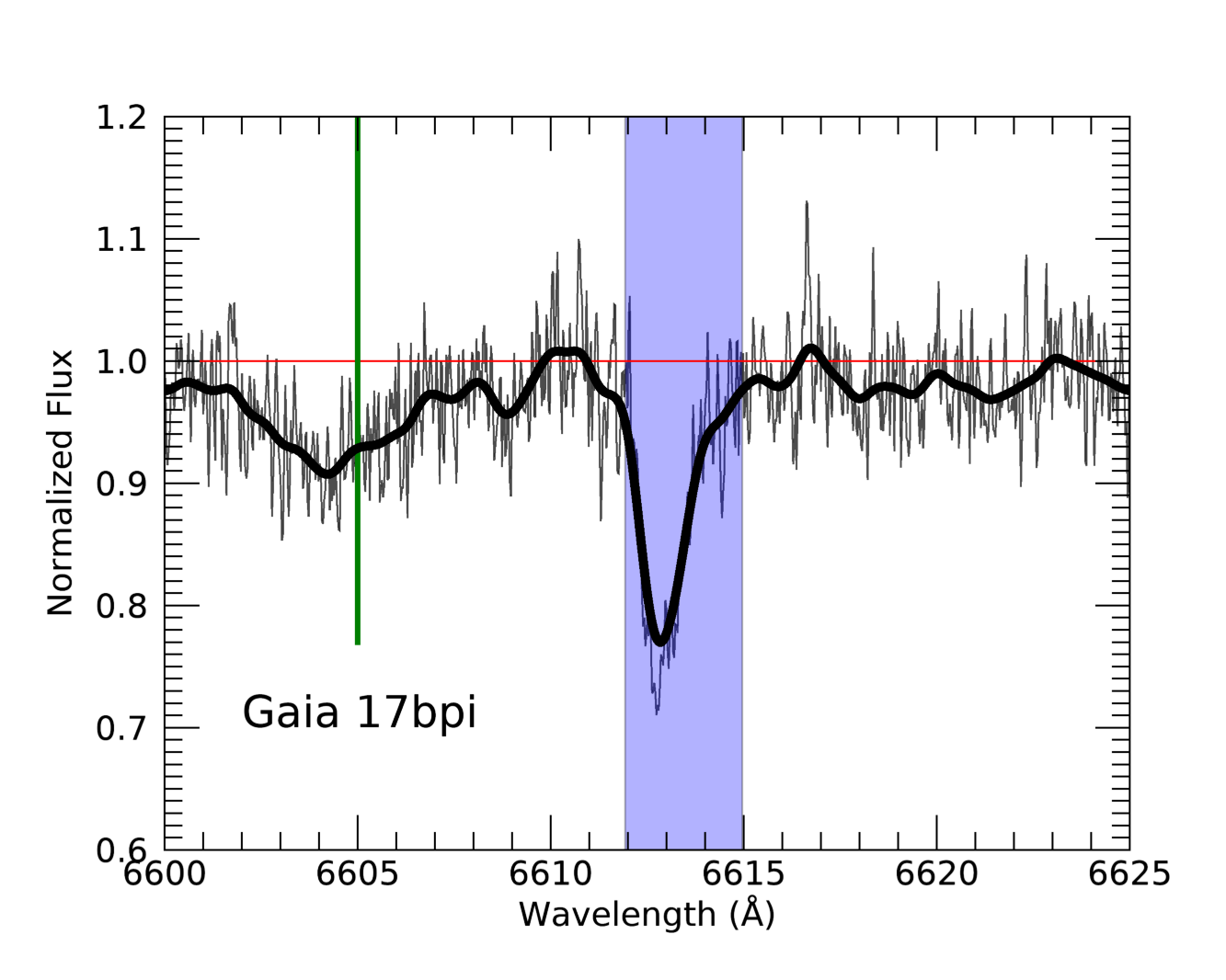}
        \includegraphics[width=0.22\linewidth]{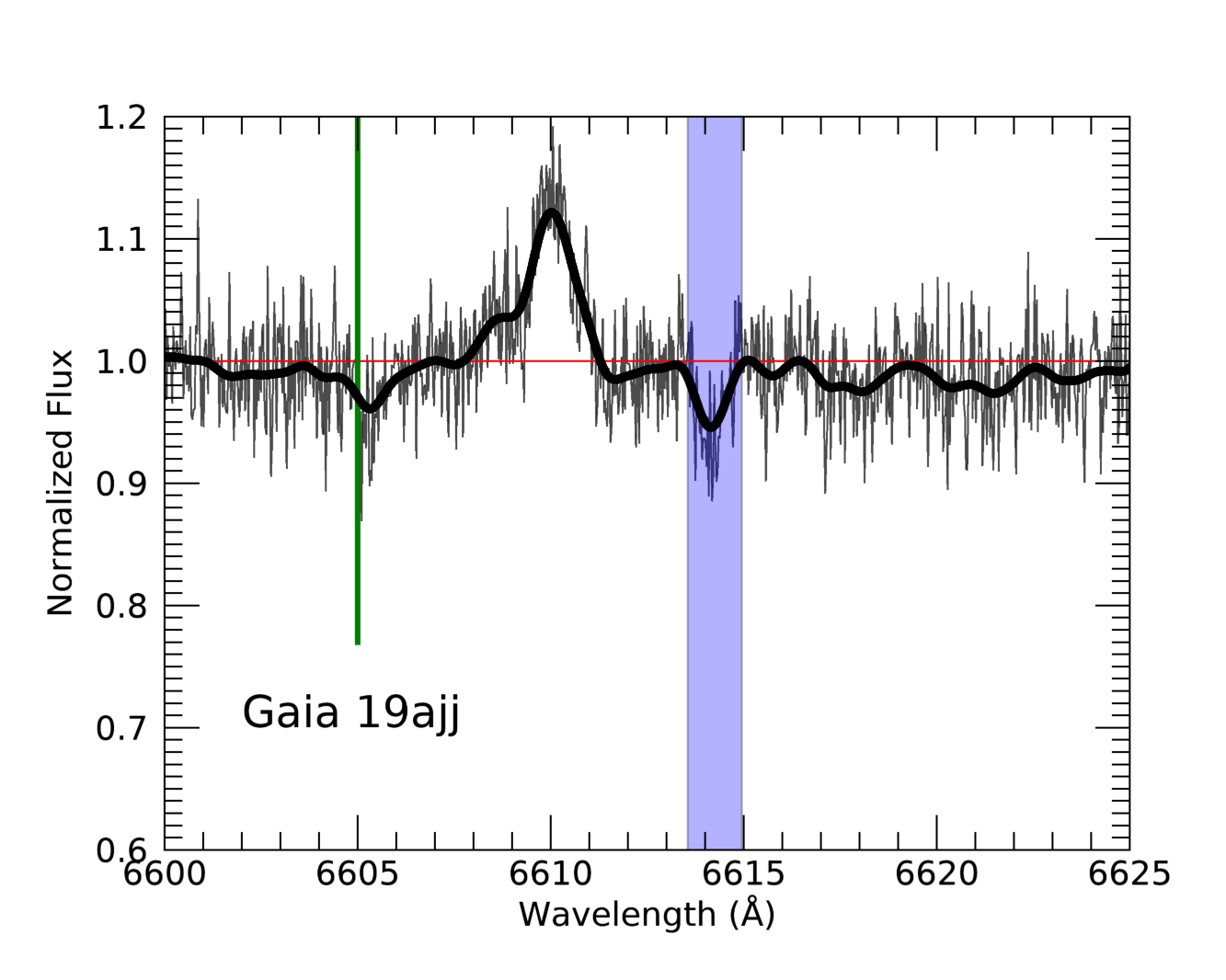}
        \includegraphics[width=0.22\linewidth]{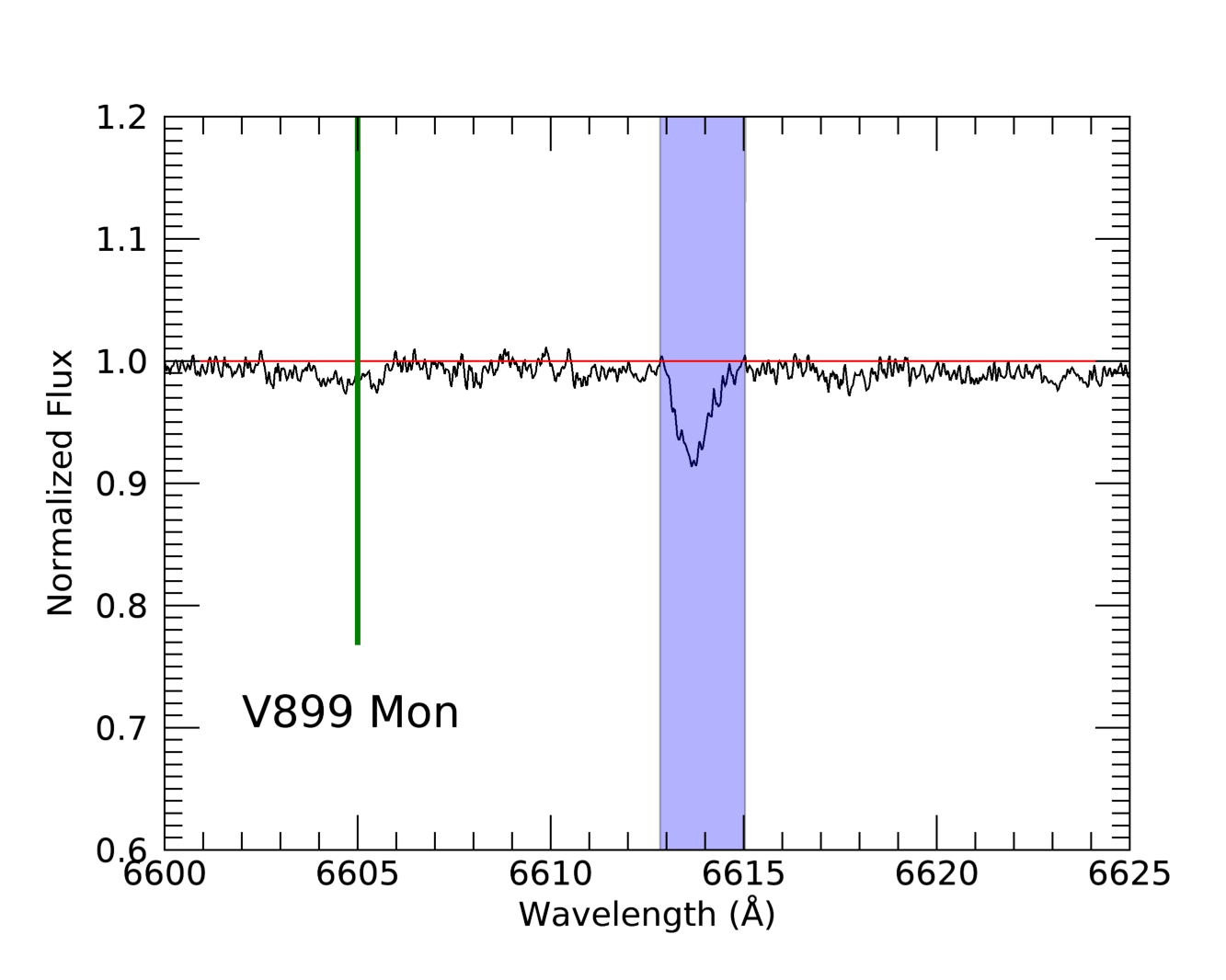}
    \includegraphics[width=0.22\linewidth]{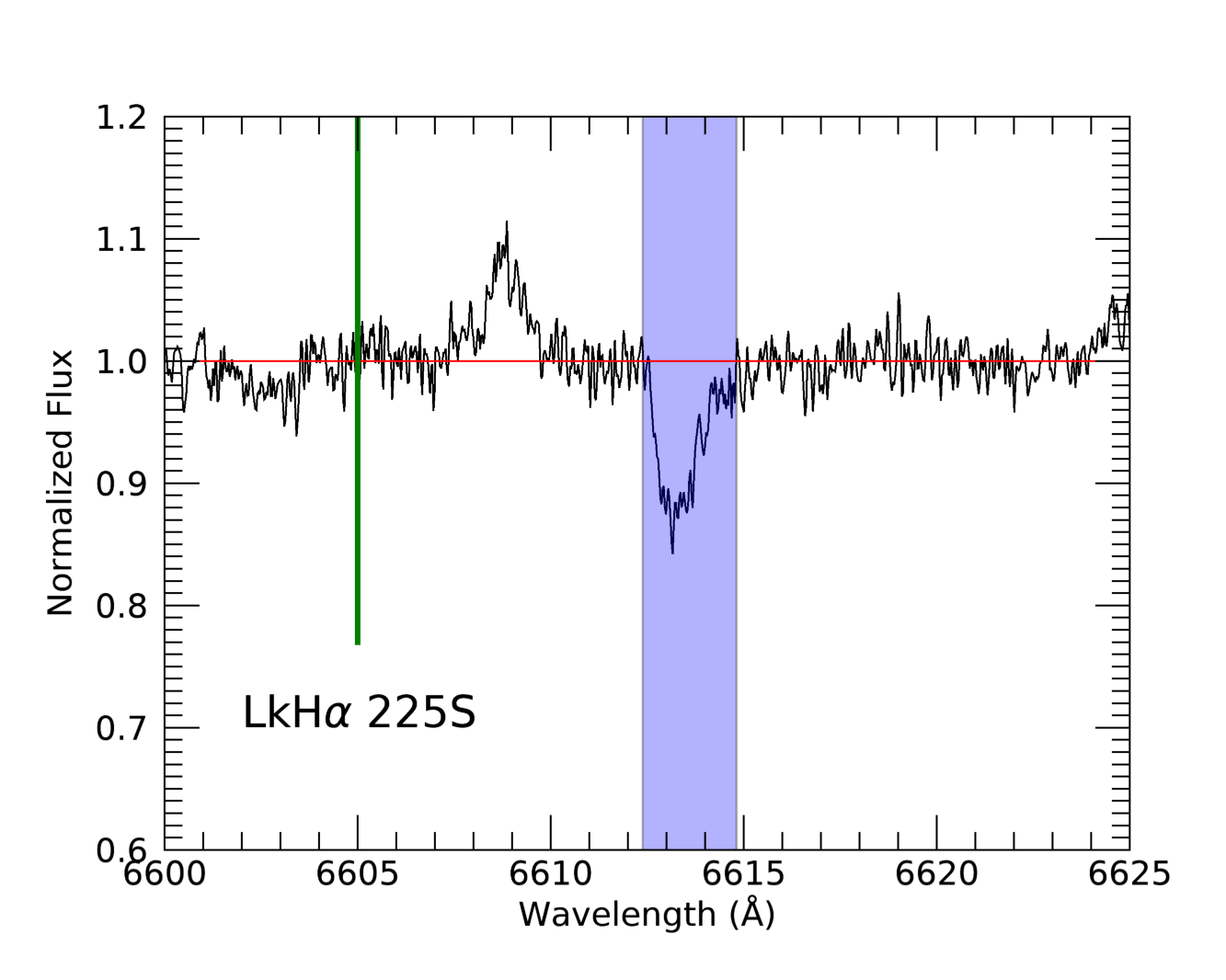}
        \includegraphics[width=0.22\linewidth]{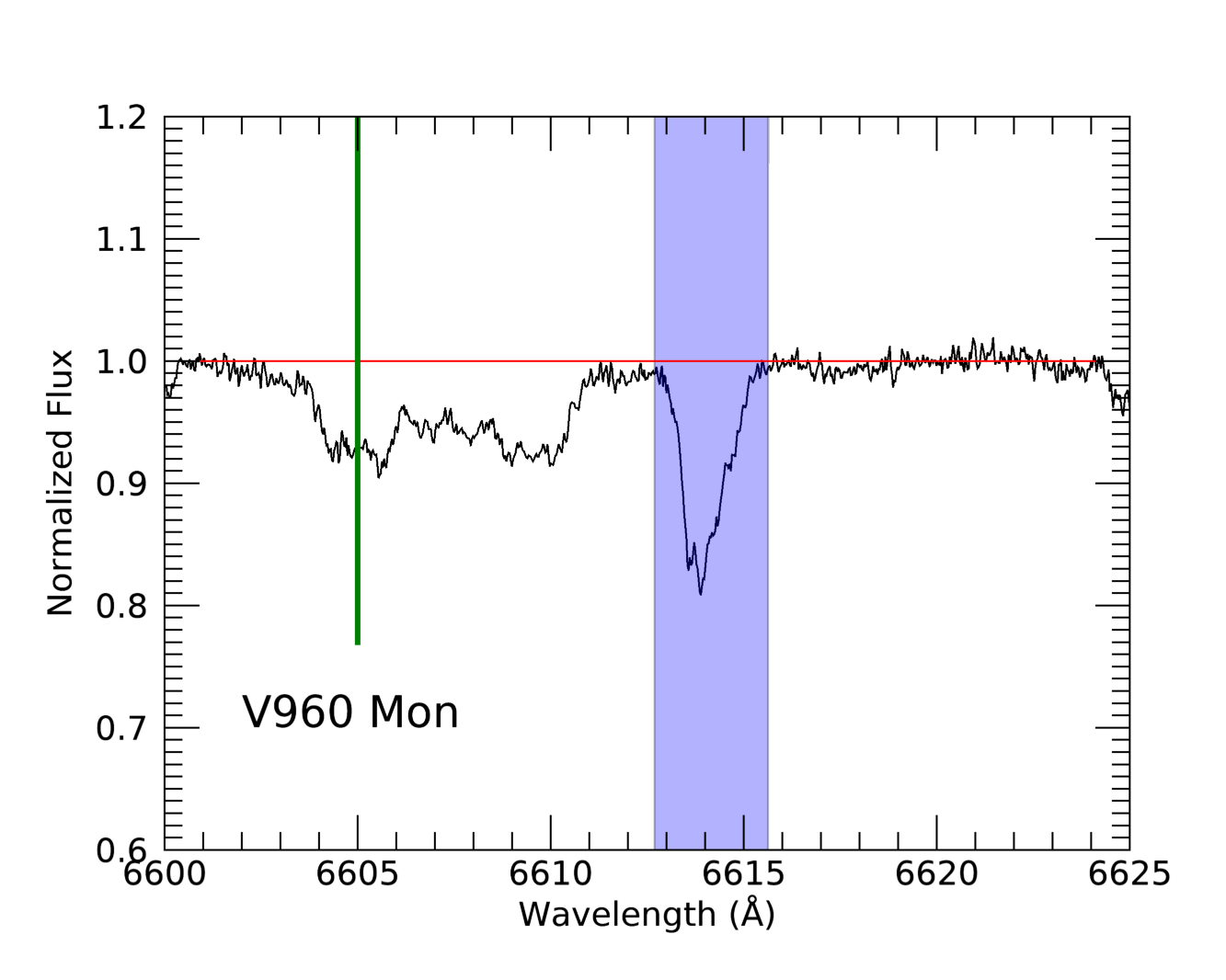}
        \includegraphics[width=0.22\linewidth]{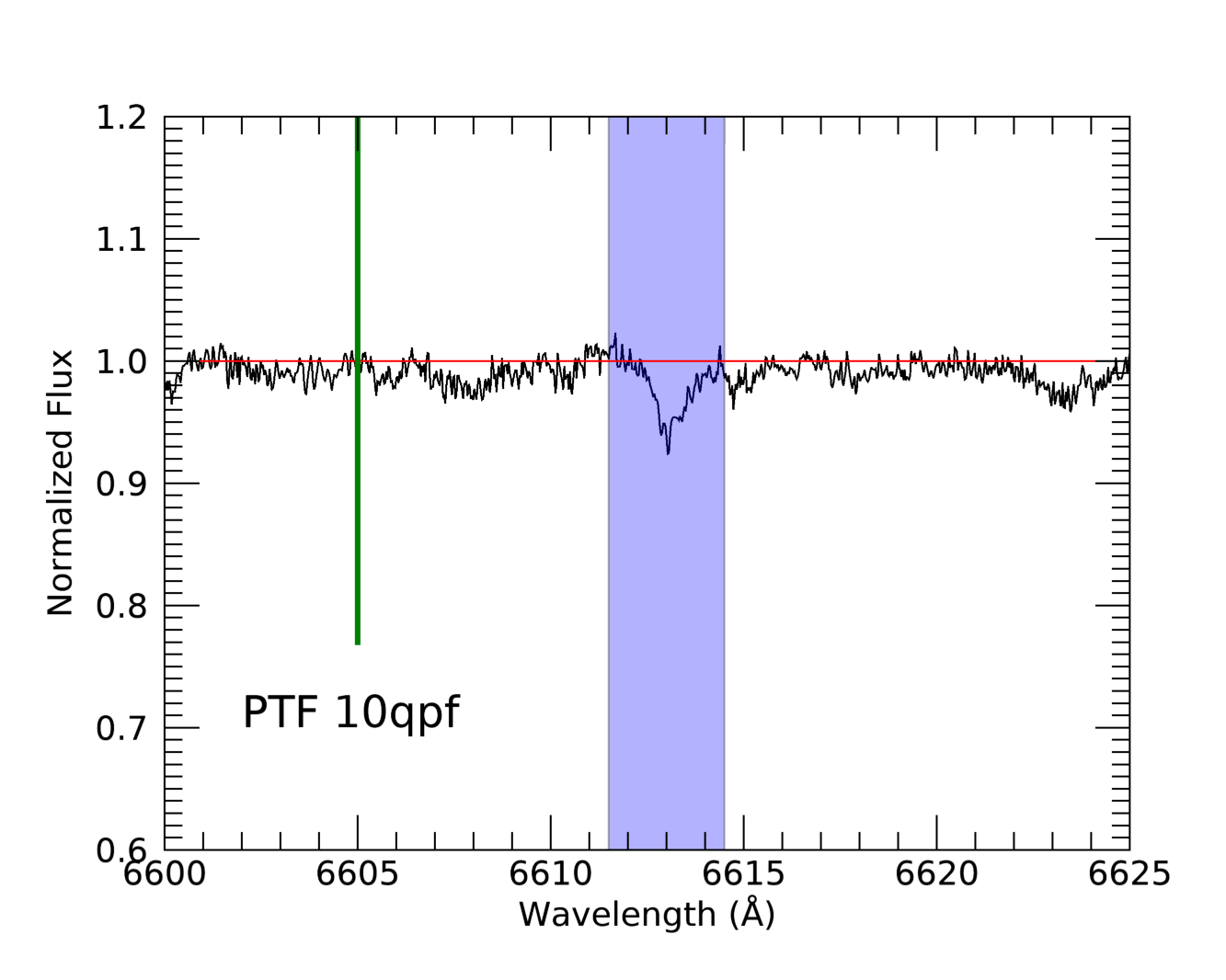}
    \includegraphics[width=0.22\linewidth]{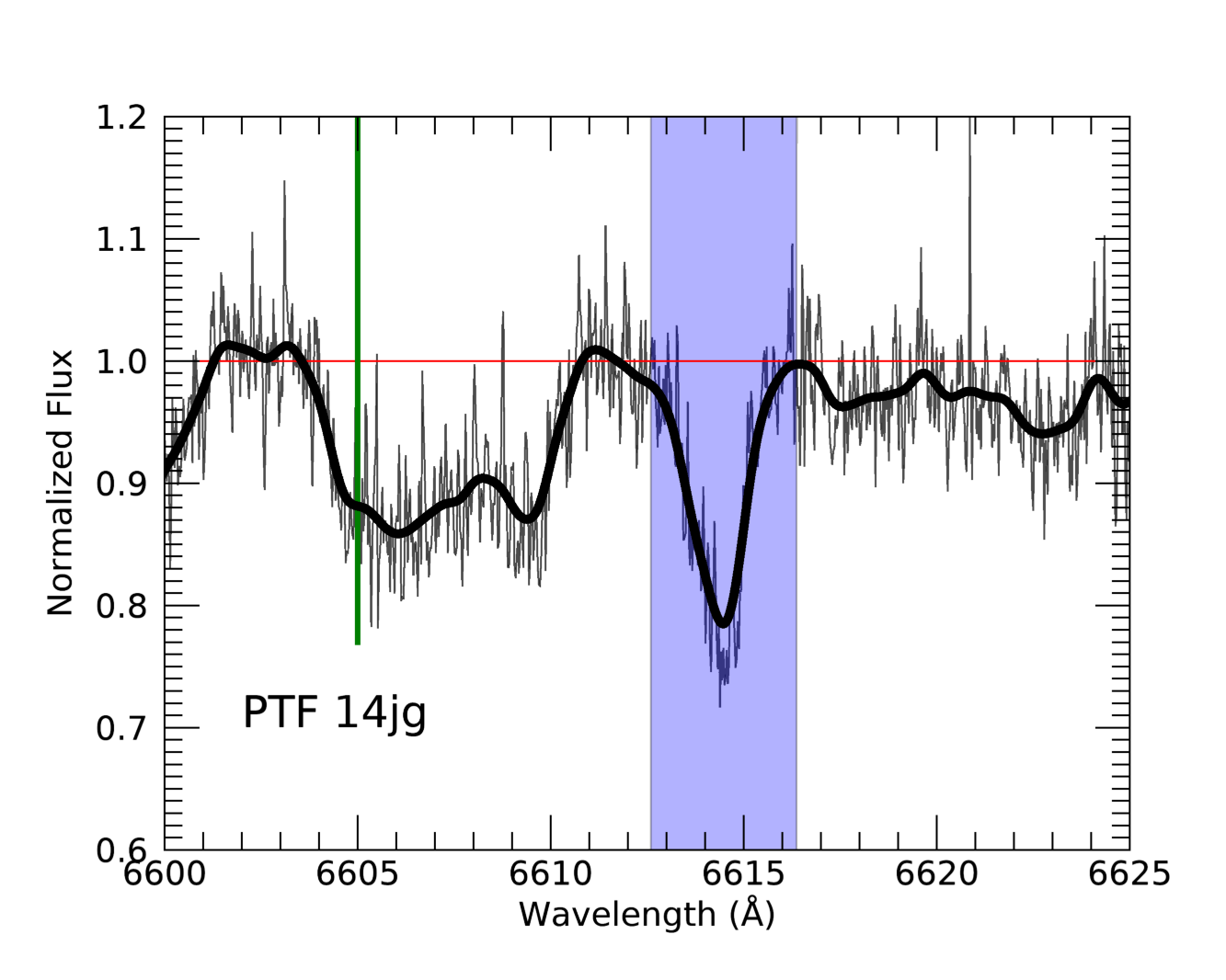}
        \includegraphics[width=0.22\linewidth]{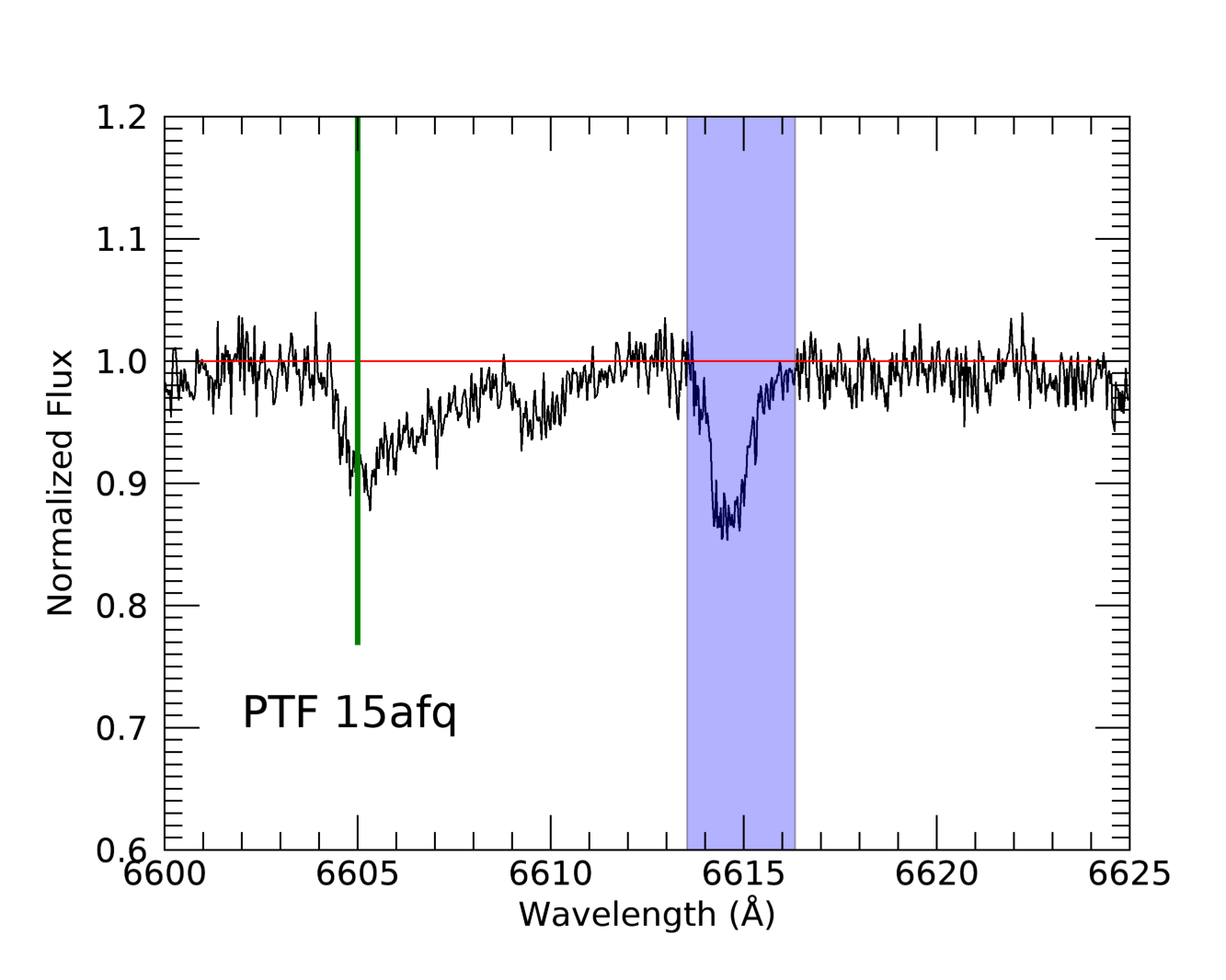}
        \includegraphics[width=0.22\linewidth]{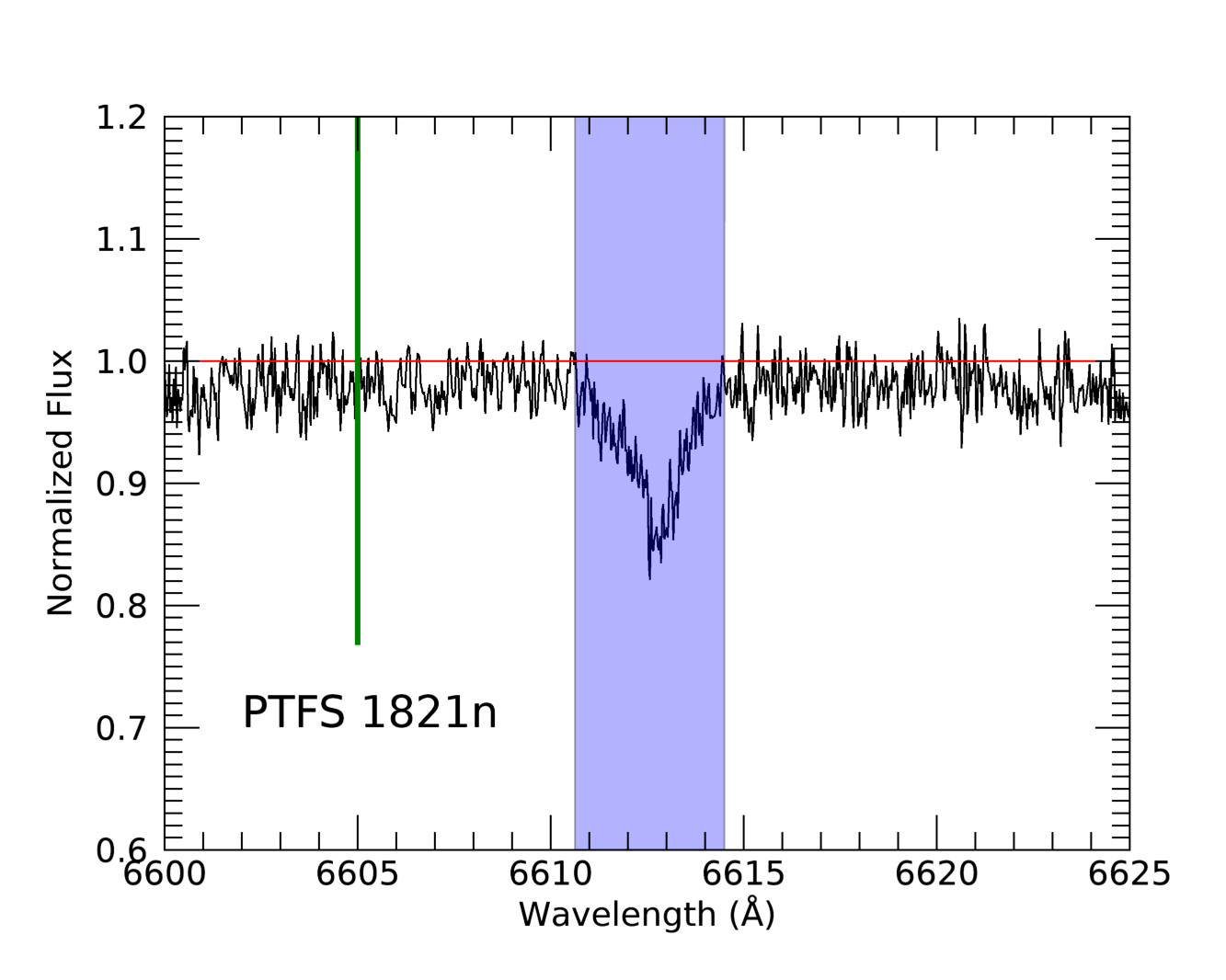}
    \includegraphics[width=0.22\linewidth]{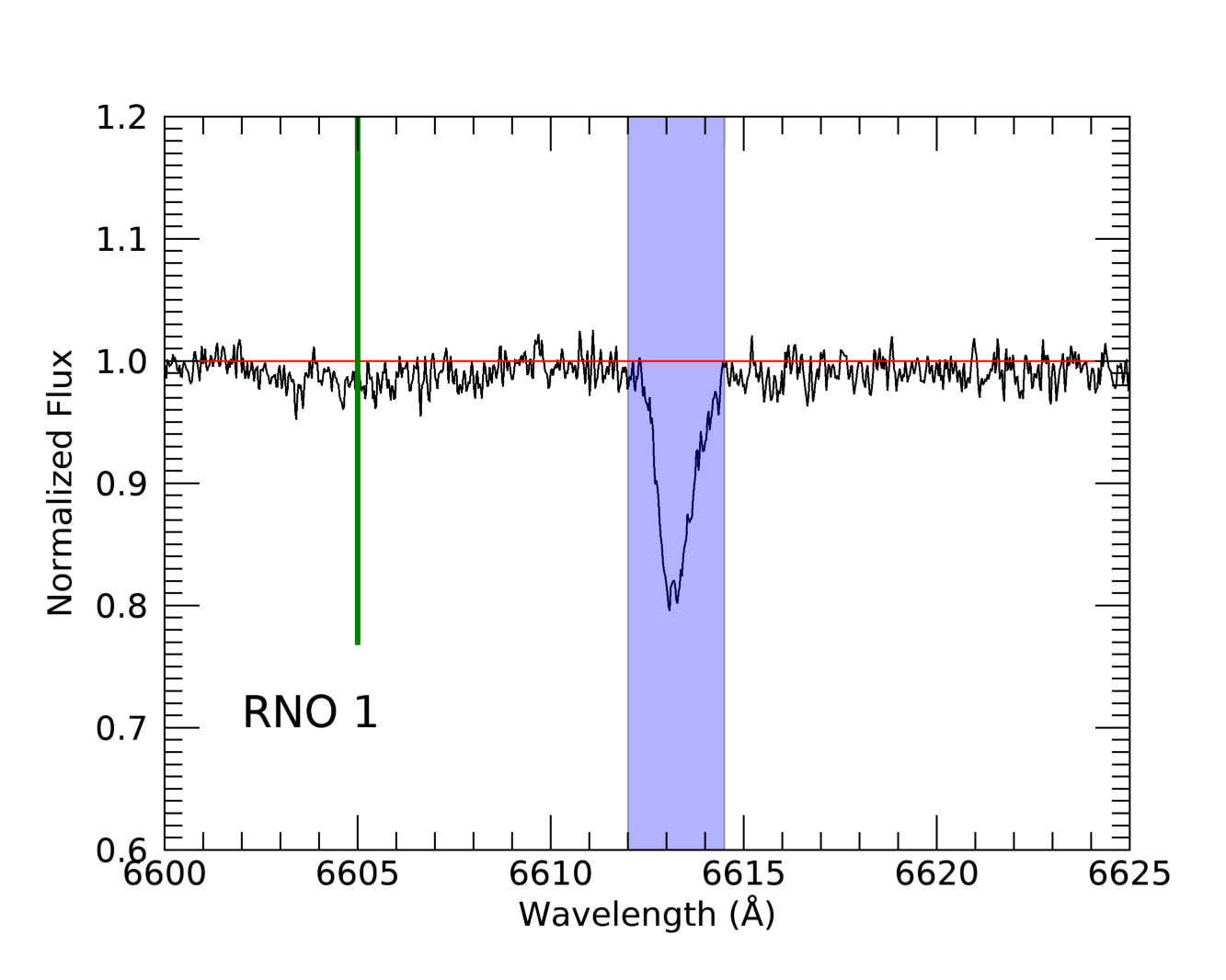}
        \includegraphics[width=0.22\linewidth]{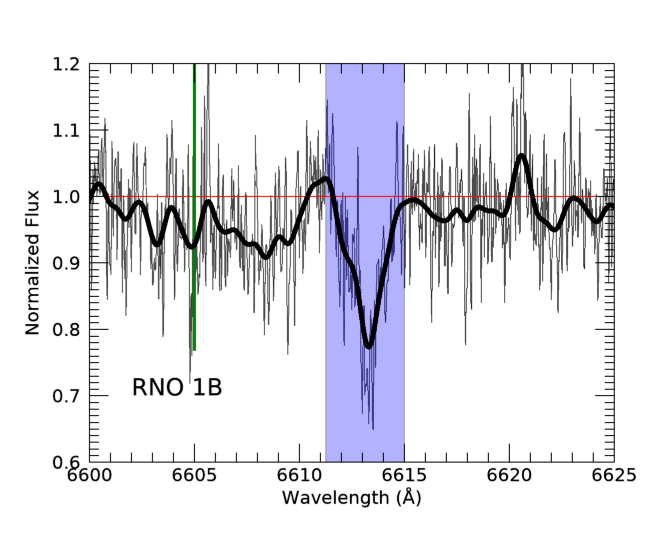}
        \includegraphics[width=0.22\linewidth]{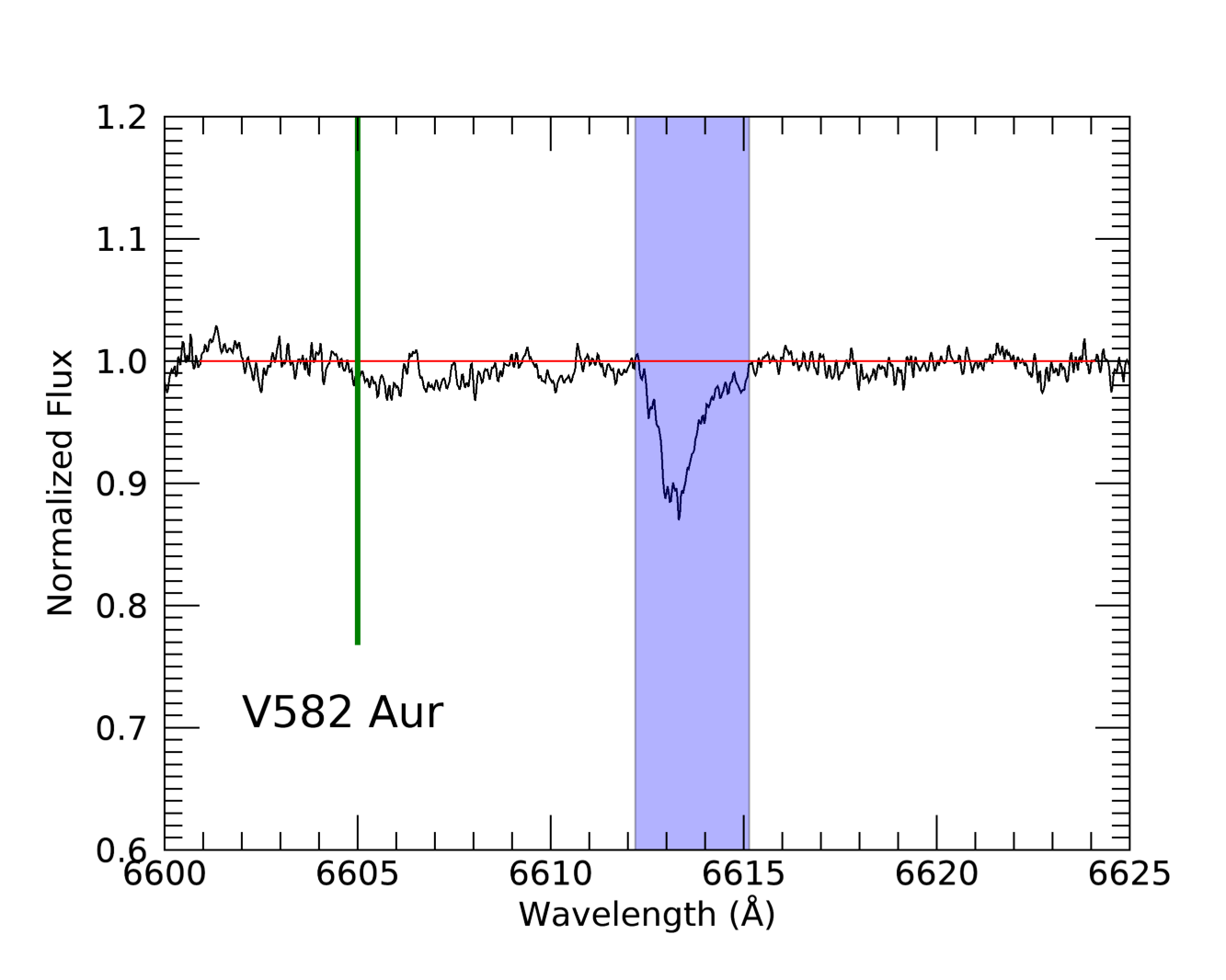}
    \includegraphics[width=0.22\linewidth]{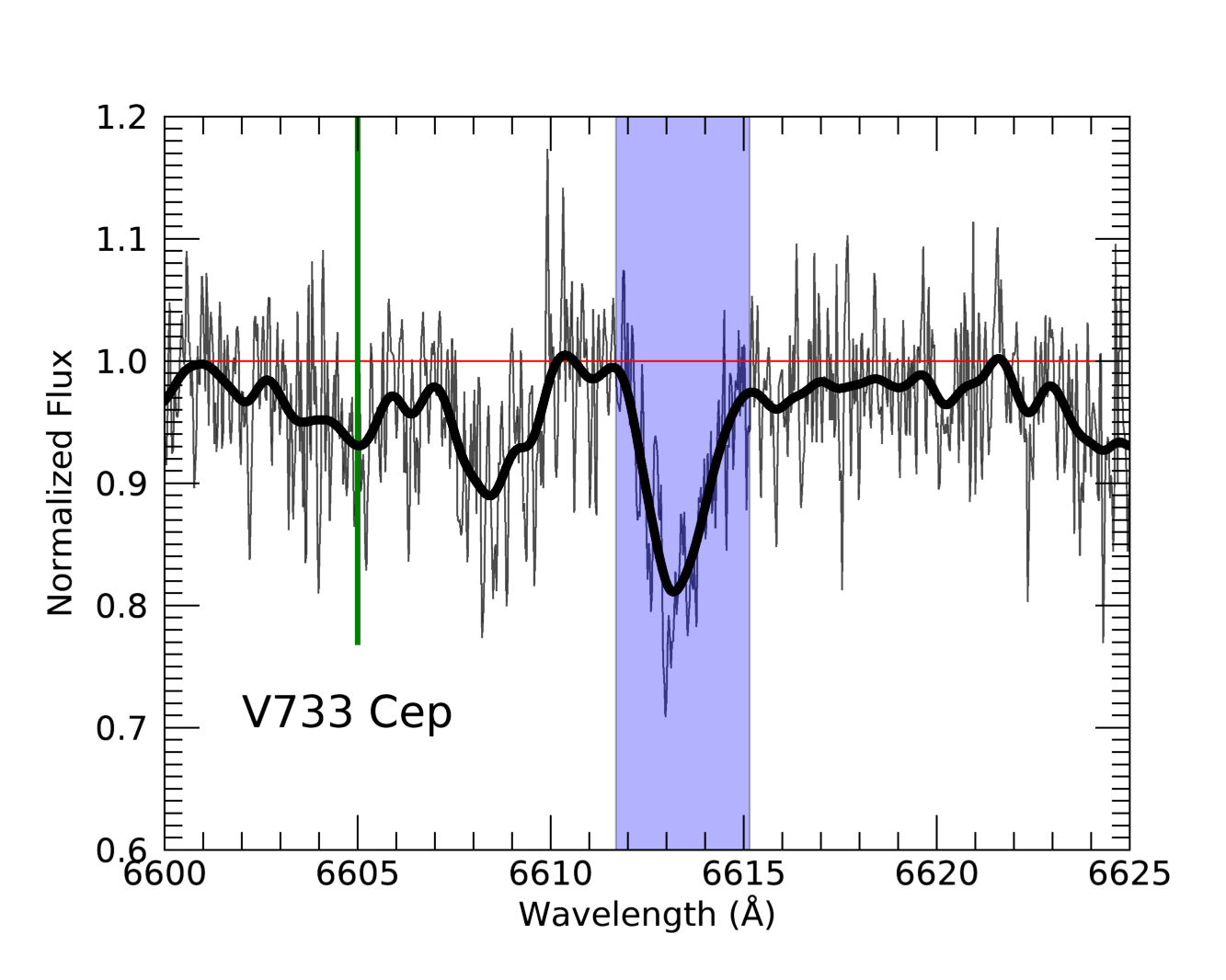}
        \includegraphics[width=0.22\linewidth]{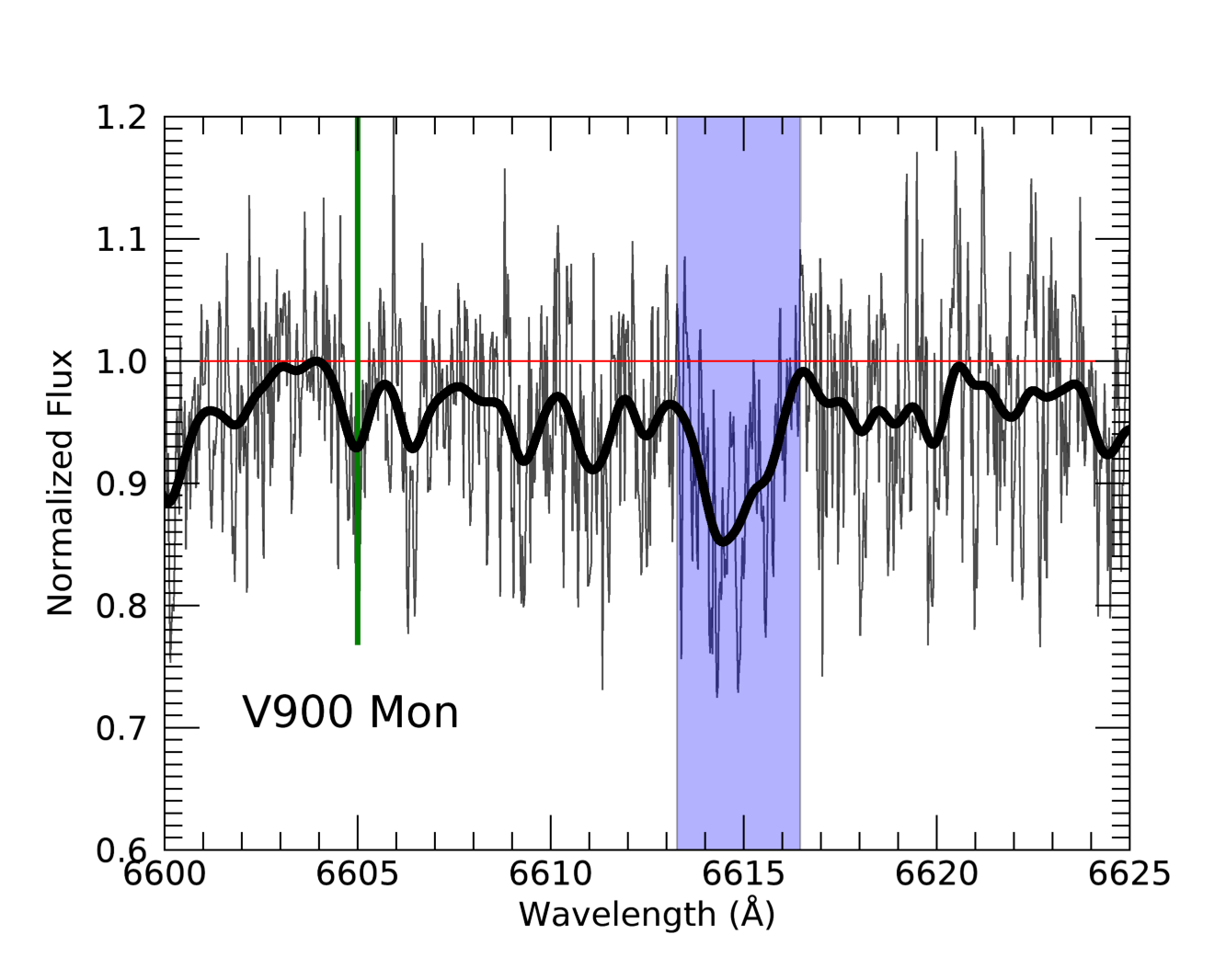}
        \includegraphics[width=0.22\linewidth]{v1057cyg_6614IntegrationSample.png}
    \includegraphics[width=0.22\linewidth]{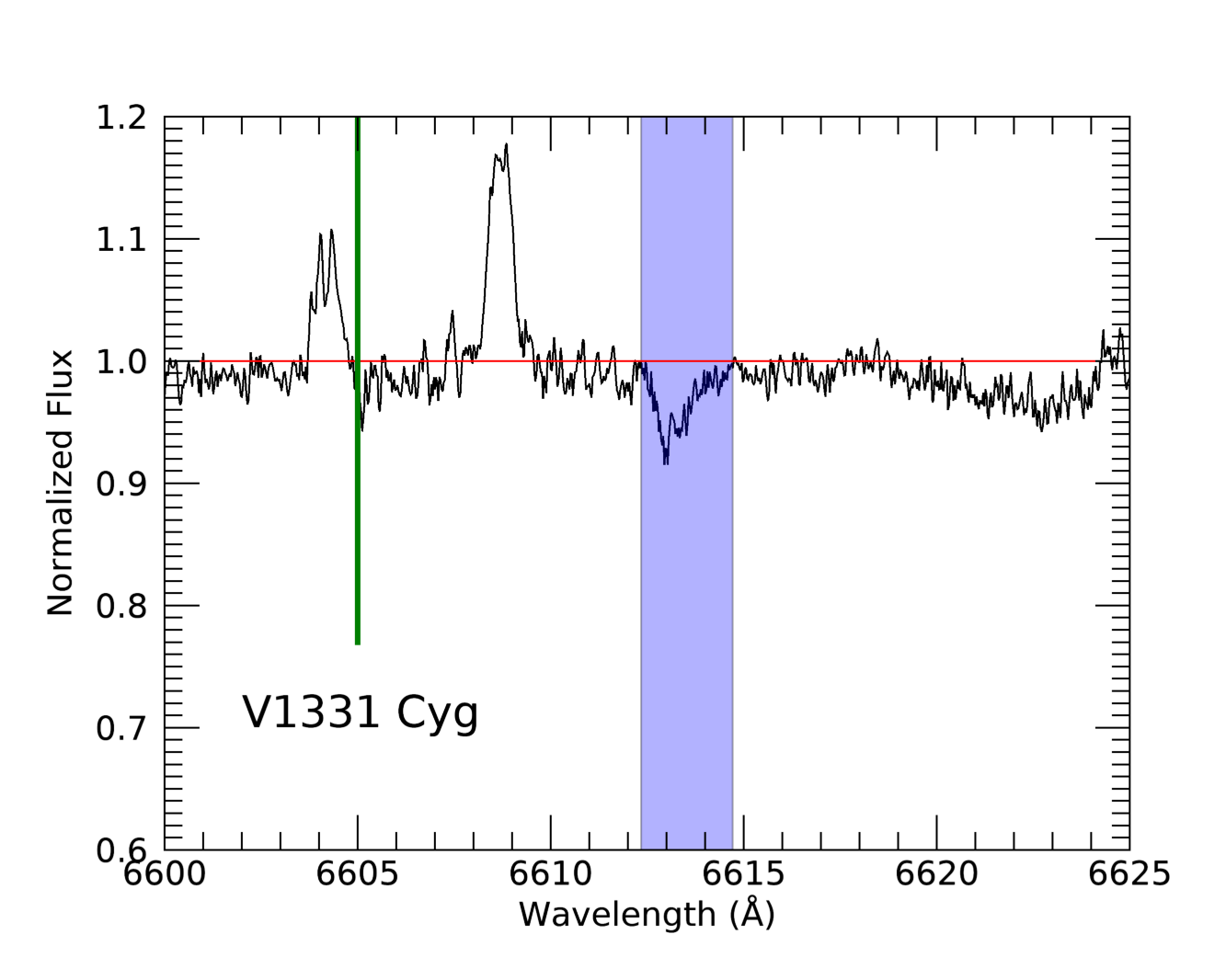}
        \includegraphics[width=0.22\linewidth]{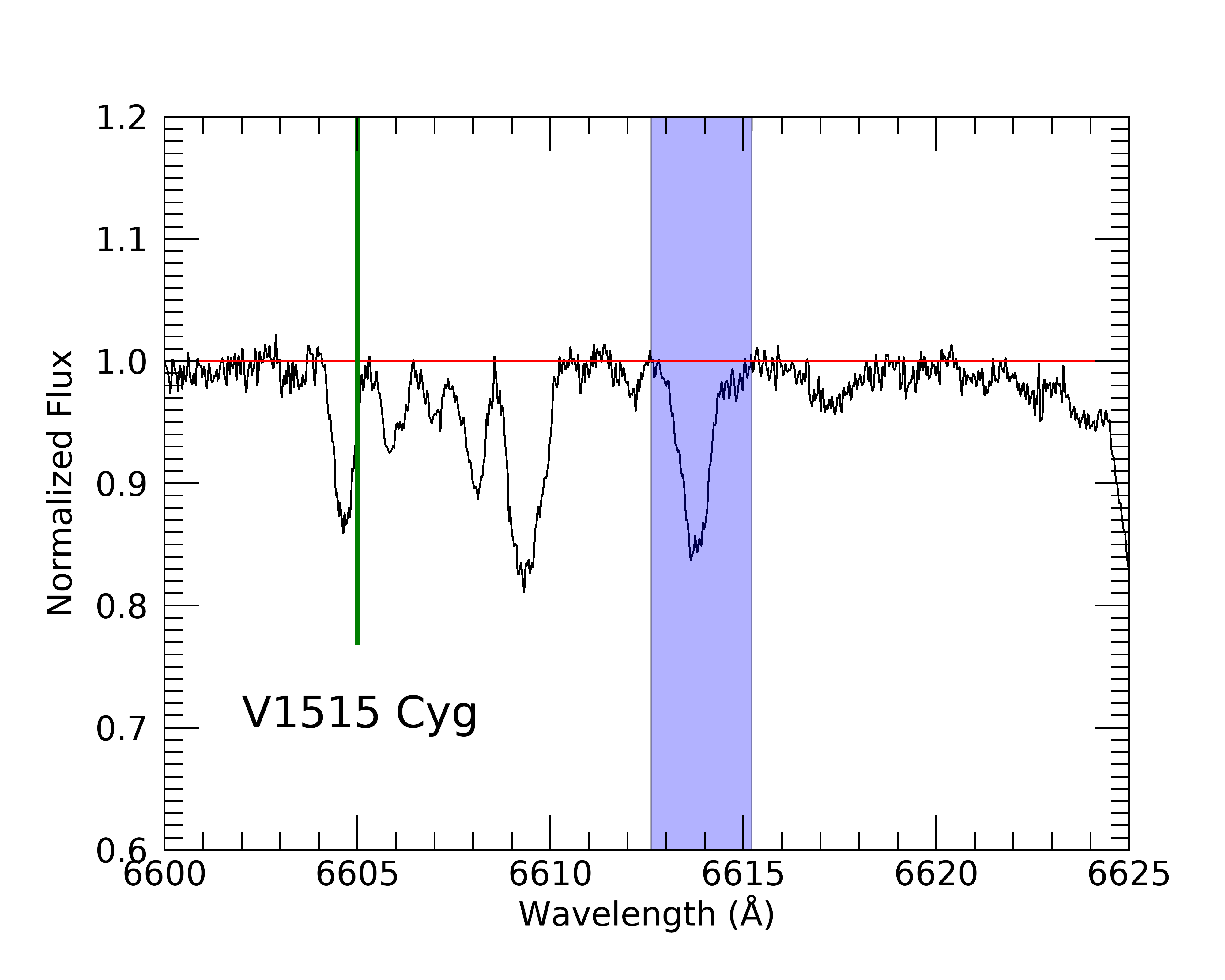}
        \includegraphics[width=0.22\linewidth]{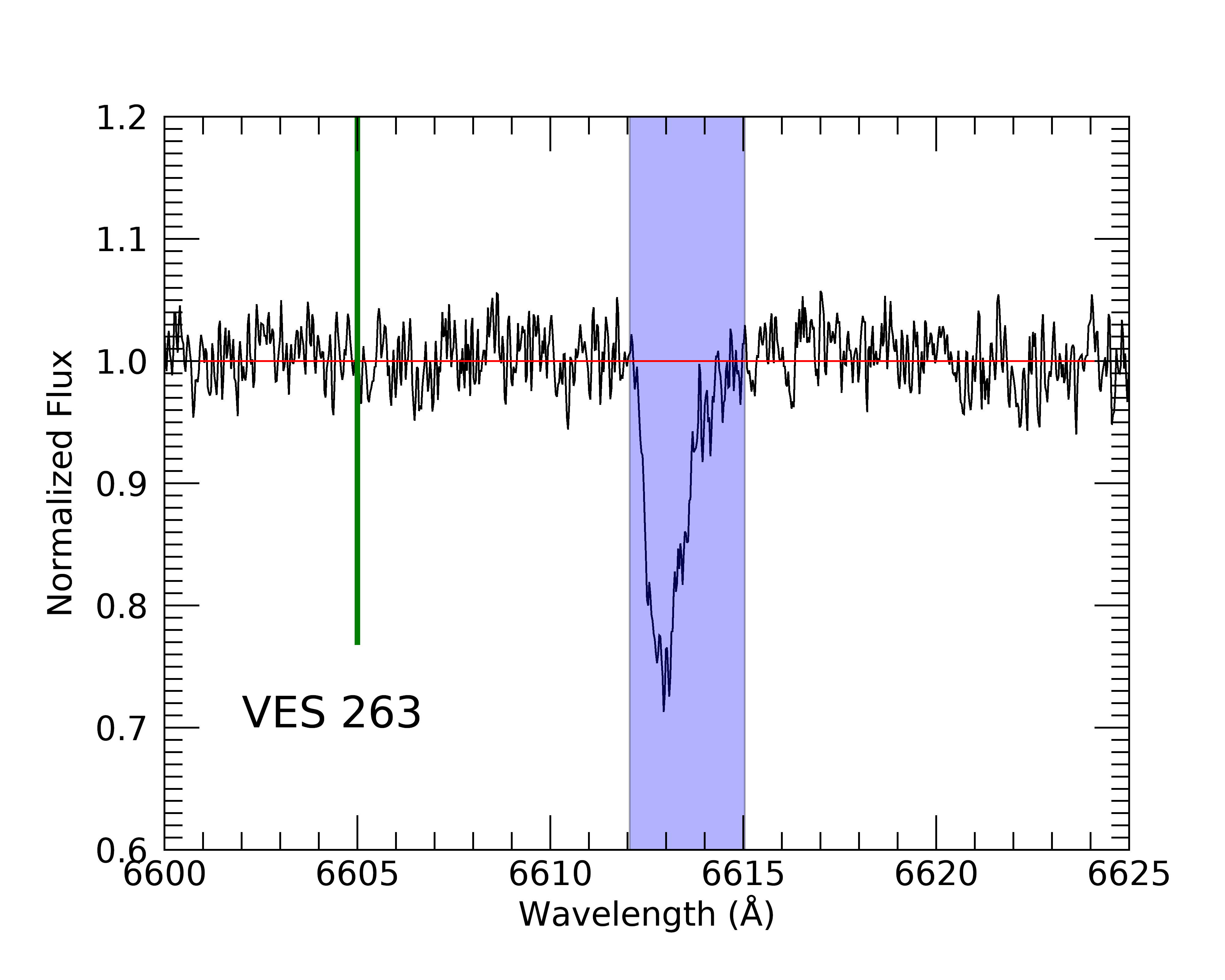}
    \includegraphics[width=0.22\linewidth]{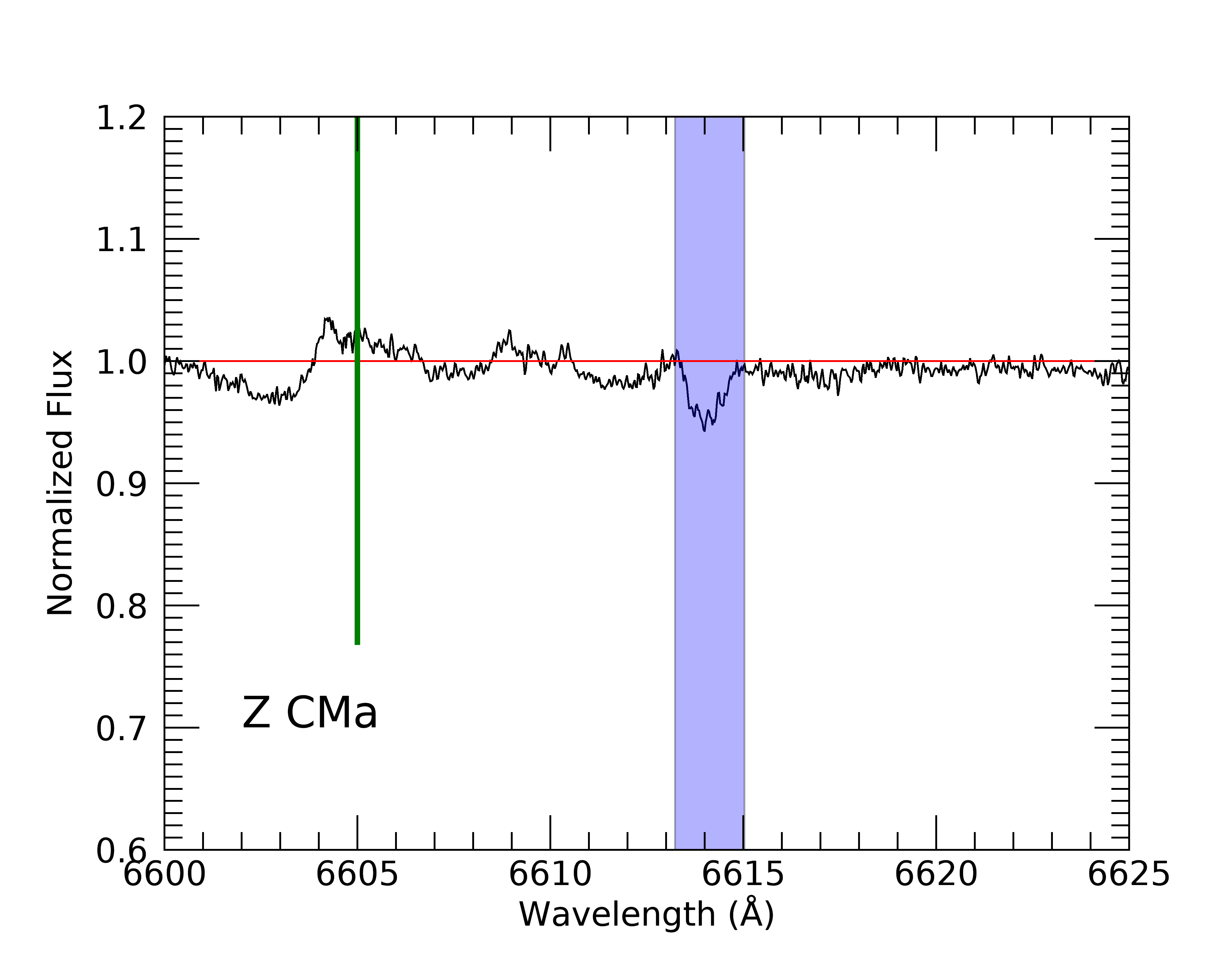}
    \caption{The $\lambda$6614 DIB in our sample. The blue region shows the region of integration for the equivalent width calculation. The red line lies at 1.0 for reference. For the noisiest spectra (Gaia 17bpi, Gaia 19ajj, PTF 14jg, RNO 1B, V733 Cep, V900 Mon) a $\sigma=10$ pixels Gaussian-smoothing of the spectrum is shown in black, with the raw data shown in grey. The green vertical line marks the center of the $\lambda$6605 band used for the stellar contamination correction.}
    \label{fig:eqWIntegration6614s}
\end{figure}

\section{The 5797 \AA\ DIB In Our Sample}\label{sec:dib5797}
In addition to the $\lambda$5780 and $\lambda$6614 DIBs, we identified the $\lambda$5797 feature consistently in our targets. However, we omit this DIB from our primary analysis because it unfortunately is located at the edge of a spectral order, where the SNR is low and the continuum normalization is less reliable. This DIB has also been shown to be more sensitive to the difference in ISM conditions along $\sigma$ or $\zeta$ sightlines, as demonstrated in \citet{kos_properties_2013} and \citet{vos_diffuse_2011}, which could add a source of uncertainty in the conversion to $E(B-V)$. We provide our measurements of $W_{DIB}(5797)$ below, and discuss their relationship to our other measurements. We also provide two more $W_{DIB}-E(B-V)$ relations which we omitted from the text above for the sake of clarity. 

\begin{figure}[!htb]
    \centering
    \includegraphics[width=0.485\linewidth]{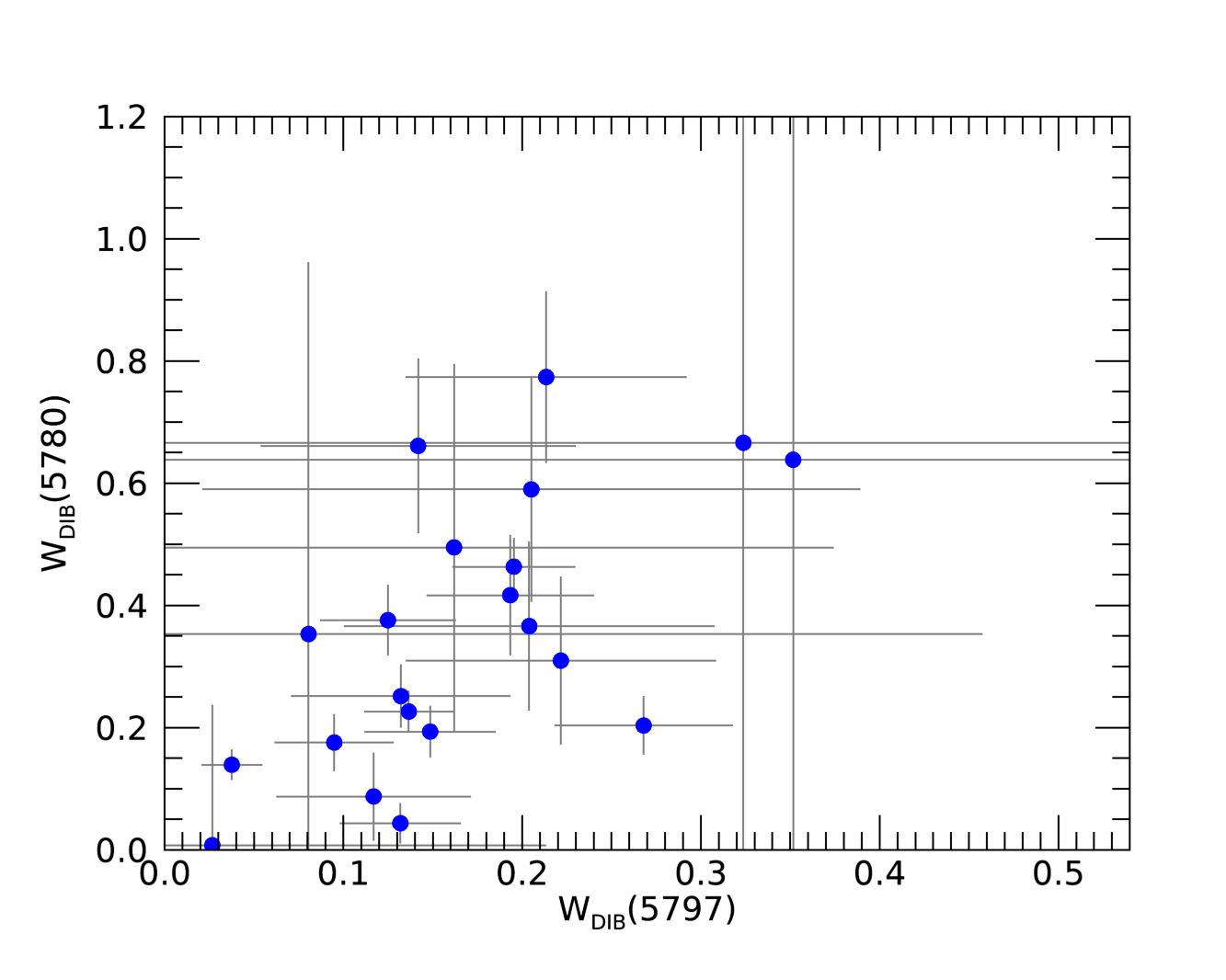}
    \includegraphics[width=0.485\linewidth]{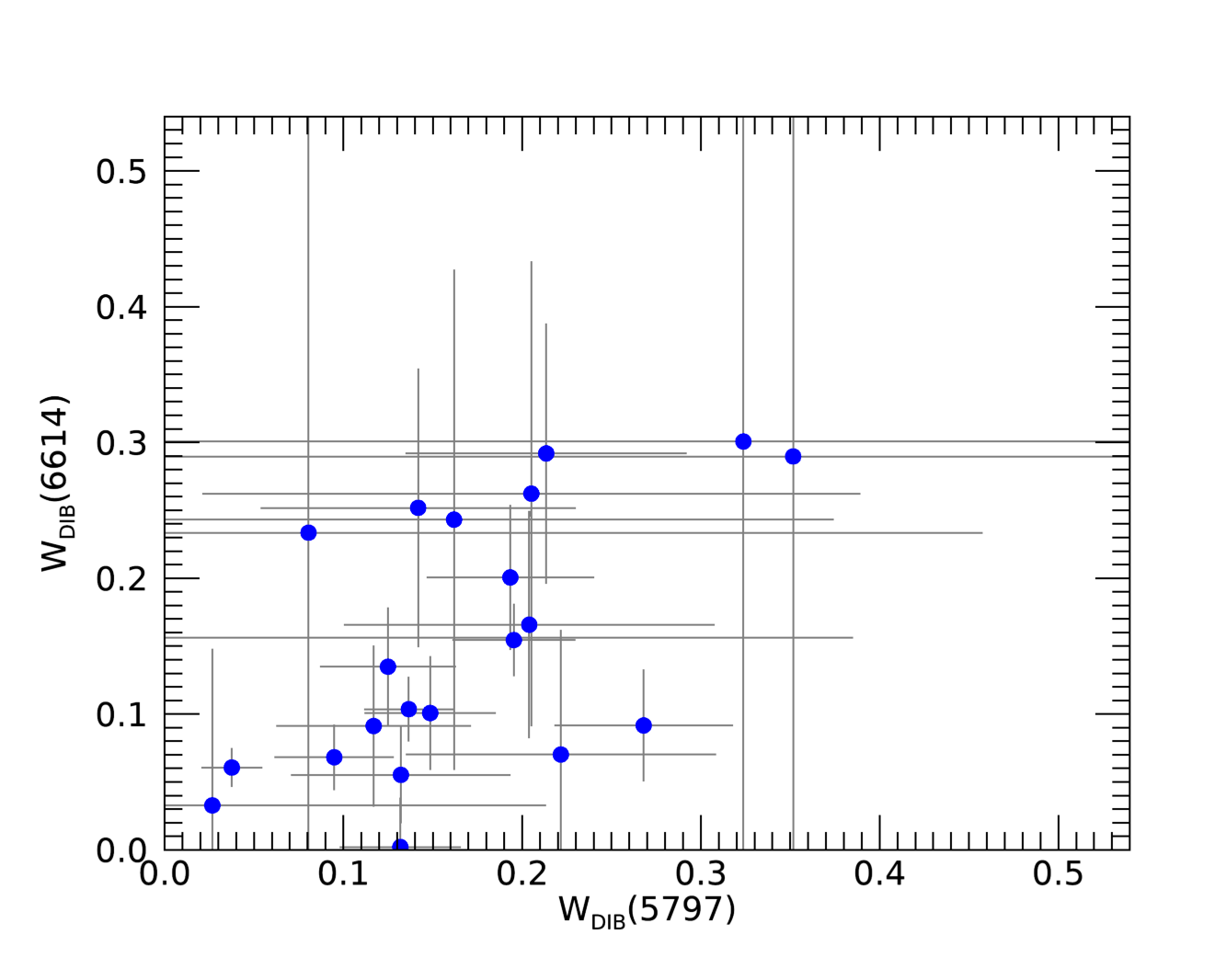}
    \caption{The $\lambda$5780 (left) and $\lambda$6614 (right) DIBs equivalent width measurements plotted against the equivalent width measurements of the $\lambda$5797 DIB. All DIBs measurements have been corrected for stellar contamination following the procedure in Section \ref{sec:StelContam}. Uncertainties on the measurements are shown in grey.}
    \label{fig:5797Corr}
\end{figure}

We computed the equivalent widths of the $\lambda$5797 DIB following the procedure described in Section \ref{sec:eqws_meas}. The stellar contamination correction is applied using the ratio of the width of the $\lambda$5785 line complex to that of the $\lambda$5797 line complex in our standards. We only present the stellar-contamination-corrected equivalent widths in this appendix. The lack of a nearby independent line complex is another reason for the omission of this DIB from our primary analysis. We also find that our measurements of the $\lambda$5797 DIB correlate poorly with our equivalent width measurements for the other two DIBs, as shown in Figure \ref{fig:5797Corr}, despite its typically strong correlation with the $E(B-V)$ in literature \citep{friedman_studies_2011, vos_diffuse_2011, kos_properties_2013}. This may be due to the greater uncertainty in our continuum normalization near the edges of the spectral orders and lack of an independent stellar contamination reference band. 

\begin{figure}[!htb]
    \centering
    \includegraphics[width=0.85\linewidth]{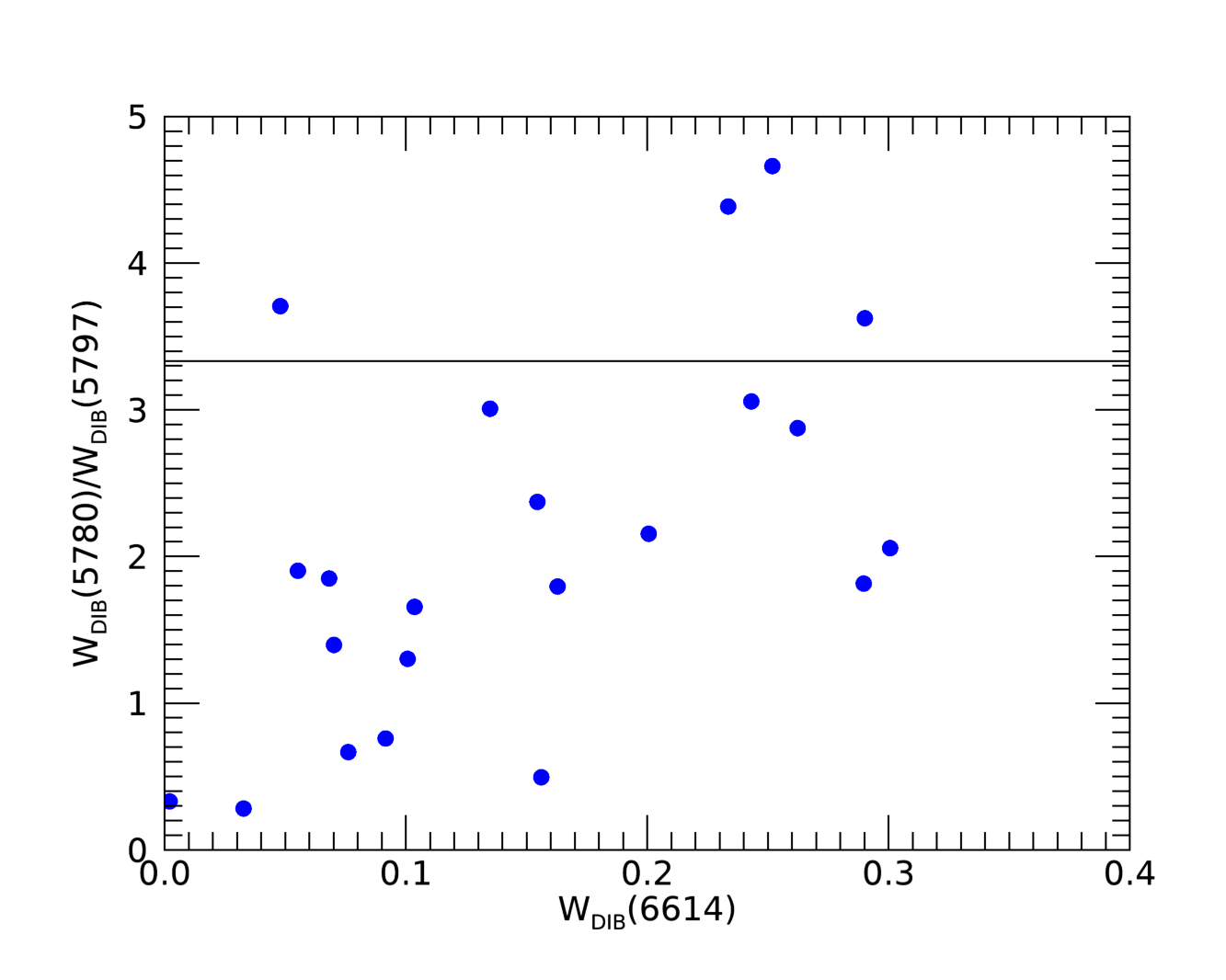}
    \caption{The $W_{5780}/W_{5797}$ ratios for our targets plotted against $W_{6614}$ as an assumed measure of $E(B-V)$. The black horizontal line at $1/0.3$ marks the division between $\sigma$ and $\zeta$ sightlines adopted in \citet{kos_properties_2013}.}
    \label{fig:SigZetRatios}
\end{figure}

Although we do not use the $\lambda$5797 DIB for our extinction measurements, we explore here the ratio of the $\lambda$5780 and $\lambda$5797 equivalent widths, shown in Figure \ref{fig:SigZetRatios}. This ratio is often used in DIBs studies to distinguish $\sigma$ from $\zeta$ sightlines. The majority of our targets lie below the $W_{DIB}(5780)/W_{DIB}(5797)= 3.3$ line, indicating they primarily lie along $\zeta$ sightlines. This is expected for the YSOs because they are found in well-shielded environments, favorable for the more UV sensitive $\lambda$5797 carrier. Also, the lower $W_{DIB}(5780)/W_{DIB}(5797)$ ratio in most of our targets may indicate a greater molecular hydrogen fraction along those sightlines \citep{fan_behavior_2017}. This is expected for star-forming environments and the molecular cores from which stars form. Although some targets lie just above the dividing line we present, the transition between $\sigma$ and $\zeta$ sightlines is smooth and continuous, so we treat those targets as lying along $\zeta$ sightlines in our primary analysis.


We also report here the best-fit model for $\sigma$ sightlines. The model is constructed from the literature data following the description in Section \ref{sec:Extincs}, but taking the $W_{DIB}$ and $E(B-V)$ values for targets with $W_{5780}/W_{5797}<3.3$. 

The resulting $\sigma$ sightline models are: 
\begin{equation} \label{eq:5780_model_sigma}
    E(B-V) = \left(-0.030 \pm 0.005 \right) + \left[(1.967 \pm 0.295) \times W_{DIB}(5780) \right]
\end{equation}
and
\begin{equation} \label{eq:6614_model_sigma}
        E(B-V) = \left( -0.041 \pm 0.009 \right) + \left[(4.479 \pm 1.03) \times W_{DIB}(6614) \right].
\end{equation}

As a helpful tool for others interested in using the $\lambda$5780 and $\lambda$6614 DIBs to compute $E(B-V)$ but are uncertain of the nature of the sightlines through which they are observing, we also provide the best-fit model to all of the data (Eqs \ref{eq:5780_model_all} and \ref{eq:6614_model_all}). 

\begin{equation} \label{eq:5780_model_all}
    E(B-V) = \left(0.005 \pm 0.001 \right) + \left[(1.961 \pm 0.412) \times W_{DIB}(5780) \right]
\end{equation}
and
\begin{equation} \label{eq:6614_model_all}
        E(B-V) = \left( 0.065 \pm 0.018 \right) + \left[(3.943 \pm 1.104) \times W_{DIB}(6614) \right]
\end{equation}

\section{The Spectral Type of RNO 1} \label{appendix:rno1}

In this Appendix, we discuss the previously assumed F5III spectral type of RNO 1 \citep{staude_rno_1991} and propose an earlier spectral type of A8 to F0, with a luminosity class between V and III, though the latter is hard to constrain.

To better identify the spectral type of RNO 1, we compare it with the spectra of HD 23326 (F3V), HD 144668 (A7III), and HD 121784 (A7V). We chose two A7 spectra to identify the gravity and luminosity class of RNO 1. We obtained the spectra of HD 144668 (Proposal 05BC17, PI Wade) and HD 121784 (Proposal 14AH14, PI Ansdell) from the Canada-France-Hawaii Telescope (CFHT) ESPaDOnS archive. The targets were observed on 24 August 2005 and 20 June 2014 respectively, using the ESPaDOnS spectropolarimeter. Both spectra were reduced at CFHT using the software package Libre-Esprit and we use the extracted Stokes I profiles. 

\begin{figure}
    \centering
    \includegraphics[width=0.85\linewidth]{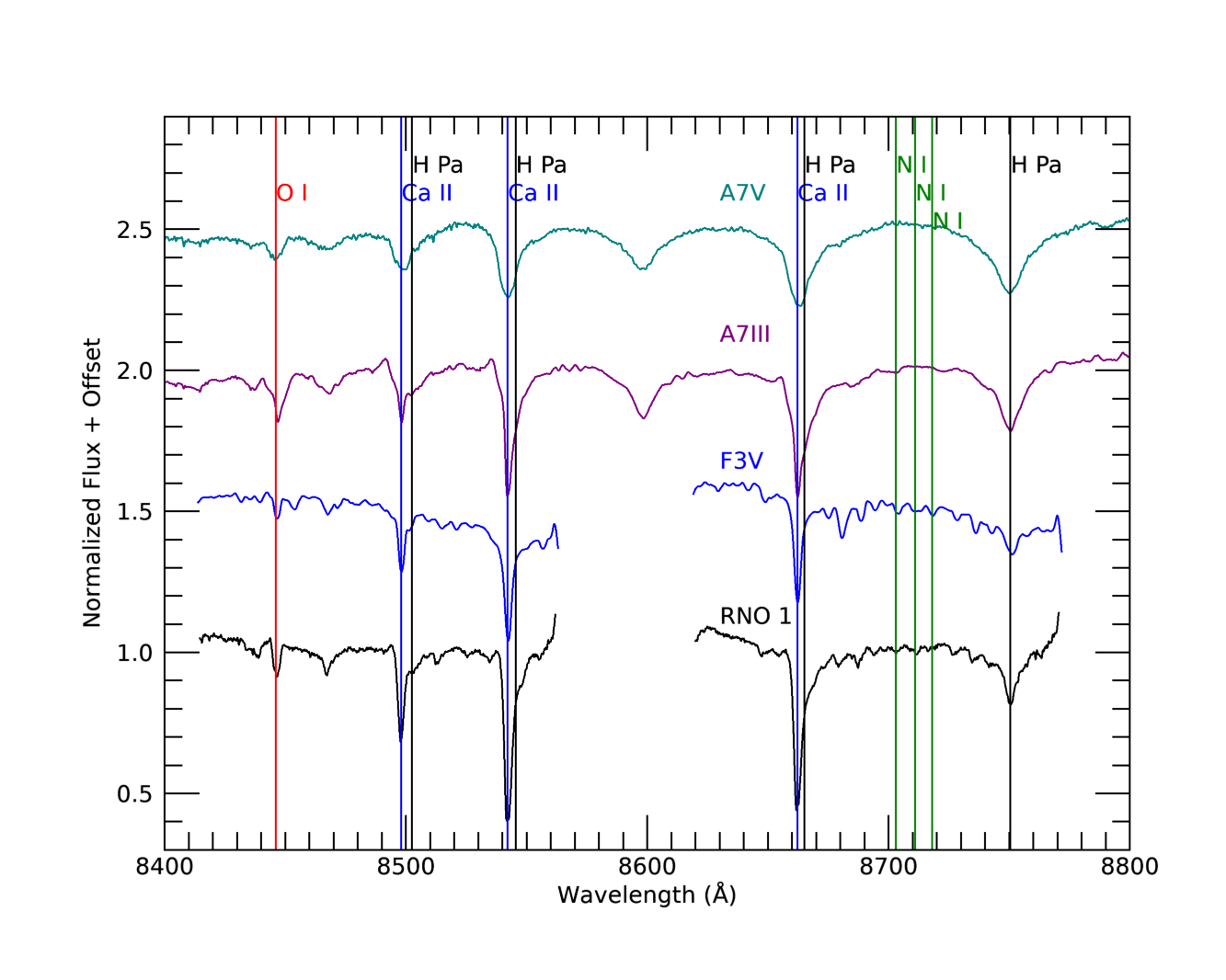}
    \caption{The 8400 - 8800 \AA\ wavelength range containing Hydrogen Paschen lines blended with \ion{Ca}{2}, \ion{O}{1}, and a \ion{N}{1} triplet. F3V, A7V, and A7III standard spectra are provided for reference. Notice the wing broadening of the Paschen lines with gravity, indicating RNO 1 is lower gravity than a main sequence star.}
    \label{fig:RNO1_PaLines}
\end{figure}

We find the presence of Paschen lines in the 8400\AA--8800\AA\ range  and their contamination of the \ion{Ca}{2} ``triplet" lines
indicates a higher temperature than a mid-F star. As can be seen in Figure \ref{fig:RNO1_PaLines}, the Paschen contributions are absent in the F3V standard. However, the Paschen lines remain weaker than the \ion{Ca}{2} lines, indicating a temperature cooler than an early-mid A type star.  The \ion{Ca}{2} lines also contain some possible red-shifted absorption, though it is difficult to distinguish from the Paschen line contamination. We also see \ion{N}{1} lines between 8700 \AA\ and 8720 \AA\ in the spectrum of RNO 1 and that of the F3V standard, but not in the A7 standards.

\begin{figure}
    \centering
    \includegraphics[width=0.85\linewidth]{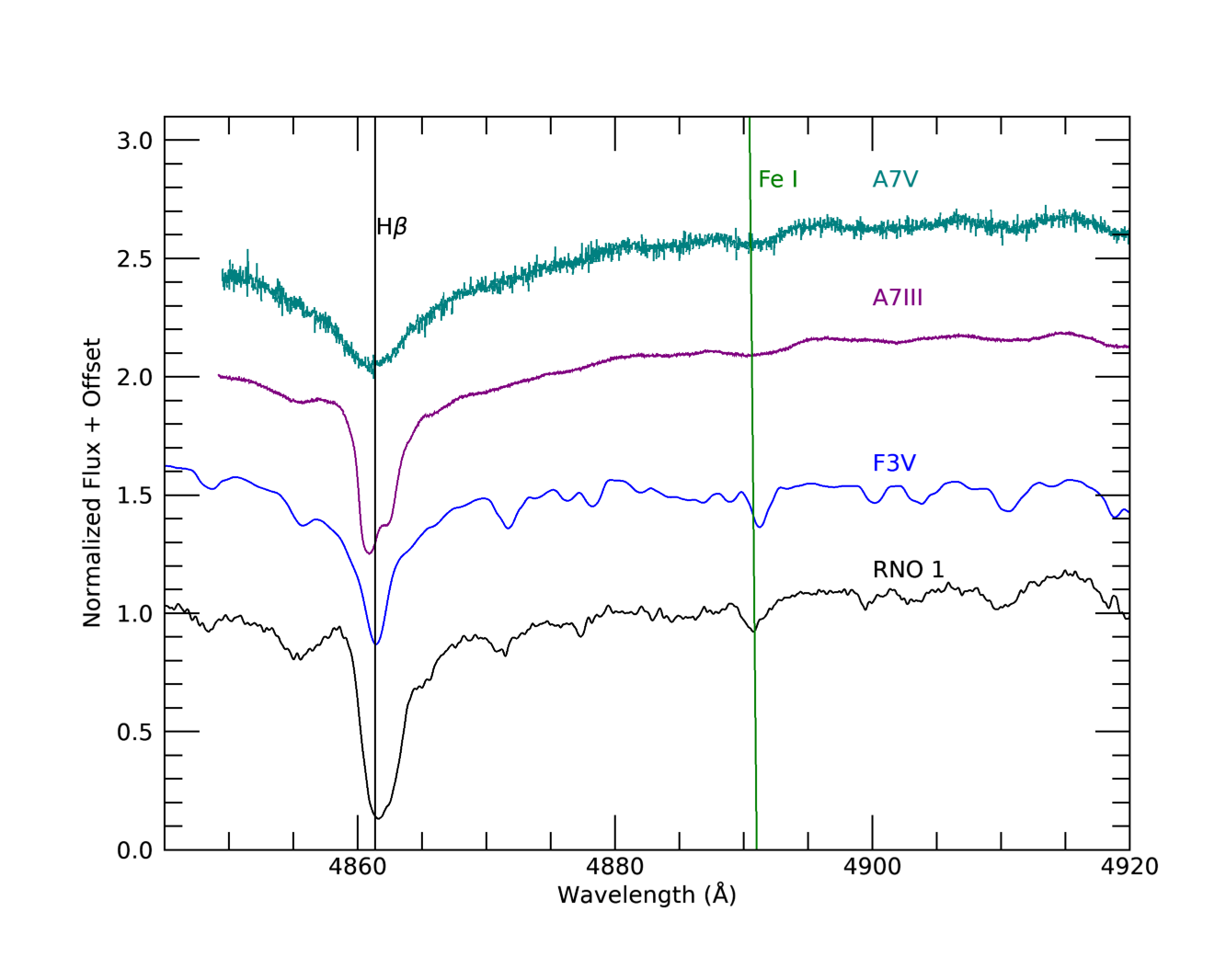}
    \caption{The 4850 - 4920 \AA\ wavelength range containing H$\beta$ line. F3V, A7V, and A7III standard spectra are provided for reference. Notice the wing broadening of the H$\beta$ line matches the trend seen in the Paschen lines in Figure \ref{fig:RNO1_PaLines}, growing broader with gravity, indicating RNO 1 is lower gravity than a main sequence star. The \ion{Fe}{1} line triplet at 4890 \AA\ more closely matches the F3V standard, indicating RNO 1 may be between a late F and early A (F0 or A8). Some redshifted absorption in the H$\beta$ line may be an infall signature. The A7III standard is an emission line object and contains an asymmetry in the H$\beta$ line.}
    \label{fig:RNO1_Hbeta}
\end{figure}

The wavelength range near H$\beta$, shown in Figure \ref{fig:RNO1_Hbeta}, also reveals a similar trend in the Hydrogen lines. We find H$\beta$ is much stronger in the spectrum of RNO 1 than in the F3V spectrum, but narrower than that of the A7V spectrum, matching the H$\beta$ line in the A7III standard well.
The lower broadening in the wings of the line in RNO 1 than in the dwarf stars, indicates RNO 1 may be lower gravity (perhaps III or IV). The H$\beta$ line also contains significant red-shifted absorption, a possible infall signature that matches the red-shifted absorption in the \ion{Ca}{2} lines. An \ion{Fe}{1} line at 4886 \AA\ also presents as a possible temperature indicator, appearing much more strongly in the spectrum of RNO 1 and the F3V standard than in the A standards. 

\begin{figure}
    \centering
    \includegraphics[width=0.85\linewidth]{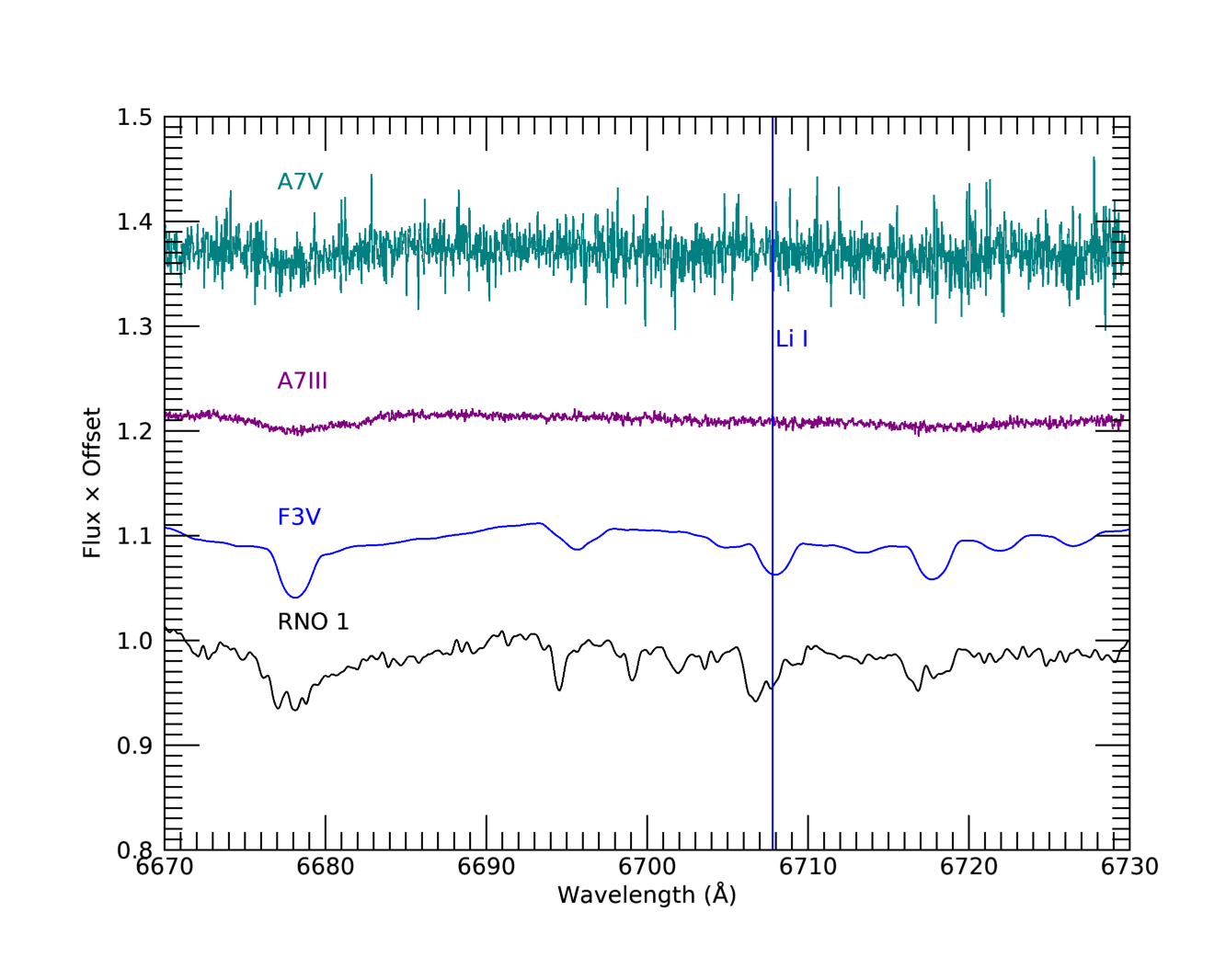}
    \caption{The 6670 - 6730 \AA\ wavelength range containing the 6707.76 and 6707.91 \ion{Li}{1} lines. F3V, A7V, and A7III standard spectra are provided for reference.}
    \label{fig:RNO1_Li}
\end{figure}

We also identify the \ion{Li}{1} 6707.8 \AA\ doublet in the spectrum of RNO 1, as shown in Figure \ref{fig:RNO1_Li}. The line is relatively strong, with an equivalent width of 0.18 \AA\, indicating the source may be a young pre-main sequence object. The line also appears in the spectrum of the F3V standard, which is a young Pleiades member, but is not apparent in the spectra of the A standards which are field stars and thus may be of comparable age. 

\listofchanges

\end{document}